\documentclass{article}

\usepackage{latexsym}
\usepackage{graphicx}
\usepackage{caption}
\usepackage{natbib}

\newcommand{\ket}[1]{|#1 \rangle}
\newcommand{\braket}[2]{\langle #1 |  #2 \rangle}
\newcommand{\braOket}[3]{\langle#1 |#2|  #3 \rangle}

\newcommand{\intsum}[1]{\sum_{#1}  \! \! \! \! \! \! \! \! \int }

\begin{document}


\title{Introduction to attosecond time-delays \\ in photoionization}
\date{}

\author{
J.~M.~Dahlstr\"om \\
\small Department of Physics, Stockholm University, \\
\small AlbaNova University Center, SE-10691 Stockholm, Sweden \\
A.~L'Huillier \\
\small Department of Physics, Lund University, \\
\small P.O. Box 118, 22100 Lund, Sweden \\
A.~Maquet \\
\small UPMC Universit\'e Paris 6, UMR 7614, \\ 
\small Laboratoire de Chimie Physique-Mati\`ere et Rayonnement, \\
\small 11 rue Pierre et Marie Curie, 75231 Paris Cedex 05, France \\
\small CNRS, UMR 7614, LCPMR, Paris, France
} 



\maketitle

\begin{abstract}
This tutorial presents an introduction to the interaction of light and matter on the attosecond timescale. Our aim is to detail the theoretical description of ultra-short time-delays, and to relate these to the phase of extreme ultraviolet (XUV) light pulses and to the asymptotic phase-shifts of photoelectron wave packets. Special emphasis is laid on time-delay experiments, where attosecond XUV pulses are used to photoionize target atoms at well-defined times, followed by a probing process in real time by a phase-locked, infrared laser field. In this way, the laser field serves as a ``clock'' to monitor the ionization event, but the observable delays do {\it not} correspond directly to the delay associated with single-photon ionization. Instead, a significant part of the observed delay originates from a measurement induced process, which obscures the single-photon ionization dynamics. This artifact is traced back to a phase-shift of the above-threshold ionization  transition matrix element, which we call the {\it continuum--continuum phase}.  
It arises due to the laser-stimulated transitions between Coulomb continuum states. As we shall show here, these measurement-induced effects can be separated from the single-photon ionization process, using analytical expressions of universal character, so that eventually the attosecond time-delays in photoionization can be accessed. 

\end{abstract}






\section*{List of abbreviations}

\begin{tabular}{ll}
APT & Attosecond pulse train \\
ATI & Above-threshold ionization \\
CEP & Carrier Envelope Phase \\
FROG- & Frequency-resolved optical gating- \\
CRAB & Complete Reconstruction of Attosecond Bursts \\
GD & Group delay \\
HHG & High-order harmonic generation \\
IR & Infrared \\
MBPT & Many-body perturbation theory \\	
RABITT & Resolution of attosecond beating \\
& by interference of two-photon transitions \\
RPA & Random-phase approximation \\ 
SAE & Single-active electron \\
SAP & Single-attosecond pulse \\
SFA & Strong-field approximation \\
SPA & Saddle-point approximation \\
TDSE & Time-dependent Schr\"odinger equation \\
TOF & Time-of-flight \\
VMI & Velocity--map imaging \\
WKB & Wentzel--Kramers--Brillouin \\
XUV & Extreme ultraviolet\\
\end{tabular}

\section{Introduction}
\label{sec:intro}


Great advances in experimental physics are being made using attosecond pulses of extreme ultraviolet (XUV) radiation, 
where it is now possible to initiate, control and probe electron dynamics in atoms and molecules in real time. 
A natural and perhaps naive question then arises: 
Is it possible to use attosecond pulses to measure the time it takes for a photoelectron to be ionized? 
It is well-known from fundamental principles in quantum theory that time is not a direct observable quantity \cite{Pauli}. 
Here, our aim is to discuss what temporal aspects of photoionization {\it can} be measured 
in state-of-the-art experiments based on attosecond XUV pulses and phase-locked infrared (IR) probe fields. 
This tutorial serves as an introduction to the theoretical 
description of radiative processes taking place in extremely short time intervals, 
with durations comparable to the temporal response of outer-shell bound electrons within atoms or molecules. 
An interesting point is that the theoretical background provides a common framework for XUV pulses 
that propagate through dispersive media, and for non-relativistic photoelectron wave packets 
that escape from the atomic potential. 
We wish to stress the similarities and differences between these two types of wave packets 
and to discuss the associated light-matter interactions occurring on the attosecond timescale.
%
To this end, we shall derive
the way in which small perturbations in the medium affect the propagation of the light pulses and 
we will detail the present ways of probing photoionization of neutral species in real time. 
In doing this, special attention must be given to the case of long-range Coulomb potentials, 
which are important for describing electron motion in photoionization.  
The theoretical methods used here span different disciplines ranging from ultra-fast optics and non-linear optics, 
to scattering theory, atomic physics and strong-field physics. 

\subsection*{Outline of Tutorial}

The content of this tutorial is divided into seven sections.
In Sec.~\ref{sec:intro} we introduce unfamiliar readers to ``attophysics'' 
and we present a short overview of the high-order harmonic generation process which is used to generate attosecond XUV pulses. 
In Sec.~\ref{sec:wplight} we review the role of the spectral phase for the propagation of XUV wave packets.  
Here, we use the so-called ``group delay'' to describe the coherent superposition of monochromatic waves in the time-domain.
Similarly, in Sec.~\ref{sec:wpelectron} we investigate the intrinsic phase of photoelectrons 
and the corresponding time-delay of electron wave packets with special attention to the long-range Coulomb potential. 
In Sec.~\ref{sec:photoionization} we perform time-dependent perturbation calculations for the creation 
and propagation of photoelectron wave packets generated by an attosecond XUV pulse.  
In Sec.~\ref{sec:STPT} we review the influence of the laser-probe field 
by performing time-dependent perturbation calculations to second order in the interaction with the light fields.
In Sec.~\ref{sec:expobs} we discuss the state-of-the art experimental efforts to measure attosecond time-delays in laser-assisted photoionization.
Finally, in Section~\ref{sec:conclusion} we present our conclusions and outlook.

\subsection{Overview of attosecond physics}
\label{sec:overview}
This section provides an ``ultra-short'' overview of the historical development and of the key concepts in attosecond physics. 
High-order Harmonic Generation (HHG), Attosecond Pulse Trains (APT) and Single Attosecond Pulses (SAP) will be introduced.
We refer the unfamiliar reader to Ref.~\cite{AgostiniRPP2004,CorkumNP2007,Kling2008,KrauseRMP2009} for more comprehensive reviews of attosecond physics. 

\subsubsection{High-order harmonic generation}
\label{sec:iHHG}
The work on attosecond light pulses stems from a process discovered in the late 1980s called HHG \cite{McPhersonJOSAB1987,FerrayJPB1988}. 
It was found that a broad plateau in the XUV range, 
containing a comb of almost equally strong, odd harmonics,  
could be produced by focusing an intense ultra-short IR laser pulse into a target of noble gas. 
A sketch of a typical HHG experiment is shown in Fig.~\ref{harmonics}~(a), 
where first an intense IR field interacts with a target of Ar-gas to generate high-order harmonics,  
then these harmonics propagate to a different chamber where they photoionize Ar-atoms 
so that photoelectrons are emitted. 
\begin{center}
	\includegraphics[width= 0.75\textwidth]{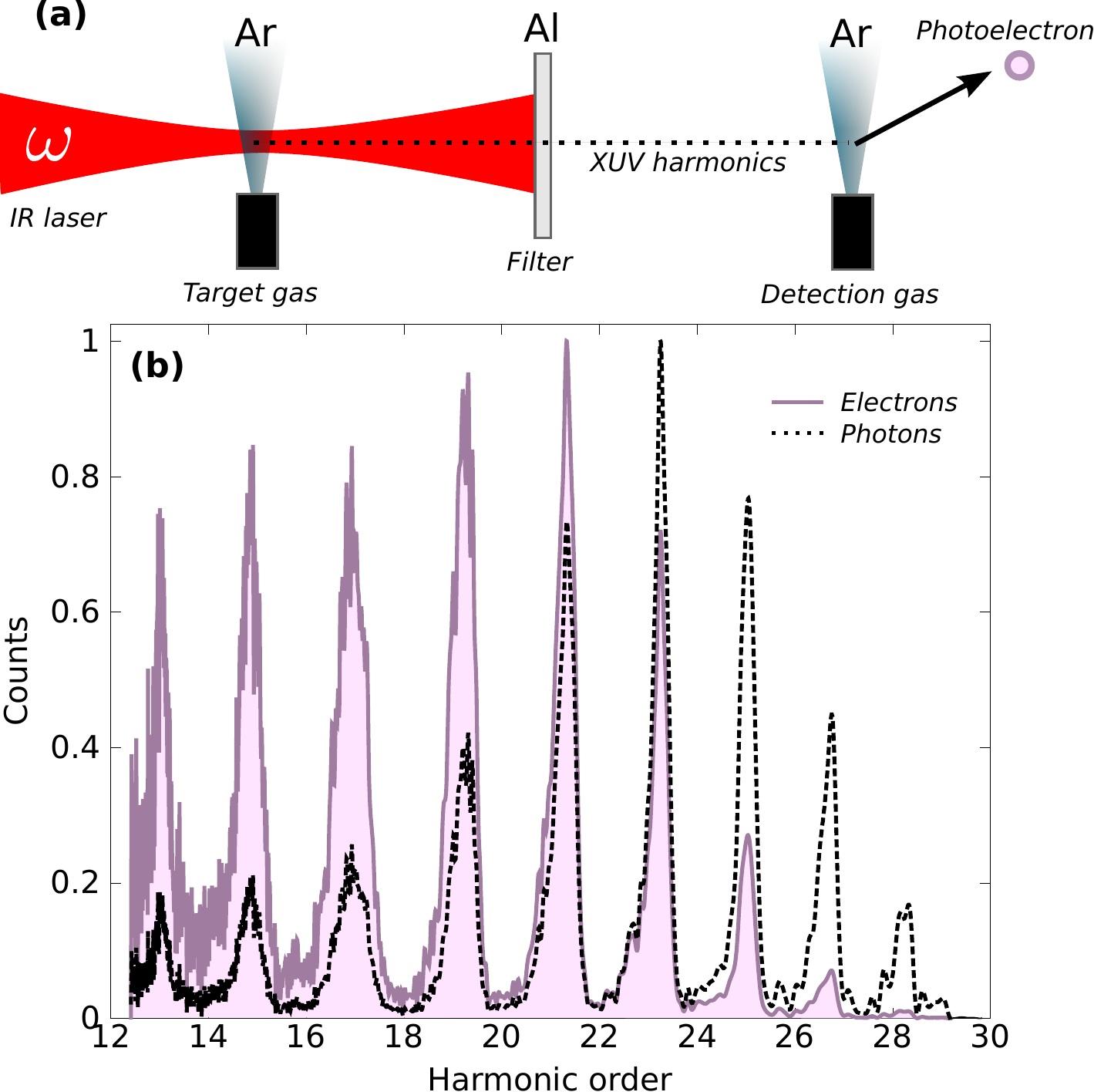}
	\captionof{figure}{
{\it A typical HHG experiment:} 
(a) A fraction of the IR laser field is converted to XUV through HHG. 
The XUV field is then filtered out and used to photoionize the detection gas.
(b) A representative photoelectron spectra (full) and XUV photon spectra (dotted) from an HHG experiment using Ar-atoms and Al-filter. 
	}
	\label{harmonics}
\end{center}
With a Titanium:Sapphire laser system (photon energy of $\hbar\omega=1.55$\,eV, corresponding to a near-IR laser wavelength of $\lambda=800$\,nm) using chirped-pulse amplification \cite{StricklandOC1985}, the typical laser intensity used for HHG is $I_L\sim 10^{14}$\,W/cm$^2$. 
Harmonic conversion efficiencies as high as $10^{-6}$ and photon energies in the soft x-ray range have been obtained  \cite{ChangPRL1997,SpielmannScience1997,SchnurerAPB2000,PaulNature2003,GibsonScience2003,SeresPRL2004}.
%
Neither the broad comb of high-order harmonics, nor their high conversion efficiency can be explained 
using standard theoretical tools of non-linear optics, where the laser field is treated as a perturbation \cite{BoydNonlinearOptics2003}. 
In the perturbative formalism, an increase in harmonic order $N$ is always accompanied by a strong reduction in conversion efficiency
as is evident from the well-known non-linear conversion formula: $I_N \propto I_L^N$, where $I_L\ll1$  is the intensity of the laser in a scaled set of variables. 
As an example, the relevant intensity scale for laser-atom interactions is the so-called atomic unit of intensity: $I_{\rm at} = 3.5 \times 10^{16} \,$ W/cm$^{2}$. As mentioned above, optimal conditions to observe  HHG processes are obtained with  Titanium:Sapphire  laser devices operated at $I_L \approx 10^{14}\,$W/cm$^{2}$, so that, in scaled units $I_L\leq 10^{-2}$.
In direct contrast with such perturbative behaviour, the harmonic comb from HHG may contain many harmonic orders of comparable intensity, $I_N \approx I_{N'} \ll I_L$, 
as seen in Fig.~\ref{harmonics}~(b). 
A \textit{cut-off law} for the maximal photon energy in the HHG plateau was found numerically to be \cite{KrausePRL1992}
\begin{equation} 
\hbar\Omega_{max}\approx I_p + 3 U_p,
\label{cutofflaw}
\end{equation}
where $I_p$ is the ionization potential of the atom, and
\begin{equation}
U_p = 
\frac{e^2 E_L^2}{4 m \omega^2} = 
\frac{e^2 \lambda^2 I_L}{8 \pi^2 \epsilon_0 c^3 m},
\label{Up} 
\end{equation}
is the so-called ponderomotive energy of the electron in the laser field \cite{BucksbaumJOSAB1987}.
The latter energy is associated with the average oscillatory (quiver) motion 
of a free electron driven by a single-frequency field.
Using Eq.~(\ref{cutofflaw}) and (\ref{Up}) in the limit $U_p \gg I_p$,  
the cut-off increases quadratically for longer laser wavelengths, $\hbar\Omega_{max}\propto\lambda^2$,
and linearly with the laser intensity, $\hbar\Omega_{max}\propto I_L$. 
This scaling has been used to increase the harmonic photon energies into the keV regime 
using laser fields with longer wavelengths (mid-IR wavelengths)  \cite{TatePRL2007,KapteynScience2007,PopmintchevOL2008,DoumyPRL2009}. 

Although HHG spectra from simple systems could be reproduced by numerical simulations based on the resolution of the Time-Dependent Schr\"odinger Equation (TDSE) \cite{KrausePRL1992}, a convenient scenario to explain the mechanism of the process is based on the so-called \textit{three-step model} \cite{CorkumPRL1993}. 
The three steps are illustrated in Fig.~\ref{threestep}.
\begin{center}
	\includegraphics[width= 0.85\textwidth]{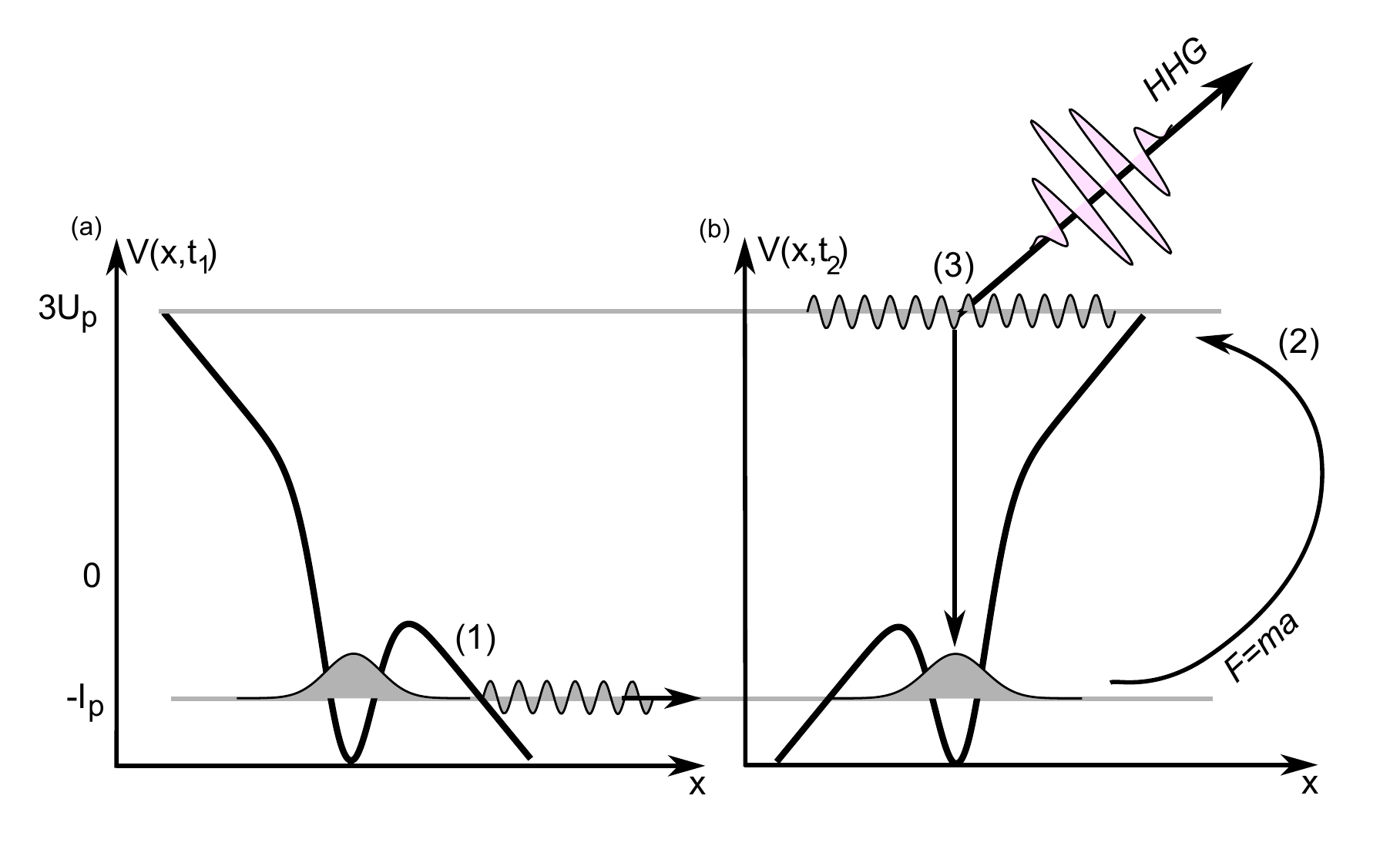}
	\captionof{figure}{
{\it The three-step model for HHG:} 
$(a)$ When the electric field of the laser is large, the electron can (1) tunnel through the Coulomb barrier into the continuum.
$(b)$ The electron wave packet will then (2) accelerate on almost classical trajectories in the laser field. 
Finally, a fraction of the electron wave packet may be driven back to the atom and (3) recombine, emitting a high-order harmonic photon. 
	}
	\label{threestep}
\end{center}
A more detailed quantum mechanical theory for HHG was published by Lewenstein and co-workers in 1994 \cite{LewensteinPRA1994}, 
showing that the three-step model could be derived from 
the Schr\"odinger equation using the Strong Field Approximation (SFA). 
In this way, a more exact cut-off law was derived analytically,
\begin{equation}
\hbar\Omega_{max}= 1.3I_p + 3.2 U_p,
\label{exactcutoff}
\end{equation}
corresponding to electrons with quasi-classical (complex) trajectories \cite{LewensteinPRA1995b,SalieresScience2001}, 
which are closely-related to Feynman's path integral approach to quantum mechanics \cite{Feynman1948}. 
This semi-classical model explained why the harmonics were of comparable strength, 
but it also revealed that several semi-classical trajectories contribute to the emission 
of each harmonic. 
This finding implied that each harmonic was emitted in two bursts per half-cycle of the laser, 
thus leading to a complicated temporal structure of the harmonic emission. 
It has been shown, as reviewed in Ref.~\cite{GaardeJPB2008}, 
that the problem of these multiple emission times can be circumvented experimentally 
thanks to the phase-matching properties of the {\it macroscopic} medium,  
leading to a ``selection'' of the emission from the shortest family of electron trajectories.  

Also, it was soon understood that such frequency combs of high-order harmonics
would correspond to light with pulse durations on the attosecond timescale, 
\textit{i.e.} shorter than any light pulses ever produced, 
under some restrictive conditions regarding their relative phases \cite{FarkasPLA1992,HarrisOC1993}.
The generation of such ``attosecond pulses'' is a challenging task especially because the experimentalists have to 
find a way of manipulating the relative {\it phase} of the harmonics over the large spectral bandwidth of the comb \cite{AntoinePRL1996}.
Another major challenge was to devise a scheme to measure the duration of the pulses.
This was because existing methods used in traditional ultra-fast optics could not be directly applied due to the short wavelength and
relatively low intensity of the harmonics.

\subsubsection{Attosecond pulses}
In 2001, more than a decade after the first HHG process was observed, 
the first Attosecond Pulse Trains (APT) were characterized using a scheme called RABITT 
(Resolution of Attosecond Beating By Interfering Two-photon transitions) \cite{PaulScience2001,VeniardPRA1996,Muller2002}. 
As measured in a set of experiments, 
the attosecond pulse duration was $\sim 250$\,as = 250$\times 10^{-18}$\,s, corresponding to $\sim 1/10$ of the laser period, and to $\sim 1/100$ of the laser pulse duration used for HHG. 
Two attosecond pulses were produced per oscillation of the laser field, 
resulting in a train of $\sim 30$ pulses in the total APT.
A typical fraction of an APT is depicted in Fig.~\ref{attotrain}. 

\begin{center}
	\includegraphics[width= 0.85\textwidth]{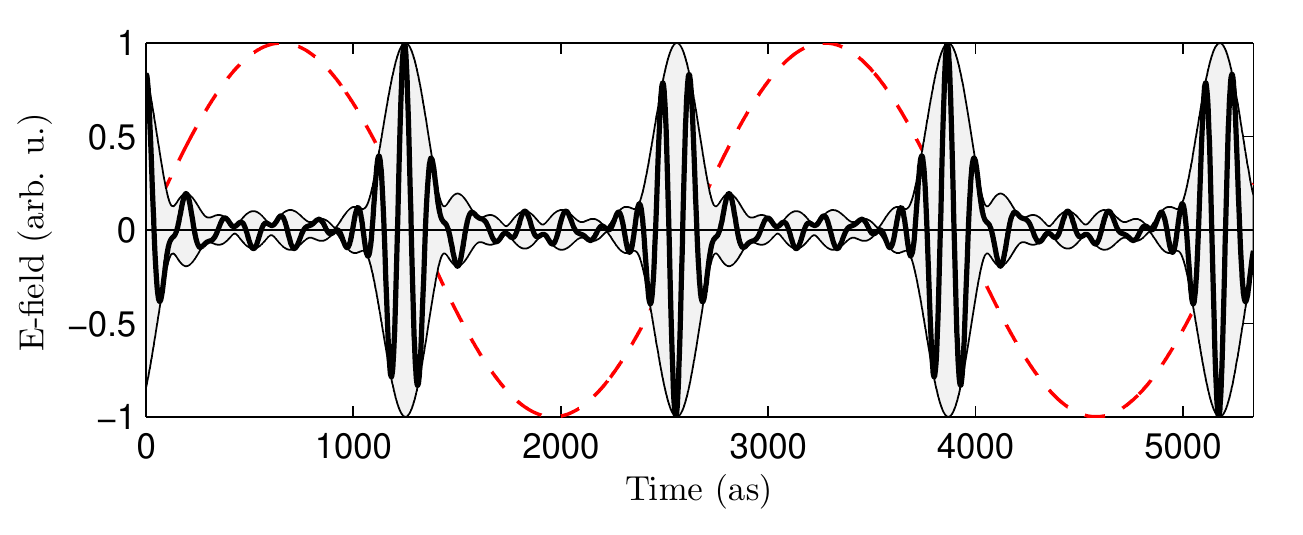}
	\captionof{figure}{
{\it Attosecond pulse train:} 
The attosecond pulse (shaded grey) duration is a fraction of the fundamental laser period (red dashed curve).  
The pulse separation is a half laser period 
and the sign of the attosecond electric field (heavy black curve) changes from pulse to pulse. 
	}
	\label{attotrain}
\end{center}

Making the IR pulse shorter leads naturally to fewer attosecond pulses.
It was soon demonstrated that a Single Attosecond Pulse (SAP) could be generated 
using a few-cycle IR pulse, $\sim 5$\,fs \cite{HentschelNature2001}. 
The basic properties of the SAP can be determined using the attosecond-streak camera \cite{ItataniPRL2002}, 
while more complete characterization requires the use of the sophisticated FROG--CRAB scheme (Frequency Resolved Optical Gating -- Complete Reconstruction of Attosecond Bursts) \cite{MairessePRA2005}. 
For such short IR pulses, the Carrier Envelope Phase (CEP) is an important parameter 
that must be controlled.
For instance, the generation of a SAP can be changed into the generation of two attosecond pulses by carefully
tuning the CEP of the laser field \cite{BaltuskaNature2003,GoulielmakisScience2008}. 
One of the most impressive applications of SAP was the ``direct measurement of light waves'' \cite{GoulielmakisScience2004}, 
where the SAP was used to map out the oscillating vector potential of a few-cycle optical laser pulse. 
Another useful way of controlling the HHG process is called ``polarization gating'' 
\cite{SolaNP2006,SolaPRA2006,ChangPRA2007,MashikoPRL2008}, 
which relies on the dependence of laser-polarization for the yield of HHG. 
In simple terms, linear polarization implies that the gate is open because attosecond pulses are being produced, 
while elliptical polarization suppresses the HHG and closes the gate. 
%


From a more fundamental standpoint, attosecond pulses are interesting because they are shorter than any pulses 
created by conventional optical lasers. 
Physically, we understand this because the duration of a light pulse can not be shorter than its own period.
Optical lasers, corresponding to visible and IR wavelengths, have a period that is longer than a femtosecond, 
leading to the so-called \textit{femtosecond barrier} for ultra-short pulses. 
The power needed to drive a laser scales strongly with the photon frequency, $P\propto \omega^5$, 
which effectively prevents the conventional laser scheme from going beyond optical wavelengths \cite{KapteynScience2007}. 
More formally, a pulse intensity envelope has a minimal duration proportional to the inverse of the supporting coherent bandwidth,  
\begin{equation}
\tau >  \frac{ C }{ \Delta \omega} \approx \frac{C}{2\pi}\frac{\lambda^2}{c\Delta\lambda},
\label{pulseduration}
\end{equation}
where $\Delta \omega$ (and $\Delta\lambda\ll\lambda$) is the intensity bandwidth of the light. In the case of a Gaussian pulse, $C=4\ln 2$ for the intensity envelope. The minimal pulse duration, called the \textit{Fourier-limited} pulse duration, occurs when all the spectral components are compressed, {\it i.e.} when the spectral phase is perfectly linear. 
The attosecond pulses produced through HHG can overcome the femtosecond barrier not only because they have a much shorter wavelength than optical light, but also because they have a broad coherent bandwidth.
It is interesting to compare the bandwidth of the laser used for HHG with  
the bandwidth of the attosecond pulses being generated.
An IR laser pulse used for generation of an APT has a typical duration of 30\,fs and a bandwidth of 60\,meV;
while the corresponding attosecond pulses are 100\,as with a bandwidth of 15\,eV. 
During the HHG process the coherent bandwidth is increased by a factor of 250 from laser to high-order harmonics.

Attosecond pulses can be used to ionize atoms and molecules at extremely well-defined times, but how do the resulting photoelectron wave packets and holes behave after the ionization event? How can the attosecond pulses be used to obtain temporal information about physical processes occurring on the atomic timescale? Using attosecond pulses to ionize target atoms or molecules does not by itself provide any high-resolution temporal information! When the photoelectrons are collected in an experiment, we can only determine the value of the momentum (or energy) they have gained in the course of the process, but not {\it how} or {\it when} it was acquired. 
This implies that the temporal ionization events must be synchronized and probed 
with the help of an external ``clock'' that ``ticks'' itself on the attosecond timescale. 
The typical period of light in the optical regime is on the femtosecond timescale, 
but one can argue that its variation occurs in the sub-femtosecond regime, 
which then justifies its common usage as a probe in attosecond photoionization experiments. 
In this way, it is possible to use a fairly long laser pulse, say tens of femtoseconds, 
to monitor attosecond photoionization events. 
Clearly, it is important that the probe-field is {\it phase-locked} with the attosecond pulses,
a requirement which is easily satisfied in HHG experiments by using fractions of the {\it same} 
laser pulse for both creating the attosecond pulses and for probing the photoionization process.  
Using attosecond pulses with phase-locked laser probes, has led to a range of experiments where time-delays between photoelectrons from different configurations and systems have been measured \cite{CavalieriNature2007,Schultze25062010,KlunderPRL2011}. 
Then, the natural question that emerges is whether or not such attosecond experiments yield direct access to the delay in photoemission, {\it i.e.} to the time it takes for an electron to photoionize, or if the observed delays should be interpreted in a different way?
Before answering these questions, we briefly review different pump-probe schemes that have been used to experimentally study light--matter interactions in real time.

\subsection{Pump--probe schemes for ultra-fast measurements}
\label{sec:attophotoexp}
\label{introattophotoexp}
In this tutorial, we provide a theoretical background for the temporal aspects of photoionization.
The attosecond pulse structure is shorter than the response time of any detector or electronic device, 
but it is possible to obtain temporal information about attosecond photoionization indirectly  
by investigating coherent cross-correlation photoelectron spectrograms 
between the attosecond pulse and a weak IR laser probe. In Fig.~\ref{pulsesketch} we present three different kinds of cross-correlation  techniques. 

\begin{center}
	\includegraphics[scale=0.40]{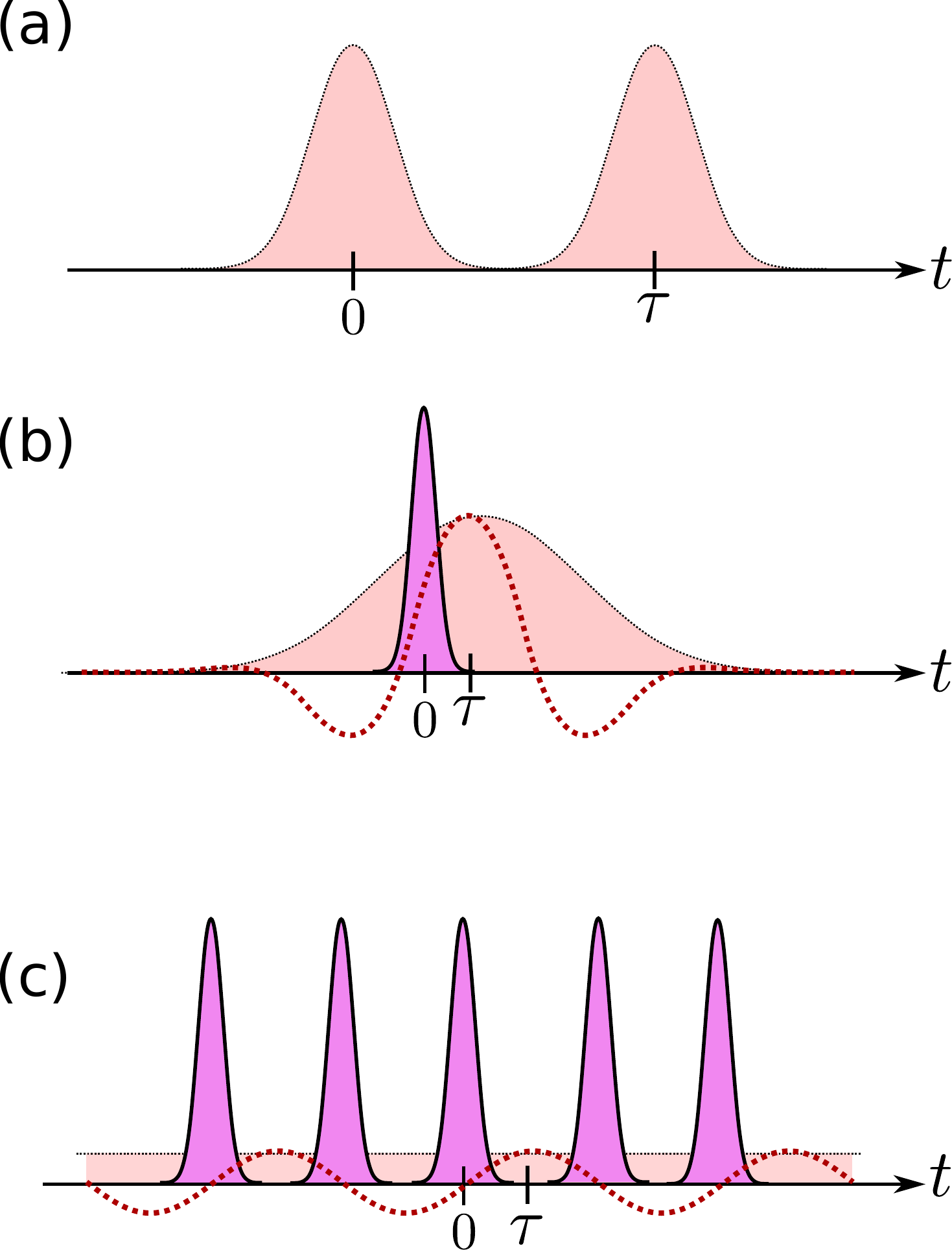}
	\captionof{figure}{
{\it Pump-probe schemes:}  
(a) Traditional pump--probe experiment with two pulses separated in time by $\tau$. 
(b) Simultaneous pump--probe experiment between a SAP and a few-cycle IR field.
(c) Simultaneous pump--probe experiment between an APT and a monochromatic IR field. 
The narrow purple area represents the attosecond XUV pulse envelope and 
the broader red area represents the one of the probing laser pulse,
while the dotted red lines indicate the corresponding E-field.  
	}
	\label{pulsesketch}
\end{center}
\subsubsection{Traditional pump--probe experiment}
A typical pump--probe scheme is illustrated in Fig.~\ref{pulsesketch}~(a).
First, a pump pulse is used to excite the system at time $t=0$.
A part of the quantum mechanical wave function is then pumped from the initial state to an excited wave packet: $\Psi^{(0)}(0)\rightarrow \Psi^{(1)}(0)$. 
The system can then evolve in a field free environment until time $t=\tau$,
when the probe pulse interacts with the excited wave packet, thus changing its state: $\Psi^{(1)}(\tau)\rightarrow\Psi^{(2)}(\tau)$.  In experiments, the modification of the wave packet leads to the change of the observable quantities 
as a function of the delay between the pump and the probe pulses.
Temporal information about the field-free propagation of the system, {\it e.g.} a molecular vibration, 
can then be extracted in real time by repeating the experiment systematically for different delays of the probe pulse. 
In this way, the intermediate steps in chemical reactions, the so-called ``transition states'', 
have been investigated in the framework of Zewail's femtochemistry \cite{Pedersen1994,Zewail1990,Zewail1988,Worner2010}.  
Clearly, the intuitive interpretation of these experiments is limited by the respective pulse durations. 
Using few-cycle optical-laser fields to pump and probe the system is adequate to study nuclear motion in molecular systems, 
typically occurring on the femtosecond timescale. Coherent, attosecond XUV  pump--probe experiments, where attosecond pulses are used for both pumping and probing, hold promise of observing the electron dynamics on the attosecond timescale, 
but they are difficult to implement experimentally due to the low probability for absorbing two photons in the XUV range,
where we assume that {\it at least} one photon must be absorbed from both the pump and the probe for a meaningful signal. 
We refer the reader to Ref.~\cite{TzallasNature2011} (and the references therein) 
for state-of-the-art experimental efforts on XUV-pump and XUV-probe experiments bordering the attosecond timescale. 
A different kind of pump-probe experiments has been carried out using a SAP as pump 
and a few-cycle laser pulse as probe, see for instance Ref.~\cite{MauritssonPRL2010}, 
where the Fourier transform of the delay-dependent spectrogram yields quantum beats,
but also interference between direct and indirect pathways in the ionization.

\subsubsection{Simultaneous pump--probe experiment using SAP: ``Streaking''}
A more commonly used attosecond pump--probe configuration is
illustrated in Fig.~\ref{pulsesketch}~(b), 
where the pump and probe pulses {\it overlap} in time.
In this situation it is not primarily the field-free system that is of interest,
but rather the temporal characterization of the pulses or the evolution of the system in the presence of the two fields. 
The pump probe is an attosecond XUV pulse, while the probe pulse is a longer (few femtosecond) IR-laser pulse. 
Strictly speaking, this is a {\it laser-assisted photoionization process}, where 
the system is simultaneously pumped and probed.  
Sub-femtosecond temporal information can be gained by repeating the experiment 
at different subcycle delays between the attosecond pump pulse and the laser field oscillation of the probe pulse. 
Clearly, it is essential that the two pulses are phase-locked  
and that the delay can be controlled with sub-femtosecond precision.

Under these conditions,  the system can absorb energy simultaneously from the two fields  {\it i.e.} while it absorbs one XUV photon, it is "dressed" by the relatively intense IR probe field. Accordingly, a non-perturbative approach is needed to account for the effect of the probe.
A simplified theory for the influence of the probe pulse on the photoelectrons is given by the {\it streak camera} formalism \cite{ItataniPRL2002}. This simplified interpretation of the photoelectron distribution relies on the use of the SFA, 
which is equivalent to assuming that the photoelectron feels the instantaneous laser field while the Coulomb potential of the ionic core is neglected. 
According to the SFA, the streaked electron momentum distribution is then shifted as 
\begin{equation}
\vec p_f(\tau) = \vec p_0 - e\vec A(\tau), 
\label{simplepshift}
\end{equation}
where $\vec p_0$ is the unshifted, probe-free momentum; $e>0$ is the elementary charge 
and $\vec A(\tau)$ is the probe-field vector potential at the time of ionization. 
This concept of ``instantaneous streaking'' 
provides a simple map from time to momentum of the streaked electrons, 
which is valid assuming that the electron is free shortly after 
the ionization event, {\it i.e.} that it either escapes from a short-range binding potential 
or is ejected with a high velocity. 
In practice, the electron is not completely free after the ionization event
due to the remaining Coulomb potential from the ion. 
Smirnova and co-workers found that 
the simultaneous action of the long-range Coulomb potential 
and the probing laser field can result in a small shift 
of the streaking process \cite{SmirnovaJPB2006}, {\it i.e.} 
an uncertainty on the absolute delay in the experiment. 
Until recently \cite{Schultze25062010}, these delays were considered too small to be accessible in streaking experiments. 

\subsubsection{Instantaneous pump--probe experiment using APT: ``RABITT''}
In Fig.~\ref{pulsesketch}~(c), a train of ``identical'' attosecond pump pulses 
are probed with a monochromatic IR-field.
This scheme is often referred to as the RABITT method \cite{PaulScience2001,VeniardPRA1996,Muller2002}.
Provided that the probe pulses are weak and 
that the fields are repeated periodically for many cycles, 
this scheme can provide equivalent information as 
the streak-camera method described above [Fig.~\ref{pulsesketch}~(b)].
We will discuss this equivalence in more detail in Sec.~\ref{sec:STPT} in terms of photoelectron wave packets,  
and we suggest that the interested reader should consult Ref.~\cite{DahlstromCP2012} for a complete derivation 
in lowest-order perturbation theory.
Higher-order effects in the probe field have been considered using the ``soft-photon approximation'' \cite{MaquetJMO2007}, 
but the high-level of accuracy that is required for attophysics remains a challenge to theory, 
especially so in the non-perturbative regime.

The shared periodicity of the APT and the probe field implies that  
the photoelectrons will appear on distinct energies corresponding 
the discrete numbers of absorbed photons.  
In Fig.~\ref{RABITT} we display a typical experimental photoelectron spectrogram 
produced by laser-assisted APT ionizing Ar gas. 
\begin{center}
	\includegraphics[width= 0.85\textwidth]{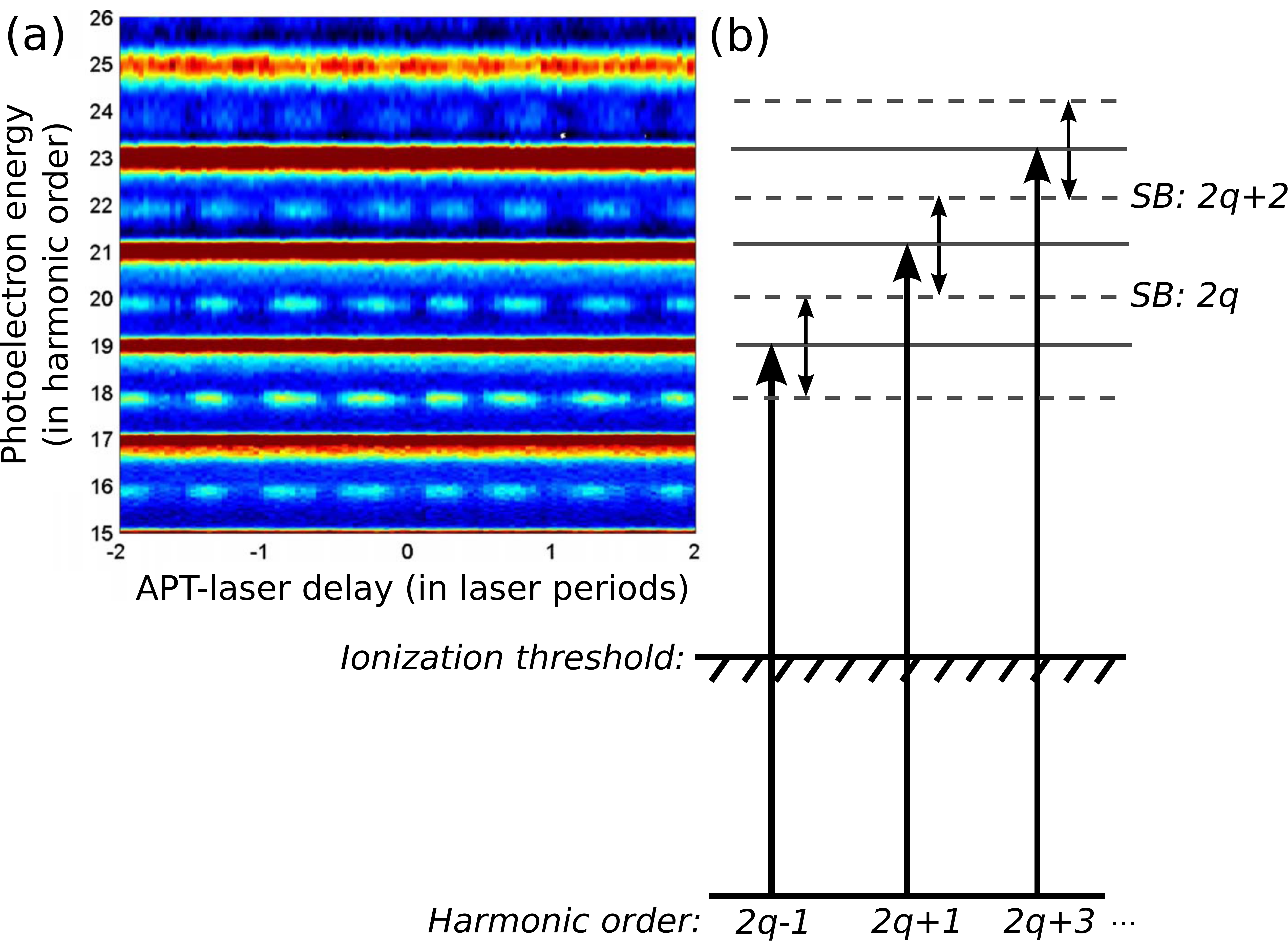}
	\captionof{figure}{
{\it RABITT method:}
(a) Photoelectron spectrogram over photon energy and delay between the APT and the IR field. 
The offset in the modulation of the sidebands contains information about the attosecond pulses
and the ionization process. (b) Schematic energy diagram over the quantum paths leading 
to the same final energy in sideband (SB) $2q$. 
The experimental data was gathered for Ref.~\cite{DahlstromPRA2009}.
	}
	\label{RABITT}
\end{center}
The advantages of the periodic time-structure of APT are rather practical: 
First, it is less demanding experimentally to produce APT and multi-cycle probe-fields, than SAP and few-cycle probe fields.
Second, the probe field can be weaker than for streaking, 
thereby, causing less side-effects on the system, 
{\it e.g.} induced polarization of the core or ionization. 
Third, the signal is read out on zero-background, {\it i.e.}
in energetic regions where no initial photoelectrons appear.  
Finally, a monochromatic probe field simplifies the analysis because  
it leads to less spectral convolutions in the experimental signal, 
as we shall discuss further in Sec.~\ref{sec:STPT}. 

In this setup, there is a phase-difference associated with the instantaneous probing by the laser field, 
which can be traced back to the phase-shift of the relevant two-photon matrix elements. 
The probability of the sideband peak modulates as
\begin{equation}
S_{2q} = {}  \alpha + \beta \cos[
2 \omega \tau - 
\Delta \phi_{2q} - 
\Delta \theta_{2q} ],
\label{S2q}
\end{equation}
where $\Delta \phi_{2q} = (\phi_{2q + 1} -\phi_{2q-1})$ is the phase difference between the consecutive harmonics $(2q+1)\omega$ and $(2q-1)\omega$; and $\Delta \theta_{2q} $ is an intrinsic atomic quantity, the so-called ``atomic phase'', associated to the difference of the phases of the transition amplitudes associated to the distinct quantum paths leading to the sideband~\cite{VeniardPRA1996,Muller2002}. 
The former phase is related to the arrival time of the attosecond pulses: $\tau_{\phi}=\Delta \phi_{2q}/(2\omega)$, 
for frequencies $\Omega\approx 2q \omega$. 
The atomic phases were studied in detail by Toma and Muller already in 2002, using lowest-order perturbation theory, for photoionization from the $3p$ state in Argon \cite{TomaJPB2002}. 
Further numerical work was performed by Mauritsson, Gaarde and Schafer 
using single-active electron (SAE) effective potentials for Helium, Neon and Argon \cite{MauritssonPRA2005}.
The two latter numerical calculations were made with the intention of making calibration curves for attosecond characterization tools such as RABITT. Our aim in this tutorial is to discuss the origin of this atomic phase in detail. We will identify its fundamental physical components in terms of time-delays and we will relate it to the above-mentioned streaking delays.

\section{Attosecond pulses of light}
\label{sec:wplight}

It it well-known that short light pulses can be described as wave packets, 
{\it i.e.} as coherent superpositions of monochromatic light waves \cite{Svelto1998}.   
The propagation of such wave packets can then be understood in terms of the {\it phase-shifts} of the individual monochromatic waves, 
which will be discussed in Sec.~\ref{sec:waveproplight}. 
An attosecond XUV pulse can be strongly stretched in time by the propagation through a dispersive medium, but conversely, 
it is also possible to compress a stretched pulse using a medium with negative (anomalous) dispersion. 
A useful theoretical tool adapted to describe these phenomena is the so-called Stationary Phase Approximation (SPA),
which is introduced in Sec.~\ref{applicationofSPA} and applied to the propagation of attosecond pulses. 
As we will show later, it applies equally well to the propagation of photoelectron wave packets.

We start our analysis of attosecond XUV pulses from the Maxwell equations in a dielectric medium 
\cite{BoydNonlinearOptics2003,Svelto1998}, and recast the wave equation into a \textit{time-independent form} 
by exploiting the orthogonality of the Fourier integral expansions of the time-dependent fields
\begin{equation}
\nabla^2\vec{E}+\frac{\omega^2}{c^2}\vec{E} ~=~ -\frac{1}{\epsilon_0}\frac{\omega^2}{c^2}\vec{P}-\frac{1}{\epsilon_0}\nabla(\nabla\cdot\vec{P}),
\label{FTwaveeq}
\end{equation}
where the electric field is expanded as a superposition of monochromatic waves
\begin{equation}
\vec{\tilde E} (t,\vec{r}) = \frac{1}{2\pi} \int d\omega ~ \vec{E}(\omega,\vec{r}) e^{-i\omega t} ,
\label{FTefield}
\end{equation}
and the polarization is given by 
\begin{equation}
\vec{\tilde P} (t,\vec{r}) = \frac{1}{2\pi} \int d\omega ~ \vec{P}(\omega,\vec{r}) e^{-i\omega t}.
\label{FTpol}
\end{equation}
where the fields in space--frequency are written without a tilde: $\vec E(\omega,\vec r)$.
Finding solutions to Eq.~(\ref{FTwaveeq}) requires knowledge of the polarization terms, $\vec P(\omega,\vec r)$.
The polarization induced by a weak attosecond XUV pulse in an isotropic medium can be approximated by the {\it linear response}:
\begin{equation}
\vec{P}(\omega,\vec{r})=\epsilon_0 \chi^{(1)}(\omega,\vec{r})\vec{E}(\omega,\vec{r}), 
\label{linearresponse}
\end{equation}
where the linear susceptibility, $\chi^{(1)}(\omega,\vec r)$, may vary in space due to changing density or composition of the medium. 
Strictly speaking, the medium may exhibit a more complicated evolution in time, 
{\it e.g.} due to ionization of the species constituting the medium or due to non-linear polarization, 
but we will not consider the latter effects here \cite{BoydNonlinearOptics2003,lhuillierJOSAB1990},  
due to the fact that the interaction between pulses with moderate intensity in the XUV range and matter are typically perturbative.
We mention, however, that such contribution may become important when studying the propagation of intense XUV or X-ray radiation
pulses, such as the ones generated by XFEL facilities, with peak intensities well above the atomic unit of intensity, 
{\it i.e.} in the range beyond $10^{16}$\ W/cm$^2$.

\subsection{Phase propagation of light in the linear regime}
\label{sec:waveproplight}
In order to describe the propagation of linearly polarized, attosecond light pulses in a medium, 
we insert the following ansatz: 
$\vec{E}(\omega,\vec{r})\propto \exp\left[i\phi(\omega,x)\right]\hat{z}$, 
into Eq.~(\ref{FTwaveeq}) to obtain a differential equation for the {\it phase} of monochromatic waves: 
\begin{equation}
i\frac{\partial^2\phi}{\partial x^2} - \left(\frac{\partial \phi}{\partial x}\right)^2+\frac{\omega^2}{c^2}=
-\frac{\omega^2}{c^2}\chi^{(1)}.
\label{ansatzeq}
\end{equation}
For slow variations of the optical properties in material, we assume $|\partial^2\phi/\partial x^2| \ll |\partial \phi/\partial x|^2$, which leads to a simpler differential equation:
\begin{equation}
\frac{\partial \phi}{\partial x}=
\frac{\omega}{c}\sqrt{1+\chi^{(1)}}.
\label{ansatzeqapprox}
\end{equation}
The phase of the electric field can be directly integrated as
\begin{equation}
\phi(\omega,x)=
\frac{\omega}{c} 
\int_{-\infty}^x dx' ~ \sqrt{1+\chi^{(1)}(x')}\equiv
\frac{\omega}{c} 
\int_{-\infty}^x dx' ~ n(x'),
\label{ansatzeqapproxphi}
\end{equation}
where $n(x)$ is the \textit{local refractive index} of the medium.
%
%
The asymptotic {\it phase-shift} obtained after propagation through the material can be written as 
\begin{equation}
\delta(\omega) = \lim_{x\rightarrow\infty} \frac{\omega}{c} \int_{-\infty}^{x}dx' ~ \left[ n(x')-1 \right],
\label{phasedifference}
\end{equation} 
which will have a finite value assuming a {\it finite} material surrounded by free space, 
{\it e.g.} that $n(x)=1$ for large enough $|x|$. 
Thus, a frequency-independent refractive index leads to a {\it linear} phase-shift in frequency. 
This linear phase-shift causes a simple time-delay, $\tau = \delta/\omega$, of the attosecond pulse in Eq.~(\ref{FTefield}),
\begin{equation}
\underbrace{
{\tilde{E}_{free}}(t-\tau,x)
}_{Shifted~wave~packet}
=\frac{1}{2\pi}\int d\omega ~ \underbrace{{E_{free}}(\omega,x)e^{-i\omega t}}_{Free~propagation}
\underbrace{\exp[i\omega\tau]}_{Phase-shift},
\label{shiftteorem}
\end{equation}
compared to free-space propagation, corresponding to the undistorted waves $E_{free}(\omega,x)$. 
Eq.~(\ref{shiftteorem}) is a direct application 
of the well-known \textit{shift-theorem} of Fourier transforms. 
Furthermore, if the medium is absorbing, 
the asymptotic phase will be a complex quantity, with 
its real part leading to the actual phase-shift 
and its imaginary part leading to an exponential reduction of the amplitude,	 
$\exp[i\delta]=\exp[i\Re(\delta)]\exp[-\Im(\delta)]$.

%
In order to illustrate the shift of the wave packet, we compute the asymptotic phase acquired by an attosecond pulse 
passing through a material of length, $L$, with a constant refractive index,  
\begin{equation}
\delta ~=~ \frac{\omega}{c}~ [n-1] L ~=~
\frac{2\pi}{\lambda}~ [n-1] L ~\approx~ \frac{\omega}{2c}~\chi^{(1)} L, 
\label{constantref}
\end{equation} 
where $\lambda=2\pi c/\omega$ is the wavelength of the light wave in vacuum, 
and where we assumed that the linear susceptibility is small. 
The corresponding delay of the pulse is 
\begin{equation}
\tau = \frac{[n-1]L}{c} \approx \frac{\chi^{(1)}L}{2c}, 
\label{simpletimedelay}
\end{equation}
where we find an intuitive linear scaling with both the length on the material and with the change in refractive index. 
\cite{Svelto1998}.

\subsubsection{Dispersion and group delays}
\label{sec:groupdelays}
So far we have assumed that the linear response is independent of the frequency of the light. 
This is an adequate approximation for narrow-bandwidth laser pulses, 
but it is not a good approximation for attosecond pulses that have a 
large spectral bandwidth \cite{lhuillierJOSAB1990}.
In a more realistic model, the material will be dispersive, \textit{i.e.} the refractive index will be frequency-dependent, $n(\omega,x)$ \cite{DalgarnoPRSL1960,Henke1988,Henke1993}. 
After passing such a material,  
the asymptotic (spectral) phase, $\delta(\omega)$, may exhibit a {\it non-linear} frequency dependence, 
which implies that the wave packet in the time domain, 
\begin{equation}
\tilde E(t,x) = \frac{1}{2\pi}\int d\omega ~ 
\underbrace{E_{free}(\omega,x) e^{-i\omega t}}_{Free~propagation} 
\underbrace{\exp[i\delta(\omega)],}_{Phase-function}
\label{Eintimedomain}
\end{equation}
may be shifted and {\it deformed} with respect to the free propagation.
It is convenient to expand the spectral phase in a Taylor series \cite{Svelto1998},
\begin{equation}
\delta(\omega) = \delta(\omega_0)+
\sum_{n=1}^{\infty}
\left.\frac{1}{n!}\frac{\partial^{n}\delta}{\partial\omega^{n}}\right|_{\omega_0}(\omega-\omega_0)^{n},
\label{taylorphase}
\end{equation}
around the central frequency of the pulse, $\omega_0$. 
%
In many practical cases it is sufficient to consider the first few terms in this expansion.
Note that we have already seen that 
the zero-order term determines the overall phase of the pulse,
and that the first order term determines the delay of the pulse.
In the optical regime, most materials have a positive (normal) dispersion, 
which means that the refractive index increases as a function of frequency. 
This implies longer delays for higher frequency pulses passing through the same medium. 
Close to resonances or close to the ionization threshold of a material, the refractive index may have a negative (anomalous) dispersion. 
The \textit{Group Delay} (GD) is defined as 
\begin{equation}
\tau_{GD}(\omega) = \frac{\partial\delta}{\partial\omega},
\label{groupdelay} 
\end{equation}
and it represents a direct generalization 
of the shift theorem in Eq.~(\ref{shiftteorem}), 
stating that the derivative of the spectral phase at {\it any} frequency $\omega$  
corresponds to a delay in the propagation of the components of the wave packet in
{\it that} spectral region. 
The GD is a rather abstract concept,  
but a useful interpretation is that it can be related to the time 
when the coherent superposition of all wave packets in the neighbourhood of $\omega$ is constructive. 
The GD is {\it not} equal to the time when the instantaneous temporal frequency, $\frac{\partial}{\partial t} \arg[\tilde E(t)] $, 
equals $\omega$ for a wave packet in general. However, we will show  in Sec.~\ref{applicationofSPA} that  
the GD does equal the time of this instantaneous frequency for wave packets that are stretched in the time domain.

In order to illustrate the GD concept, we now consider an XUV wave packet after travelling through a plasma of length $L$, where 
\begin{equation}
n_p(\omega) ~=~ 
\sqrt{1-\left(\frac{\omega_p}{\omega}\right)^2}~\approx~ 
1-\frac{1}{2}\left( \frac{\omega_p}{\omega} \right)^2, 
\label{plasman}
\end{equation} 
is the refractive index of the plasma, 
$\omega_p=\sqrt{N e^2 / (\epsilon_0 m)}$, 
is the plasma frequency, 
and $N$ the concentration of free electrons in the plasma \cite{Kittel1996}. 
Using Eq.~(\ref{constantref}), the asymptotic phase becomes negative 
\begin{equation}
\delta_p(\omega)~ =~ 
\frac{\omega}{c}\left[n_p(\omega)-1\right]L
~\approx~ 
-\frac{1}{2}\frac{\omega_p^2}{c}\frac{L}{\omega}, 
\label{plasmad}
\end{equation} 
which implies that the phase-velocity, $v_{phase}=c/n(\omega,x)$, of the monochromatic field 
is \textit{faster} than $c$ (superluminal), because the local refractive index is smaller than one! 
The corresponding GD is, however, positive 
\begin{equation}
\tau_{GD}=\frac{\partial\delta_p}{\partial \omega} ~\approx~ 
\frac{1}{2}\frac{\omega_p^2}{c}\frac{L}{\omega^2}, 
\label{plasmadGD}
\end{equation} 
which shows that any light pulse is delayed by the plasma, 
and that it travels at a speed less than $c$.
These effects prove to be practical for phase-matching of non-linear processes, 
where control of the superluminal phase velocity is achieved by simply increasing or decreasing the intensity of the laser field.
This is because it indirectly alters the density of the plasma due to a changing ionization rate,
which affects in turn the electron density in the plasma. 
%

\subsubsection{Attosecond pulse trains}
Our discussion of delays is very general and it clearly applies to the propagation of both SAP and APT. 
APT arise naturally from the HHG process when the driving laser field is many periods long.
The train contains two identical attosecond pulses per period of the fundamental field,
with electric fields having opposite signs. 
In this context, uncompensated dispersion will lead to a {\it temporal} overlap 
and interference between consecutive attosecond pulses in the APT, 
when the individual pulse duration increases beyond a half-period of the fundamental laser field. 
An \textit{ideal} APT can be written as a Fourier sum, 
\textit{i.e.} as a discrete version of Eq.~(\ref{FTefield}), over all high-order harmonics
\begin{equation}
\tilde{E}(t)=\sum_{q} |E_{2q+1}| \exp\left[ i\delta_{2q+1} - i(2q+1) \omega t \right],
\label{APT}
\end{equation}
where $2q+1$ labels the odd harmonics that are generated from HHG. 
We write ``ideal'' to indicate that the APT in Eq.~(\ref{APT}) corresponds to an infinite number of pulses in the time-domain,
while an actual APT will have a finite duration comparable to that of 
the fundamental IR laser pulse  \cite{VarjuJMO2005,VarjuPRL2005}.
A short APT is likely to exhibit strong pulse-to-pulse variations.
The subcycle behaviour of the ideal APT can, however, be understood as 
an average subcycle pulse structure in a long APT composed of many attosecond pulses.

\subsubsection{Compression of attosecond pulses using metallic foils}
As a general rule, attosecond pulses are generated with a positive intrinsic chirp from the HHG process \cite{LewensteinPRA1995b}.
However, they can be compressed in time using thin metallic foils in the setup depicted in Fig.~\ref{harmonics}~(a)  \cite{MartinezJOSAB1984,LopezMartensPRL2005,GustafssonOL2007}.
An example of such XUV pulse compression is shown in Fig.~\ref{attofoils}, where the experimental data corresponds to propagation through a 400\,nm thin Al-foil.

\begin{center}
	\includegraphics[width= 0.85\textwidth]{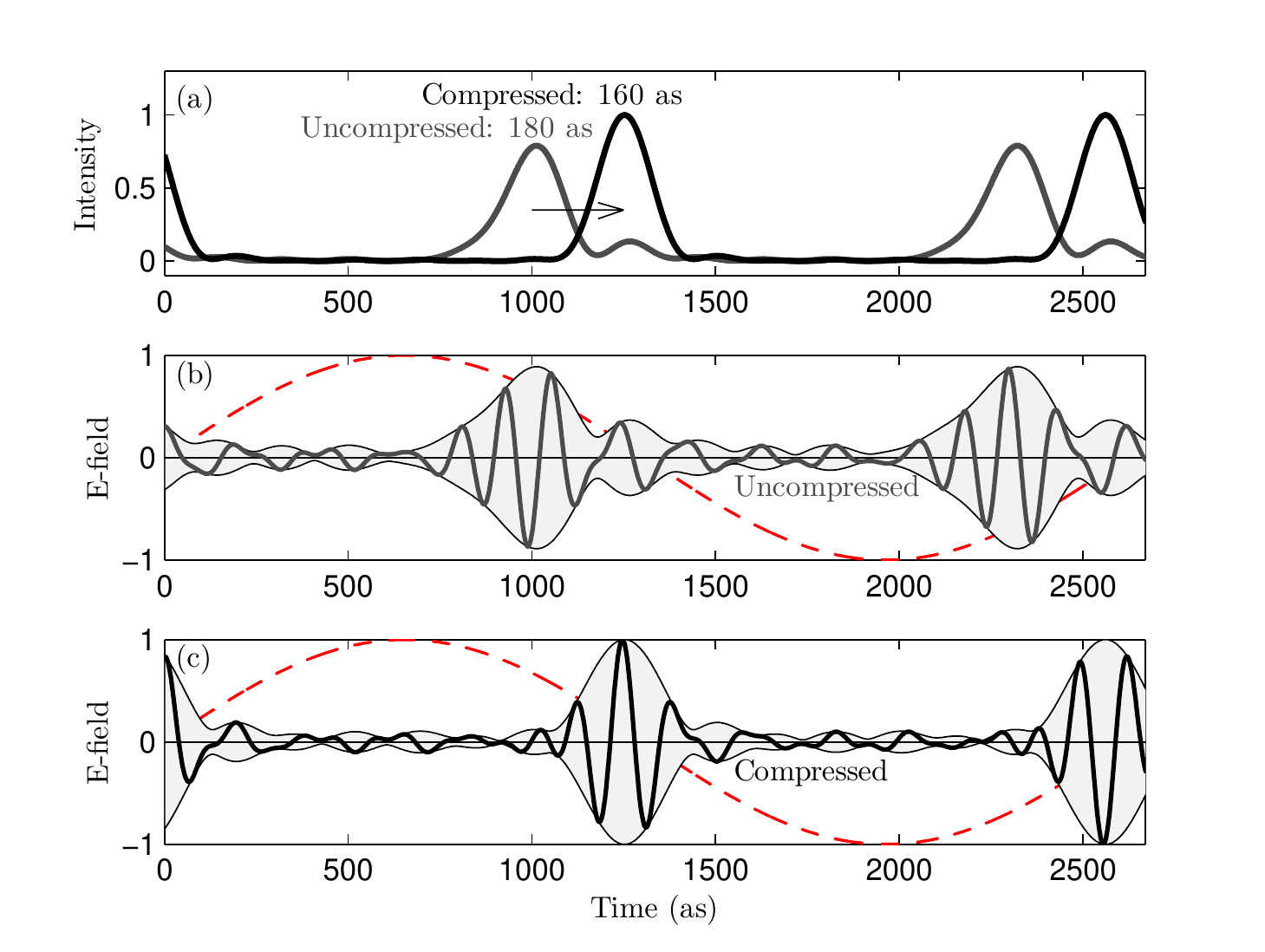}
	\captionof{figure}{ {\it Attosecond pulse compression:}
(a) Attosecond pulses generated in Ar before (grey curve) and after (black curve) passing through a 400\,nm Al-foil. 
In addition to a delay of the pulse of ~$\sim250$\,as, the pulse shape changes due to the foil. 
(b) The uncompressed pulse is 180\,as and it is asymmetric with small post pulses.
(c) After the foil, the compressed pulse is 160\,as and more symmetric.
The CEP has been set to be ``cos-type'' and  
the IR laser field (dashed curve) is plotted for comparison.
Amplitude effects due to the foil are neglected for simplicity.
	}
	\label{attofoils}
\end{center}

The Al-foil provides negative dispersion in the low-energy part of the transmission window, 
while the high-energy part provides positive dispersion. 
In experiments, this property of Al-foils is used to generate close to
Fourier-limited attosecond pulses  
by compensating for the intrinsic positive chirp of the HHG process by the negative dispersion  
provided by the low-energy part of the metallic transmission window.
Notice the structure of the APT, with two pulses per period of the fundamental laser field.
We mention again that the electric field of the two adjacent pulses have opposite phases, {\it i.e.} that they are $\pi$-shifted.

\subsection{Stationary Phase Approximation}
\label{applicationofSPA}
The SPA is an essential theoretical tool in attophysics because it can be used to evaluate integrals with complex valued integrands, 
such as the ones encountered when computing time-frequency Fourier transforms. 
It has proven itself especially successful for understanding the HHG process  
\cite{LewensteinPRA1994,LewensteinPRA1995b}, 
and it remains a useful tool for gaining better understanding of strong field dynamics, 
see for instance Ref.~\cite{IvanovJMO2005,Chirila2010,DahlstromJPB2011}. 
In the next subsections: Sec.~\ref{spaex1} and \ref{spaex2}, we will apply the SPA to evaluate XUV wave packets.
Further on, in Sec.~\ref{generalizationFirstOrder}, we will apply the SPA to the propagation of electron wave packets
ionized by attosecond pulses. 

In simple terms, the SPA is related to the well-known integral of a Gaussian \cite{WeberArfken2004},
\begin{equation}
\int_{-\infty}^{\infty} d\omega ~ \exp\left[ -\beta \omega^2 \right] ~=~ \sqrt{\frac{\pi}{\beta}},
\label{gaussian}
\end{equation}
which has a finite value provided that $\beta > 0$. 
As an example, we plot the integrand and integral for $\beta=1$ in Fig.~\ref{expx2}\,(a) and (b) respectively. 
The integral value in (b) approaches $\sqrt{\pi}\approx 1.7725$ as expected from Eq.~(\ref{gaussian}).

\begin{center}
	\includegraphics[width= 0.85\textwidth]{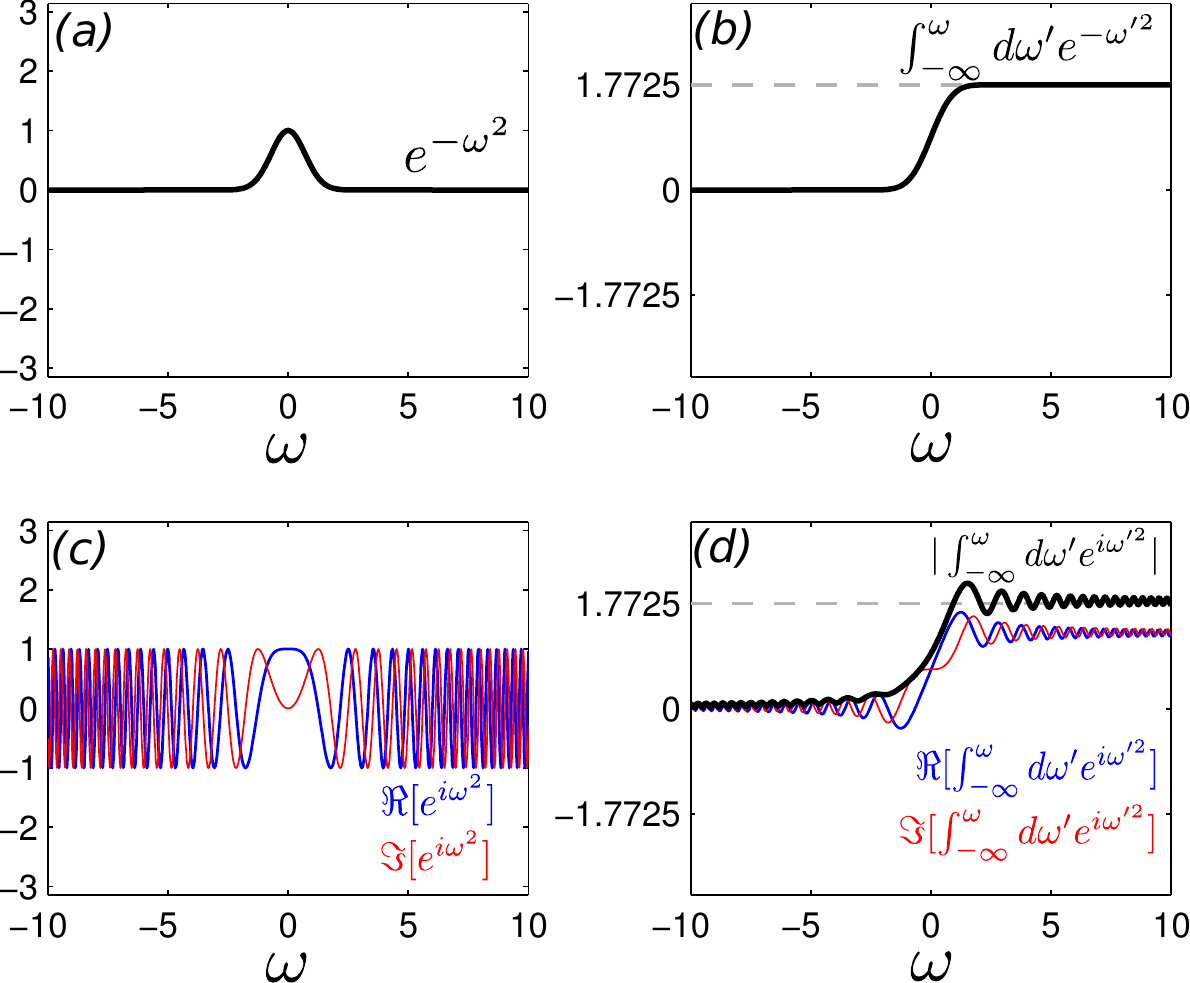}
	\captionof{figure}{ {\it Gaussian integrands and integrals:}
(a) A real Gaussian function ($\beta=1$). 
(b) The real Gaussian integral equals $\sqrt{\pi}\approx 1.7725$. 
(c) A complex Gaussian function ($\beta=-i$). Notice the fast complex oscillations for large values of $|\omega|$.
(d) The complex Gaussian integral equals $\sqrt{\pi}\exp[i\pi/4]$.
	}
	\label{expx2}
\end{center}


\subsubsection{Method of steepest descent}

In more detail, the SPA belongs to a class of techniques of wide use to compute integrals with integrands containing the exponential of a complex analytical function, $f(z) = u(x,y) + i v(x,y)$, with $z=x+iy$. It is closely related to the so-called {\it steepest descent} method which deals with integrals of the following form \cite{WeberArfken2004}:
\begin{equation}
K = \oint_{C} dz \exp\left[ f(z) \right] ,
\label{steepest descent}
\end{equation}
where  the contour $C$ is chosen so that the ends of the path do not contribute significantly. 
In the vicinity of a complex {\it saddle-point} $z_0$, where $f'(z_0) = 0$, the complex function can be expanded as
\begin{equation}
f(z) \approx f(z_0)  + {1 \over2}(z-z_0)^2 \: f''(z_0). 
\label{expansion}
\end{equation}
Inserting the expanded function into the exponential yields the approximate integral 
\begin{equation}
K \approx \oint_{C} dz \exp\left[  f(z_0)  + {1 \over2}(z-z_0)^2 \: f''(z_0)   \right],
\label{steepest descent1}
\end{equation}
where the zeroth-order, $f(z_0)$, and the second-order,  $f''(z_0)$, coefficients are complex. 
In order to evaluate this integral, we parametrize the contour: $z(\rho)=z_0+\rho e^{i\varphi}$, 
with $\rho$ being a real parameter and $\varphi$ being the angle 
at which the contour crosses the saddle-point at $z_0$ in the complex plane.
After this change of variable we get 
\begin{eqnarray}
K &\approx& \exp\left[ f(z_0)+i\varphi\right] \nonumber \\  
& \times & \int_{-\infty}^{\infty} d\rho  ~ \exp\left[ {\rho^2 \over 2}  ~ \exp[i2\varphi]  f''(z_0)  \right],
\label{steepest descent2}
\end{eqnarray}
where the integral runs over the real parameter, $\rho$, but the integrand remains complex. 
Using Eq.~(\ref{steepest descent2}), it is, however, simple to choose an angle for the contour  
so that the exponent becomes negative and real, $2\varphi+\arg[f''(z_0)]=\pi$,  
which turns the integrand into a simple real Gaussian with $\beta=|f''(z_0)|/2$ in analogy with Eq.~(\ref{gaussian}). 
A technical point is that we must have $-\pi/2<\varphi<\pi/2$ for the contour to run in the ``right'' direction, 
{\it i.e.} with the real part of $z$ increasing, but this is typically the case in physical problems.
The final result simplifies to the following beautiful expression  	
\begin{eqnarray}
K &\approx& i \exp\left[ f(z_0) \right]  \sqrt{\frac{2\pi}{f''(z_0)}},
\label{steepest descent3}
\end{eqnarray} 
which we can interpret as the ``height'' $\exp[f(z_0)]$ times the ``width'' $\sqrt{2\pi/f''(z_0)}$ 
of the integrated function close to the saddle-point.  
%
%

It is now of interest to compare the simple integration of a Gaussian, Eq.~(\ref{gaussian}),
with the result of our complex analysis, Eq.~(\ref{steepest descent3}), and conclude that 
the integrand in Eq.~(\ref{gaussian}) may be complex.  
The integral converges provided that $\Re(\beta) > 0$, but also 
in the special case of $\Re(\beta) = 0$, provided that $\Im(\beta)\neq 0$. 
In the latter case, the convergence can be understood as being due to 
rapid complex oscillations for large values of $|\omega|$ as shown in Fig.~\ref{expx2}\,(c),
leading to a finite complex integrated value in Fig.~\ref{expx2}\,(d) of $\sqrt{\pi}\exp[i\pi/4]$. 
%


\subsubsection{General equation for SPA}

Inspired by the results above,
we now consider a typical frequency domain integral of a more general function: $E(\omega)\exp(if(\omega))$, 
where $E(\omega)$ is a smooth amplitude function and $f(\omega)$ is a smooth phase function. 
The phase function may contain several saddle-points, also called ``stationary'' points. 
In regions where $f(\omega)$ is varying fast, we expect the contribution to the integral 
to be small in analogy with Fig.~\ref{expx2}\,(c) and (d).
Conversely, we expect significant contributions from regions close to the stationary frequency,  
$\omega_s$, where the first derivative of the phase function vanishes: 
$\partial f/\partial \omega |_{\omega_s}\equiv f'(\omega_s)\equiv f'_s=0$.  
Then, using the truncated Taylor expansion in Eq.~(\ref{expansion})  
in the vicinity of the stationary points and replacing
the exact integrand by the approximate local Gaussian function, one gets 
the final result for the SPA approximation \cite{WeberArfken2004}:
\begin{equation}
\int_{-\infty}^{\infty} d\omega ~ 
E(\omega)
\exp\left[
if(\omega)
\right]
~\approx~
\sum_s
E_s
\sqrt{\frac{2\pi i}{f''_s}} 
\exp\left[ i f_s \right],
\label{spapprox}
\end{equation}
where $f_s=f(\omega_s)$, $f''_s \equiv \left.\partial^2 f/\partial \omega^2\right|_{\omega_s}$ and $E_s\equiv E(\omega_s)$. 
The sum runs over all the roots of  
the stationary-phase equation:  
\begin{equation}
\frac{\partial f}{\partial \omega}~=~0,
\label{stationary}
\end{equation}
where the roots $\omega_s$ are labelled by the index $s$.
The number of solutions depends on the properties of the phase factor, $f(\omega)$.
In the simplest cases, which we will consider here, there is only one solution. 
The SPA is derived assuming that the second-order term is dominant as compared to 
higher-order phase terms, {\it e.g.} that 
$(\omega-\omega_s)^2f''_s/2 \gg (\omega-\omega_s)^3f'''_s/6$, in the neighbourhood of each saddle point, $|\omega-\omega_s|<\pi$. 
If the second-order phase term happens to be zero, $f''_s/2=\beta=0$, 
this would correspond to an artificial divergence in Eq.~(\ref{spapprox}).   
This divergence is most likely non-physical [and it may happen close to the cut-off in the HHG] and the result would be finite 
if the third-order phase term in the Taylor expansion of the exponent was used in the integral. 
Having derived the SPA, Eq.~(\ref{spapprox}),   
we now need to demonstrate that it can help us to make physical arguments about the propagation of XUV wave packets.

\subsubsection{Stretched XUV wave packets}\label{spaex1}
It is well-known from ultra-fast optics that uncompensated dispersion 
can lead to stretched pulses in the time domain \cite{StricklandOC1985}. 
It is illustrative to use the SPA to calculate the properties of such stretched pulses.
Consider an attosecond pulse, as expressed in Eq.~(\ref{Eintimedomain}),  at $x = 0$,
with a rather broad Gaussian envelope,
\begin{equation}
E(\omega)=E_0 \exp[-g(\omega-\omega_0)^2],  
\label{F0}
\end{equation}
and with a phase factor $f(\omega) = -\omega t + \delta(\omega)$,  
where the spectral phase is 
\begin{equation}
\delta(\omega) \approx \delta_0 + (\omega-\omega_0) \delta'_0  + {1 \over 2}(\omega-\omega_0)^2 \delta''_0,  
\label{delta012}
\end{equation}
with a large quadratic chirp, $|\delta''_0|/2\gg g$.	
The SPA expressed in Eq.~(\ref{spapprox}) then implies that the magnitude of the amplitude factor will be dominated by the contribution of a local value, 
$E_s = E_0\exp[-g(\omega_s-\omega_0)^2]$, occurring at the stationary frequency, $\omega_s$, 
as determined by solving Eq.~(\ref{stationary}).
This is quite similar to the way in which a delta function operates, selecting a specific value of a function. 
Here it arises from the small constructive window in the complex integral that occurs around the stationary frequency. 

The next task is to find the stationary frequency, $\omega_s$,
which can be seen as the frequency that makes the dominant contribution to the wave packet in a given point in space-time. 
The stationary-phase equation, Eq.~(\ref{stationary}), becomes first-order:
$f'(\omega)\approx\delta'_0+(\omega-\omega_0)\delta''_0-t=0$, 
which we use to find the stationary frequency for a given time, $t$: 
\begin{equation}
\omega_s(t)=\omega_0+\frac{1}{\delta''_0}~ (t-\delta'_0).
\label{ws}
\end{equation}
We can rewrite Eq.~(\ref{ws}) to see that, due to the chirp induced from propagation, 
any frequency of the pulse, $\omega=\omega_s(t)$, 
``arrives'' at a {\it unique} time, $t= \delta'_0+[\omega_s-\omega_0]\delta''_0 \approx \delta'_s$, 
which equals the GD as defined in Eq.~(\ref{groupdelay}).
Furthermore, one can show that the stationary frequency equals the instantaneous temporal frequency of the wave packet:  
\begin{equation}
\omega_s(t)= \frac{\partial}{\partial t}\left[ \arg \left\{ \tilde E(t)  \right\} \right],
\label{intantaneousfreq2}
\end{equation} 
which provides a simple link between time and frequency for stretched pulses. 
In the derivation above, it is only required that the spectral envelope  
is ``broad'' as compared to the spectral width induced by the quadratic chirp.  
Any other broad envelope, $E(\omega)$, will also work, leading to a mapping of the intensity 	 
from the spectral domain to the temporal  domain, 
\begin{eqnarray}
\underbrace{|\tilde{E}(t)|^2}_{\textrm{Temporal}} &~ \propto ~&
\underbrace{\frac{1}{|\delta''_0|}}_{\textrm{Constant}}~
\underbrace{\left|E(\omega_s(t))\right|^2}_{\textrm{Spectral}},
\label{envelopelongpulse}
\end{eqnarray}
because the stationary frequency increases linearly in time as shown in Eq.~(\ref{ws}).  
The mapping of the spectral envelope to the temporal envelope 
is scaled by the inverse spectral chirp, $1/\delta''_0$, so that an increased spectral chirp leads to a more
stretched pulse in the time domain. 
At the end of Sec.~\ref{generalizationFirstOrder}, we will see that the properties of these stretched light pulses are 
similar to the way in which non-relativistic electrons are dispersed through propagation in vacuum.

\subsubsection{Ultra-short Gaussian pulse}\label{spaex2} 
Consider again an attosecond pulse, Eq.~(\ref{spapprox}), 
with a Gaussian envelope, Eq.~(\ref{F0}), 
but this time the envelope is not ``broad'' compared to the quadratic spectral chirp, 
instead, as in the cases of interest in attoscience, one have XUV pulses so that $g \approx |\delta''_0|/2$.   
Then, for the SPA to be meaningful, the Gaussian envelope must be \textit{included} in the phase factor 
so that it becomes complex:
\begin{equation}
f(\omega)=\delta(\omega)-\omega t+ig(\omega-\omega_0)^2. 
\label{fastphaseampl}
\end{equation}
Using the asymptotic phase to second order, Eq.~(\ref{delta012}), 
the stationary-phase equation, Eq.~(\ref{stationary}), remains first-order, with complex and time-dependent coefficients
\begin{equation}
f'(\omega)\approx\delta'_0+\delta''_0(\omega-\omega_0)-t+i2g(\omega-\omega_0)=0,
\end{equation}
with the root: 
\begin{equation}
\omega_s(t)
= \omega_0 + \frac{\delta''_0-i2g}{\delta''_0+(2g)^2} ~ (t-\delta'_0).
\end{equation}
We mention that it is common to obtain complex stationary points in the SPA 
in the case of steep variations of the spectral amplitude.  

The real part of $\omega_s$ is identified as the instantaneous frequency, as was done in Sec.~\ref{spaex1}; 
while the imaginary part of $\omega_s$ limits the intensity duration (or probability distribution) 
of the attosecond pulse in time. 
At the origin, $x=0$, the intensity of the attosecond pulse will be large only at a time, $t$, close to the group delay:
$t\approx\delta'_0$, where the imaginary part of the stationary frequency vanishes. 
Inserting $\omega_s(t)$ into the right-hand-side of Eq.~(\ref{spapprox}) yields the result:
\begin{equation}
\tilde E(t)=\frac{E_0}{2\pi}\sqrt{\frac{\pi}{\beta}}\exp\left[i(\delta_0-\omega_0 t)-\frac{(\delta'_0-t)^2}{4\beta}\right],
\label{exactGaussian}
\end{equation} 
where $\beta=g-i\delta''_0/2$.
In this particular case, the result of the SPA is exact because it was applied to a Gaussian pulse, but the  
method is more general and it can be applied also to other spectral envelopes. 
This is achieved by replacing the third term in Eq.~(\ref{fastphaseampl}) by the more general expression: $\ln[E(\omega)]/i$.  

In the special case of a {\it Fourier-limited} pulse, $\delta''_0=0$, 
the real part of the stationary frequency becomes the central frequency of the pulse at {\it all} times;
while the imaginary term remains time-dependent: 
$\omega_s(t) = \omega_0 + i(\delta'_0-t)/2g$.
This is closely related to the Fourier-limited pulse shape, which has a constant carrier frequency equal to the central frequency. 
The imaginary part of the stationary frequency imposes the finite duration of the pulse in time.

\subsection*{Summary of attosecond pulse propagation}
\label{attosecondpulsesummary}
In this section, we have reviewed the basic properties of light pulses that propagate 
through a medium in one dimension.
Attosecond XUV light pulses are a special kind of light pulse with a large spectral bandwidth.
The larger the bandwidth, the more sensitive the pulses are to the dispersion induced by the medium.
Using an appropriate medium, it is possible to correct for spectral chirps and reduce the
attosecond pulses to their shortest possible duration.
Once the attosecond pulses are adjusted, they can be used to ionize atoms at extremely well-defined times \cite{MauritssonPRL2008}. 
We now review the properties the properties of non-relativistic photoelectron wave packets.




\section{Photoelectron wave packets}
\label{sec:wpelectron}
%
Attosecond XUV pulses have been used to initiate photoelectron wave packets at well-defined times, 
but how does such a wave packet evolve after the ionization event? 
In this section, we perform an analysis of the evolution of these wave packets and we shed light on the similarities with the propagation of attosecond XUV pulses, Sec.~\ref{sec:wplight}.
Although our final goal is to describe photoionization as a time-dependent process, 
we start our study with a theoretical treatment of positive energy photoelectron wave packets in the presence of a static binding  potential. More details on the actual attosecond-XUV transition are given in Sec.~\ref{sec:photoionization}, 
while the important features related to laser-probing in long-range Coulomb potentials are given in Sec.~\ref{sec:STPT}.  

In the absence of an external field, the time-evolution of a non-relativistic photoelectron is described by the time-dependent Schr\"{o}dinger equation (TDSE) \cite{Friedrich1994}
\begin{equation}
H_0 \bigl|\Psi(t)\bigl>~=~
[T+V_0] \bigl|\Psi(t)\bigl>~=~
i\hbar \frac{\partial }{\partial t}\bigl|\Psi(t)\bigl>,
\label{TDSE}
\end{equation}
where $H_0$ is a time-independent Hamiltonian, and 
$\Psi=\Psi(t,\vec{r})=\bigl<\vec r\bigl|\Psi(t)\bigl>$ is the time-dependent wavefunction.
The Hamiltonian consists of two parts:   
$T=-\hbar^2\nabla^2/(2m)$ being the kinetic energy operator, 
and $V_0=V_0(\vec{r})$ being a time-independent, single-electron potential describing the binding of the electron to the atom. 
For simplicity, we limit ourselves to systems that can be described by a single electron, namely the SAE approximation.
A common approach to solving the TDSE, Eq.~(\ref{TDSE}),  
is to first find the solutions to the {\it time-independent} Schr\"{o}dinger equation,
\begin{equation}
H_0\psi_{\alpha}(\vec{r})= \epsilon_\alpha \psi_{\alpha }(\vec{r}),
\label{SE}
\end{equation}
where $\epsilon_{\alpha}$ is the energy (eigenvalue) of the solution (eigenstate), $\psi_{\alpha}(\vec r)$
labelled by a set of quantum numbers $\alpha$.
In the following, these time-independent solutions are referred to as \textit{states}.
They form the basis that spans the complete space in which the electron moves.
%

For a spherical potential, $V(r)$, 
the states can be written on a spherical basis, 
$\psi_{\alpha}(\vec{r})=R_{n,\lambda}(r)Y_{\lambda,m}(\theta,\phi)$,
where $R_{n,\lambda}$ is the radial wavefunction and $Y_{\lambda,m}$ is the spherical harmonic, 
for quantum numbers: $\alpha=[n,\lambda,m]$, with spin neglected for simplicity.
It is possible to reduce the Schr\"{o}dinger equation 
into separate radial equations, each corresponding to 
a specific angular momentum quantum number, $\lambda$. 
%
To emphasize the similarities and differences between 
the one-dimensional propagation of light wave packets  [Sec.~\ref{sec:wplight}],
the \textit{effective radial} electron states, $u_{n,\lambda}(r)=rR_{n,\lambda}(r)$, are used. 
These effective radial states can be interpreted as the solutions of 
a {\it one} dimensional Schr\"odinger equation for $r>0$  
with an effective potential
\begin{equation}
[T+\underbrace{V_r(x)+V_\lambda(x)}_{V_0(x)}] u_{n,\lambda}(x) =  \epsilon_\alpha u_{n,\lambda}(x),
\label{SE1D}
\end{equation}
where  
$T=-\hbar^2/(2m)\partial_{xx}$ is the radial kinetic energy,
$V_r(x)=-e^2/(4\pi\epsilon_0 x)$ is the radial potential (explicit for hydrogen) and 
$V_{\lambda}(x)= \lambda(\lambda+1)\hbar^2/(2mx^2)$ 
is the centrifugal potential that depends the angular momentum through $\lambda$.
Starting with Eq.~(\ref{SE1D}), the radial variable $r$ is relabelled $x$ 
to ease the comparison between the propagation of one-dimensional light pulses and of photoelectron wave packets. 
The boundary condition for all
states is $u_{n,\lambda}(0)=0$, 
and the states are  taken to be real
%
%
A photoelectron wave packet, in angular subspace $\lambda$, 
can then be can be written as an integral--superposition of continuum states,
\begin{equation}
\Psi_{\lambda}(t,x)=\int_0^{\infty} d\epsilon ~a(\epsilon,t)u_{\epsilon,\lambda}(x)\exp\left[-i\epsilon t/\hbar\right],
\label{superposition}
\end{equation} 
where $a(\epsilon,t)$ is the complex amplitude for the energy-normalized radial state, $u_{\epsilon,\lambda}(x)$, at energy $\epsilon>0$ . 

\subsection{Phase propagation of photoelectrons}
\label{sec:WKB}
The propagation of a  photoelectron wave packet, Eq.~(\ref{superposition}),  
is governed by the way in which
the phases of the continuum states $u_{\epsilon,\lambda}(x)$ vary as a function of energy, $\epsilon$. 
The Wentzel--Kramers--Brillouin (WKB) approximation is well suited for studying 
continuum states within a semi-classical framework \cite{Friedrich1994}. 
The WKB approximation relies on a similar ansatz as in Sec.~\ref{sec:waveproplight} for XUV light. 
For an effective potential which vanishes at infinity, $\lim_{x\rightarrow\infty}V_0(x)=0$, 
the effective radial wavefunction is taken to be 
\begin{equation}
u_{\epsilon}(x) \propto \Im\left\{\exp[i\phi(\epsilon,x)]\right\}=\sin[\phi(\epsilon,x)],
\label{u2phase}
\end{equation} 
where reference to $\lambda$ is dropped for compactness. 
Inserting the WKB ansatz into the Schr\"{o}dinger equation, Eq.~(\ref{SE1D}), leads to a 
differential equation for the phase of the state,
\begin{equation}
-i\frac{\hbar^2}{2m}\frac{\partial^2\phi}{\partial x^2} + 
\frac{\hbar^2}{2m}\left(\frac{\partial\phi}{\partial x}\right)^2
=\epsilon-V_0(x).
\label{WKBeq1}
\end{equation}
Further, assuming that the {\it local momentum} of the photoelectron varies slowly, 
$|p(x)/\hbar| \equiv |\partial \phi/\partial x| \gg |\partial^2 \phi / \partial x^2|^{1/2} $, the above equation can be simplified, 
\begin{equation}
\frac{\partial\phi}{\partial x}
=\frac{1}{\hbar}\sqrt{2m[\epsilon-V_0(x)]},
\label{WKBeq2}
\end{equation}
and then integrated, 
\begin{eqnarray}
\phi(\epsilon,x,x_0)
&~=~&\frac{1}{\hbar} \int_{x_0}^{x}dx' ~ \sqrt{2m[\epsilon-V_0(x')]} \nonumber \\
&~\equiv~&\frac{1}{\hbar} \int_{x_0}^{x}dx' ~ p(x'),
\label{WKBeq3}
\end{eqnarray}
where $p(x)$ is the local momentum of the electron, and $x_0$ is from where the electron ``starts''. 
In applications, the parameter $x_0$ is associated to a classical turning point, see below. 
The local momentum corresponds to the local kinetic  energy, which is the total energy minus the local potential energy. 
The phase of the wavefunction varies faster as the local momentum is increased.
%
Eq.~(\ref{ansatzeqapproxphi}) and~(\ref{WKBeq3}) are now compared to 
highlight the differences between the propagation of XUV light and photoelectrons. 
In a medium with negligible dispersion, the phase of light is approximately proportional to the photon momentum, 
$\omega/c$, leading to a simple translation of the wave packet in space-time according to the shift theorem, Eq.~(\ref{shiftteorem}). 
Similarly, in regions of space where the action of the potential is negligible, $V_0(x)\ll \epsilon$, 
the phase of the electron is proportional to the electron momentum, 
but this corresponds to the square root of the energy and frequency, $ p(x) \propto \sqrt{\epsilon}$, 
in contrast to the linear frequency dependence of light.
This implies that the electron wave packet will {\it not} maintain its shape under propagation. 
This non-linear phase dependence leads to a broadening effect referred to as ``quantum diffusion'' 
and it originates physically from the fact that electrons of different energies travel at different speeds. 
In the more general case with dispersion, we may rewrite Eq.~(\ref{WKBeq3}) and define an ``electron-susceptibility'' ${\chi^{(e)}}=-V_0(x)/\epsilon$
in analogy with the linear susceptibility phase of light, $\chi^{(1)}$.
The electron phase is real as long as $\epsilon>V_0$, 
but it will be imaginary if $\epsilon<V_0$, which implies that the electron must tunnel through some barrier
and that the transmission amplitude will be exponentially damped.

When writing the WKB solution, it is natural to include $|\partial^2 \phi/\partial x^2|$ to leading  order \cite{Friedrich1994}. 
This leads to the well-known form for the WKB state 
\begin{equation}
u^{(WKB)}_{\epsilon,(\pm)} \propto \frac{1}{\sqrt{p(x)}} \exp \left[ \pm\frac{i}{\hbar}\int_{x_0}^{x}dx'~ p(x') \right] ,
\label{WKBstate}
\end{equation}
where the two complex solutions $(\pm)$ describe  
an outgoing and incoming electron, respectively. 
The probability density, $\rho(x)=|u^{(WKB)}_{\epsilon,(\pm)}(x)|^2\propto 1/|p(x)|$, 
is smaller where the electron moves rapidly.
Physically, we understand this because the electron will spend a shorter time 
in a place when moving with a greater velocity. 
Alternatively, we can form an energy normalized and real WKB solution,
\begin{equation}
u^{(WKB)}_\epsilon(x) =
\sqrt{\frac{2m\hbar}{\pi p(x)}} 
\sin \left[ \frac{1}{\hbar}S_\epsilon(x,x_0)+\phi_{x_0} \right],
\label{realWKBstate}
\end{equation}
which is valid for $x>x_0$, 
where $x=x_0$ is a {\it classical turning point}, 
{\it i.e.} it separates a classically forbidden region, $\epsilon<V_0(x)$ for $x<x_0$, 
and a classically allowed region, $\epsilon>V_0(x)$ for $x>x_0$.
The phase of the state is a sum of the semi-classical action,
\begin{equation}
\frac{1}{\hbar}S_\epsilon(x,x_0) ~ = ~\frac{1}{\hbar} \int_{x_0}^{x}dx'~ p(x'),
\label{Swkb}
\end{equation}
plus the quantum mechanical reflection phase, $\phi_{x_0}$.
Depending on the type of potential and on the energy of the electron, 
the quantum mechanical reflection at the classical turning point 
can be regarded as either ``hard'' or ``soft'' \cite{TrostPL1997}. 
Photoelectrons in a pure Coulomb potential, with no angular momentum, 
will suffer a hard reflection at $x_0=0$, with a reflection phase of $\phi_{x_0} = 0$, 
so that the wavefunction is exactly zero at the reflection point, {\it i.e.} 
$u^{(WKB)}_\epsilon(0)\propto \sin(0)=0$. 
Photoelectrons with angular momentum, $\lambda>0$, will reflect in an effective potential
composed of the Coulomb potential and the centrifugal potential at $V_0(x_0)\equiv V_r(x_0)+V_\lambda(x_0)=\epsilon$.
This combined barrier typically leads to reflection points further out from the nucleus, $x_0>0$,  
and to softer reflections. 
A soft reflection implies that the photoelectron wavefunction can extend a bit 
into the classically forbidden region, $x<x_0$. 
The soft reflection limit, {\it i.e.} when the effective potential is slowly varying compared to the wavelength of the electron, 
can be calculated using a linear potential so that the wavefunction becomes an Airy function with a reflection phase of $\pi/4$ 
\cite{Friedrich1994}, but this is typically not a good approximation for photoelectrons that we consider here. 
An accurate determination of the reflection phase, $\phi_{x_0}$, 
requires a fully quantum mechanical theory beyond the WKB approximation
and {\it ab initio} calculations of such phases present a challenge to theory for large atomic systems or complex molecular systems. 

\subsection{Quantum diffusion and Wigner delays}
\label{sec:wignersmith}
As an example, consider the situation illustrated in Fig.~\ref{WKBbarrier},
where an electron passes through a short-range attractive potential.
%
%
\begin{center}
	\includegraphics[width= 0.85\textwidth]{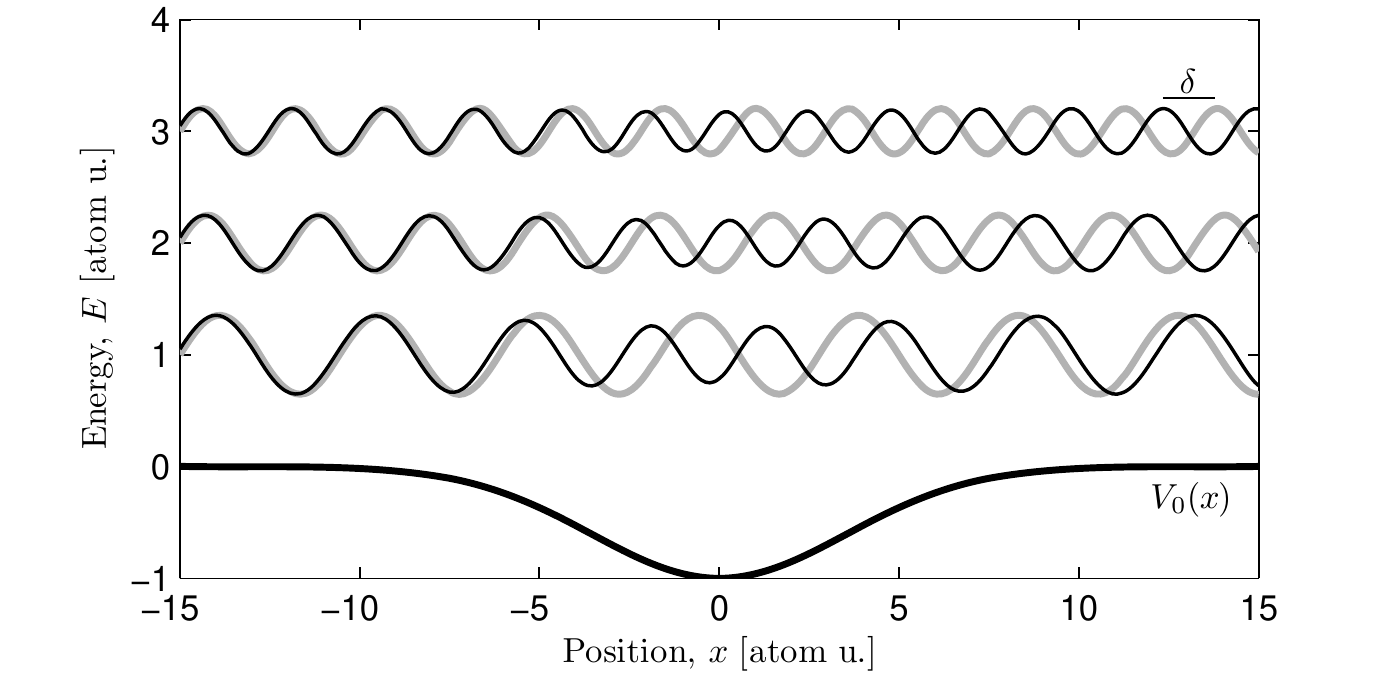}
	\captionof{figure}{
Electrons passing through an attractive potential, $V_0(x)$. 
The real WKB solutions (thin black curves), Eq.~(\ref{realWKBstate}), 
are compared to the real free electron states (thick grey curve) 
at three different energies: $\epsilon=1,2,3$ atomic units of energy (27.2\,eV).
The phase of the electron wavefunction varies more rapidly in the potential, 
which leads to an accumulated phase difference, $\delta$, compared to the free case.
(Note that the modulations of the wavefunctions 
should be interpreted in the third dimension of the graph, and not 
as an energy modulation.)	
	}
	\label{WKBbarrier}
\end{center}
The electron is classically allowed to pass through the potential, 
but as it does so, it will acquire a quantum phase.
The asymptotic phase difference, between an electron wave 
propagating through a short-range potential, Eq.~(\ref{WKBeq3}), and free wave propagation,
is defined in analogy with Eq.~(\ref{phasedifference}): 
\begin{equation}
\delta(\epsilon) = \frac{1}{\hbar}
\lim_{x\rightarrow\infty} 
\int_{-\infty}^{x}dx' ~ \left[ p(x) - p_0 \right],
\label{ephasedifference}
\end{equation}
where $p_0=\sqrt{2m\epsilon}$ is the free (asymptotic) momentum of the electron.
The GD concept, Eq.~(\ref{groupdelay}), can also be applied 
to electron wave packets using Eq.~(\ref{ephasedifference}):
\begin{equation}
\tau_{\lambda}(\epsilon)
=\frac{\partial \delta}{\partial \omega}
=\hbar \frac{\partial\delta}{\partial \epsilon}.
\label{wignerdelay}
\end{equation} 
The electron GD then results only from the interaction with the potential, 
since the intrinsic delay due to quantum diffusion has been subtracted.
Using the asymptotic phase to calculate the delay of electron wave packets was first done by 
Eisenbud, Wigner and Smith  \cite{WignerPR1955,SmithPR1060,CarvalhoPR2002}, 
hence the name Eisenbud--Wigner--Smith delay, (or Wigner delay for short).
In the simple scattering example given above, the asymptotic phase shift equals twice the radial phase shift:
once for the incoming electron ($0>x$) and once for the outgoing electron ($0<x$). 
Regarding photoionization, it can be considered as a \textit{half-collision} because the electron only moves out of the atom ($0<x$). 
As compared to free propagation, the delay of the wave packet in photoemission is, therefore, the derivative of the radial asymptotic phase-shift.
%

For a weak potential, $|V_0| \ll \epsilon$, the local momentum is $p(x)\approx p_0\left[ 1 - V_0(x)/2\epsilon \right]$, which leads to an asymptotic phase
\begin{equation}
\delta(\epsilon) 
\approx 
-\frac{1}{\hbar}
\sqrt{\frac{m}{2\epsilon}} \int_{-\infty}^{\infty}dx' ~ V_0(x')
\equiv
-\frac{1}{\hbar}
\sqrt{\frac{m}{2\epsilon}} ~ I_V,
\label{ephasedifferenceweak}
\end{equation}
where the potential integral can be written as $I_V \approx \epsilon_V L_V$, 
in terms of an effective ``height'' $\epsilon_V$ and ``range'' $L_V$ of the potential well.
As expected, in this lowest-order approach, the phase acquired by the electron 
increases linearly with both the height and range of the barrier.
The Wigner delay is 
\begin{equation}
\tau_\lambda=\sqrt{\frac{m}{8}}\frac{I_V}{\epsilon^{3/2}}
\approx\sqrt{\frac{m}{8}}\frac{\epsilon_V}{\epsilon^{3/2}}L_V.
\label{wignerweak}
\end{equation}
It takes a shorter time, $\tau_\lambda<0$, for an electron to pass a weak attractive potential, $\epsilon_V<0$, than through free space. 
Physically, this is due to the higher local velocity of the electron in the potential valley.
The opposite is true for a weak repulsive potential, where the electron is slowed down on the potential hill.
Furthermore, the timing of slow electrons is affected more than the timing of fast electrons, 
 due to the longer total time that the slow electrons spend in the potential.
%
%
In terms of numbers, this simple analysis leads to a delay of $5.9$\,as per eV and nm of the potential barrier, given an electron with one unit of atomic energy ($27.2$\,eV).

\subsection{Phases and delays in Coulomb potentials}
\label{sec:coulomb}
Photoelectrons created from neutral atoms are submitted to the {\it long-range} Coulomb potential of the remaining ion, 
$V_C=-C/x$, where the constant is $C=Ze^2/(4\pi\epsilon_0)$ with $Z=1$ for a singly charged hydrogen ion.
In order to study this long-range interaction in detail, 
we consider the WKB approximation, Eq.~(\ref{realWKBstate}). 
Then, the local momentum can be expanded as 
\begin{equation}
p(x) = \sqrt{2m\left[\epsilon+\frac{C}{x}\right]} \approx p_0\left[1+\frac{C}{2\epsilon x}\right], 
\label{coulombplocal}
\end{equation}
with $p_0=\sqrt{2m\epsilon}$
being the asymptotic momentum 
in the remote region of space where  
$\epsilon \gg |C/x|$.
Using Eq.~(\ref{Swkb}) and (\ref{coulombplocal}), 
the total phase of the real WKB state is asymptotically 
\begin{eqnarray}
\frac{S_\epsilon(x,x_0)}{\hbar}+\phi_{x_0}
~& \approx &~ 
kx +\frac{C}{2\epsilon}\ln(x) - kx_0 -\frac{C}{2\epsilon}\ln(x_0)+\phi_{x_0}  
\nonumber \\
~&\equiv&~
k x ~+~ \underbrace{\frac{\ln(2 k x)}{k a_0}~+~\eta_{k,\lambda}~-~\frac{\pi\lambda}{2} }_{\Phi_{k,\lambda}(x)}  
\label{coulombphase}
\end{eqnarray}
where $k=p_0/\hbar$ is the wave number; $a_0$ is the Bohr radius; and $Z = 1$.  
Line 1 of Eq.~(\ref{coulombphase}) corresponds to the WKB notation with an approximate reference to the reflection point $x_0$,  
while line 2 follows standard notation from atomic physics with reference to 
the absolute scattering phase, $\eta=\eta_{k,\lambda}$.
The real WKB state varies as 
\begin{eqnarray}
u_{k,\lambda}(x)  &\propto&
\sin\left[k x+\frac{\ln(2k x)}{k a_0} + \eta_{k,\lambda} -\frac{\pi\lambda}{2}  \right] 
\label{coulombfunction}
\end{eqnarray}
for $x\rightarrow\infty$.
Interestingly, the phase does not settle into free particle behaviour, 
instead, the phase diverges logarithmically! 
What does this space-divergent phase mean for the delay of the photoelectron wave packet?
When applying the frequency derivative to the the logarithmic term, 
we find a delay that depends on the {\it spatial} position of the wave packet, 
\begin{eqnarray}
\tau_{LR}(k,x)&\equiv&
\hbar \frac{\partial}{\partial \epsilon}
\left[
\frac{\ln(2 k x)}{k a_0}
\right]  \nonumber \\&\approx&
\frac{m}{\hbar a_0 k^3 }\left[
1 - \ln\left(\frac{ 2\hbar k^2 }{ m } \tau \right)
\right],
\label{wignercoulomb}
\end{eqnarray}
where we have replaced this spatial-position of the electron, $x$, by the approximate classical position, 
$x_{cl}(\tau)\approx v\tau =\hbar k \tau /m$, 
with $\tau=t-t_0$ being the time that has passed since the electron was ionized by the attosecond pulse at $t=t_0$.
Interestingly, $\tau_{LR}$ also diverges logarithmically as $\tau$ increases and it cannot be neglected.   
As a consequence, $\tau_{LR}$ will completely dominate over any (short-range) Wigner delay, 
$\tau_\lambda  \ll \tau_{LR} \ll \tau$, as $\tau\rightarrow\infty$.  
This implies that there is  {\it no absolute} delay of the Coulomb wave packet with respect to free propagation.
To define a physical delay between two photoelectron wave packets of different energy is already difficult for free propagation, 
due to the different velocities of the photoelectrons as discussed with Eq.~(\ref{wignerdelay}), 
but it is even more difficult for Coulomb wave packets due to the spatial dependence of the phase.
In the special case of two {\it different} Coulomb wave packets of the {\it same} energy, 
the logarithmic delays cancel and their relative delay then equals the difference in Wigner delay.
In this way, we can talk about the {\it delay} of a Coulomb wave packet, 
but only relative to a Coulomb reference, {\it e.g.} the hydrogenic system.      
To summarize this analysis, for the propagation of photoelectron wave packets, we find that 
the approximate position is predominately given by the linear, free propagation relation, $p_0x - \epsilon \tau=0$.
The logarithmic, long-range phase is of intermediate importance, and it dominates over the absolute, short-range phase.
On the other hand, only the short-range phase contain detailed information about the atomic potential. 

As already mentioned, 
the absolute determination of the short-range asymptotic phase-shift, $\eta$, is a difficult problem,  
because it depends on the detailed atomic potential, which is modified by electron--electron interactions at short range. 
In the special case of hydrogen system of nuclear charge $Z$, and for a given wave number, $k$, and angular momentum, $\lambda$, the phase is known analytically \cite{Friedrich1994},
\begin{equation}
\eta_{k,\lambda}^{(H)}\equiv\sigma_{k,\lambda}\equiv\arg\{\Gamma[\lambda+1-iZ/(k a_0)]\}, 
\label{etaH}
\end{equation}
with $\Gamma(z)=\int_0^{\infty}dt~t^{z-1}\exp[-t]$ being the complex Gamma function.
This result can be generalized for any spherical atom, 
\begin{equation}
\eta_{k,\lambda}\equiv\sigma_{k,\lambda}+\delta_{k,\lambda},
\label{etaAny}
\end{equation}
where $\delta_{k,\lambda}$ is the phase difference compared to hydrogen that is induced by the short-range deviation from the pure Coulomb potential.
The WKB states are compared with the \textit{exact} Coulomb states for hydrogen in Fig.~\ref{coulombstates}.
\begin{center}
	\includegraphics[width= 0.85\textwidth]{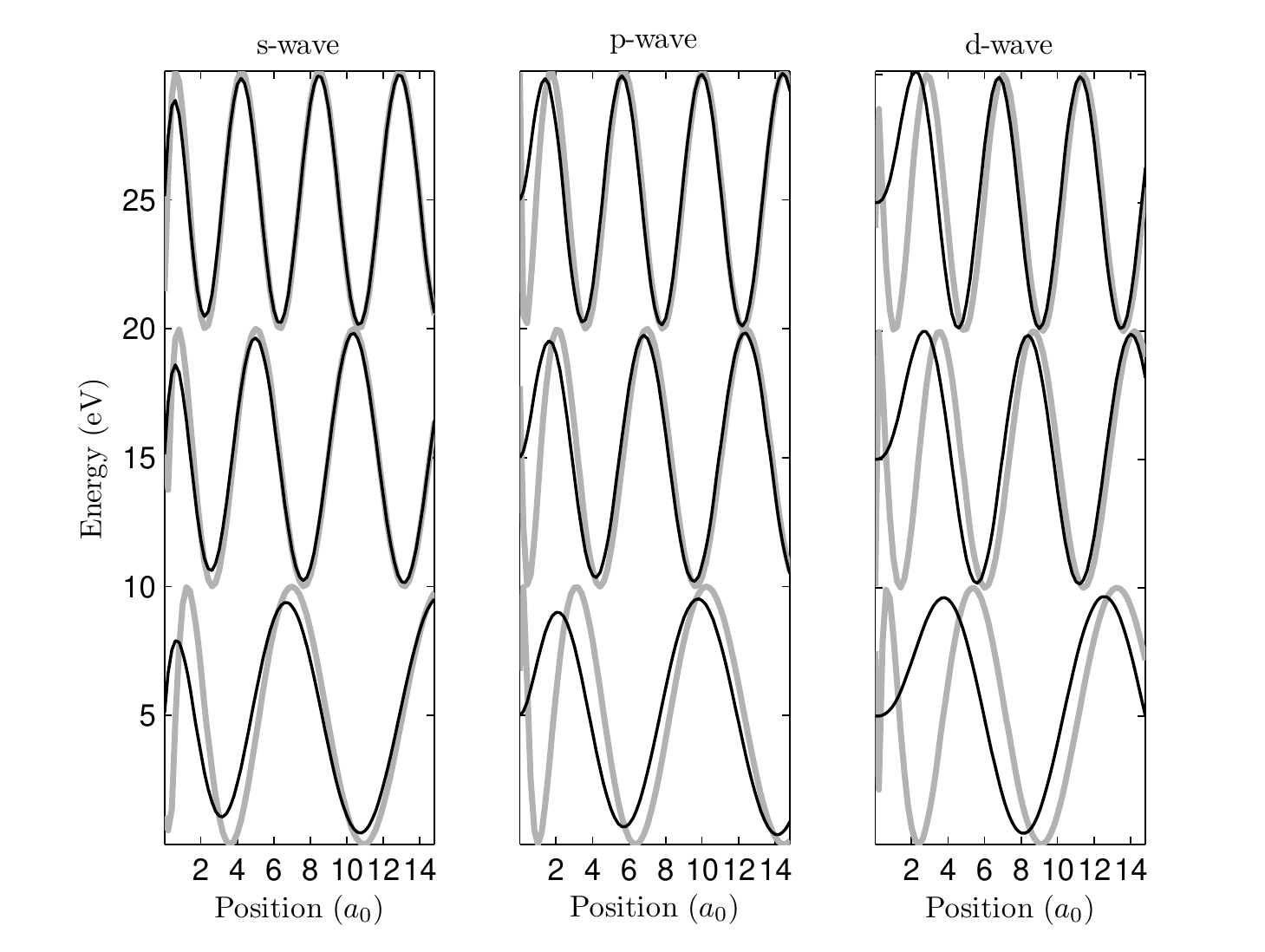}
	\captionof{figure}{
Exact hydrogen wavefunctions (thin black curve) for $\lambda=0$, $1$ and $2$, 
corresponding to s-, p- and d-waves in the continuum at three different energies 5, 15 and 25\,eV.
The {\it asymptotic approximation} (grey thick curve) is shown for comparison. 
It will be explained further in Sec.~\ref{sec:photoionization} and \ref{sec:STPT},
but we mention here that it  
utilizes the long-range WKB phase matched to the correct asymptotic phase-shift,
while the amplitude prefactor is taken to be constant [see Eq.~(\ref{u2phase})]. 
(Note that the modulations of the wavefunctions 
should be interpreted in the third dimension of the graph, and not 
as an energy modulation.)
	}
	\label{coulombstates}
\end{center}
The exact analytical states of the hydrogen atom are called the regular Coulomb functions of the first kind \cite{Abramowitz,MathWorld}.
The WKB states converge quickly to the exact solution, if the correct scattering phase is known and applied \cite{TrostPL1997}.



\section{Photoionization on the attosecond timescale}
\label{sec:photoionization}
In this section, we will discuss, within the framework of lowest-order time-dependent perturbation theory \cite{SakuraiMQM1994,Landau}, 
the photoionization of atoms in the presence of attosecond XUV pulses (SAP or APT).
Photoionization is a fundamental process in light-matter interaction, 
which has been studied extensively using time-independent methods, 
in parallel with the implementation of synchrotron radiation sources,  
see for instance the seminal papers about Cooper minima 
and photoelectron phase-shifts \cite{CooperPR1962,KennedyPRA1972}. 
Here we start by studying photoionization in the time-domain, 
but as we proceed we shall recover the more traditional, spectral representation 
for the determination of the {\it phase-shifts} of the photoelectron states, 
from which one can recover {\it relative delays} between photoelectron wave packets.

Although attosecond pulses of high-order harmonics  have relatively high peak intensities, 
as compared to conventional sources of XUV radiation, 
single-photon absorption remains the dominant mechanism for atomic ionization. 
Then, due to this {\it fluence} of XUV or X-ray photons 
being sufficiently small, the interaction between attosecond pulses and atoms can be approximated using 
first-order, time-dependent perturbation theory \cite{SakuraiMQM1994,Landau}.
%
The field-free TDSE, as written in Eq.~(\ref{TDSE}), 
is satisfied by the time-dependent {\it field-free} states of the atom,  
\begin{equation}
\bigl|\tilde \alpha(t)\bigl>~\equiv~ \bigl|\tilde \alpha\bigl>~\equiv~ \bigl|\alpha\bigl>~\exp[-i\omega_\alpha t],
\label{excitedstates}
\end{equation} 
where $\alpha$ labels the quantum numbers for a given energy, $\epsilon_\alpha=\hbar\omega_\alpha$. 
The interaction between the XUV field, 
$\tilde V_I(z,t)$, 
and the unperturbed, initial state of the atom, $\bigl|\tilde i(t)\bigl>$ with energy $\epsilon_i=\hbar\omega_i$ and a binding energy of $I_p=|\epsilon_i|$,  
leads to the creation of a photoelectron wave packet 
\begin{eqnarray}
\bigl|\Psi(t)\bigl> & \approx & 
\bigl|\tilde i(t)\bigl>~+~
\intsum{\alpha} ~
a_\alpha^{(1)}(t)~\bigl|\tilde{\alpha}(t)\bigl>,
\label{firstwp}
\end{eqnarray}
where the first-order complex amplitudes are \cite{SakuraiMQM1994}
\begin{eqnarray}
a^{(1)}_\alpha(t) 
~& = &~ 
\frac{1}{i\hbar} \int_{-\infty}^{t} dt'~ 
\bigl<\tilde{\alpha}(t')\bigl|\tilde V_I(z,t')\bigl|\tilde{i}(t')\bigl>.
\label{firstorderpert}
\end{eqnarray} 
In the case of photoionization, the interaction Hamiltonian,
$\tilde V_I(z,t)=e z ~\tilde E(t)$, 
is the dipole-interaction operator for the attosecond XUV field, 
here given in length gauge for linear polarization along $\hat z$. 
For the case of short, coherent pulses, {\it e.g.} attosecond pulses, 
the interaction can be expanded as 
\begin{eqnarray}
\tilde V_I(z,t)
~ &=&~ e z ~ \frac{1}{2\pi} \int d\Omega' ~ E(\Omega') \exp[-i \Omega' t],
\label{dipoleoperator}
\end{eqnarray}
where $E(\Omega')$ are the electric-field, spectral components of the attosecond pulse.
After the interaction is over, when the attosecond pulse has passed, the first-order complex amplitude, 
or {\it S-matrix element} (not to be confused by the electron action), 
is proportional to both the dipole matrix element, $z_{\alpha i}=\bigl<\alpha\bigl|z\bigl|i\bigl>$, 
and to the Fourier transform of the electric field, 
\begin{eqnarray}
S_{\alpha/i}^{(1)}~&\equiv&~
\lim_{t\rightarrow\infty}~a^{(1)}_\alpha(t) \nonumber \\ 
~&=&~
\frac{e}{i\hbar}~z_{\alpha i} 
\int dt~
\tilde{E}(t)\exp\left[i(\omega_\alpha-\omega_i) t\right] 
\nonumber \\
~&=&~
\frac{e}{i\hbar}~z_{\alpha i} ~ E(\Omega),
\label{firstorderover}
\end{eqnarray}
at the energy-conserving frequency, $\Omega=\omega_\alpha-\omega_i$, where $\ket{\alpha}$ is a continuum state 
under the single-active electron (SAE) approximation.  
This shows that the Fourier components of the attosecond pulse  
are continuously mapped on the \textit{complex transition amplitude} associated to the photoionization process.
Most attosecond pulse-characterization schemes take advantage of this 
mapping from XUV light to photoelectron states, 
see for instance Ref.~\cite{MairessePRA2005} and \cite{VeniardPRA1996,Muller2002}
for the attosecond streak camera and the RABITT method respectively, 
which we will discuss further in Sec.~\ref{sec:STPT}. 
%

More generally, in order to determine the outcome of the interaction, {\it e.g.} the momentum distribution, 
the complete photoelectron wave packet can be projected on a momentum state, 
\begin{equation}
S_{\vec k / i} = \lim_{t\rightarrow\infty}\braket{\vec k}{\Psi(t)},
\label{scatterstate}
\end{equation}
of the field-free Hamiltonian with energy, $\epsilon_k=\hbar^2k^2/(2m)$.  
The real expectation values, $|S_{\vec k/i}|^2$, yield then the corresponding probability density of the momentum states.
These momentum states are {\it not} plane waves, but they do approach 
a plane-wave like behaviour in the asymptotic limit, $r\rightarrow\infty$. 
Their wavefunction, $\varphi_{\vec k}(\vec r) = \braket{\vec r}{\vec k}$,
can be expanded on partial-waves as \cite{Landau}
\begin{equation}
\varphi_{\vec k}(\vec r) =
(8\pi)^{3/2}\sum_{L,M} i^L e^{-i\eta_L(k)}Y^{*}_{L,M}(\hat k) Y_{L,M}(\hat r) R_{k,L}(r),
\label{planewaveexpanded}
\end{equation}
where the inserted scattering phases, $i^L e^{-i\eta_L(k)}$, are designed to form  
an {\it unified phase front}, of all partial waves $(L,M)$, in the forward direction, $\hat k$. 
For electrons ejected from a neutral system, outgoing waves have the following generic form 
(as we will discuss further in Sec.~\ref{afofpwp})
\begin{equation}
u_{k,L}^{(out)}(r) \propto \exp \left\{i\left[kr+\frac{\ln(2kr)}{ka_0}+\eta_L(k)-\frac{\pi L}{2}\right]\right\}, 
\label{asympR}
\end{equation}
where the asymptotic phases clearly cancel with the phase factors in Eq.~(\ref{planewaveexpanded}) for each partial wave.   
However, there remains a logarithmic term due to the {\it long-range} Coulomb potential, 
as derived in Sec.~\ref{sec:coulomb}, 
which shows that the asymptotic behaviour of the momentum state is {\it not} exactly that of a simple plane wave. 
Furthermore, it is only the phase {\it difference} between the partial waves that is {\it physical}, 
so that the total phase of the state can be set arbitrarily.

In this way, the spectral intensity distribution of attosecond XUV pulses can be determined 
using electron spectrometers, such as time-of-flight (TOF) or velocity-map imaging (VMI), 
to count the number of photoelectrons reaching the detector, 
and then correcting for the dipole matrix element, {\it i.e.} for the ``quantum efficiency''
\cite{KruitJPE1983,EppinkRSI1997,VrakkingRSI2001}, as shown in Fig.~\ref{harmonics}. 
We stress, however, that in either case, 
information about the temporal structure of the XUV light pulse 
is not accessible because the phase of the complex amplitude is not measured. 

\subsection{Snapshots of photoelectron wave packet}
By considering a simple ``flat-top'' ($\sqcap$) XUV attosecond pulse, we can observe 
``snap-shots'' of how the photoelectron wave packet builds up with time.
The flat-top attosecond pulse is of a given duration, $\tau_I$, and it has
a central frequency of $\hbar\Omega>I_p$.
%
Fig.~\ref{outgoing} illustrates the evolution of the complex amplitudes and the 
corresponding reconstructed photoelectron wave packets, $\bigl|\Psi^{(1)}(t)\bigl>$, 
for increasing interaction durations, $\tau_I$.

\begin{center}
	\includegraphics[width= 0.85\textwidth]{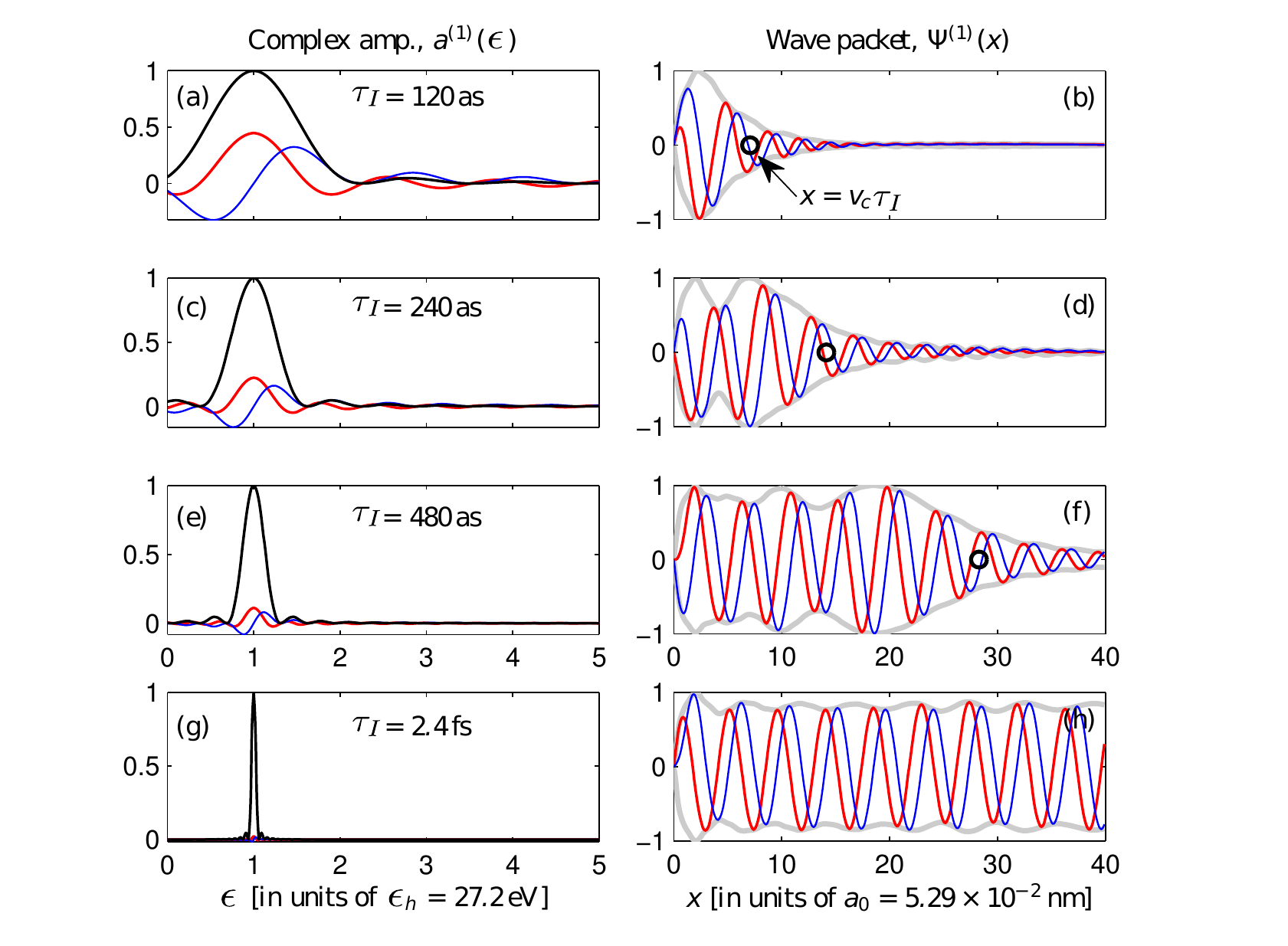}
	\captionof{figure}{
Complex amplitudes (left) and ``snapshots'' of the corresponding photoelectron wave packets (right) 
after photoionization of increasing pulse duration, $\tau_I$: 
(a,b) 120, (c,d) 240, (e,f) 480 and (g,h) 2400 as. 
(a,c,e,g) The probability distribution over energy (black line) 
becomes narrower as $\tau_I$ increases. 
(b,d,f,h) In space, this corresponds to an expanding electron wave packet.
The size of the photoelectron wave packet is well estimated by the classical distance $x=v_c \tau_I$ (black circle). 
This is a one-dimensional model calculation where the continuum states are plane waves 
and the dipole transition matrix element is approximated as constant. 
The carrier frequency of the light is $\hbar\Omega = 41$ eV and the binding energy is $I_p=13.6$ eV, 
resulting in a central energy, $\epsilon_c=\hbar\Omega-I_p=27.1$ eV = 1 atomic unit, of the photoelectron.
The photoelectron wave packet is complex;  the real and imaginary parts being shown in red and blue, respectively. 
	}
	\label{outgoing}
\end{center}

From standard time-dependent perturbation theory, 
we know that the total transition probability  
of a perturbative excitation increases linearly with time, 
$P = \int d\epsilon_\alpha~|a_\alpha(t)|^2 \propto \tau_I$,
as long as the depletion of the initial state can be neglected. 
The photoelectron wave packet, in Fig.~\ref{outgoing}, 
behaves indeed as expected  
since it extends linearly in space as a function of the interaction time, $\tau_I$. 
The spatial extent of the wave packet is approximately $x \approx v_c \tau_I$, 
where $v_c=\sqrt{2\epsilon_c/m}$ is the classical velocity of the photoelectron after ionization
and $\tau_I$ is the duration of the interaction.
Inside this ``classical'' extent of the wave packet, the probability density is roughly constant.

Because the electron is ionized at a ``well-defined time'', 
namely during the brief interaction with the attosecond pulse,
the uncertainty principle 
dictates that its energy content must be very broad. 
This can also be verified in Fig.~\ref{outgoing}~(g,e,c,a) for progressively shorter interactions. 
It may be of interest to know how long the photoelectron wave packet remains ``localized'' in time and space.
At time $\tau$ after the ionization, 
the electron wave packet will have spread in space to an extent of $\Delta r$, 
which can be estimated from a classical ensemble of velocities, 
\begin{equation}
\Delta r = 
\left(v_{max}-v_{min}\right)\tau \approx 
\frac{\Delta \epsilon}{m v_c}\tau
\label{spatialextent}
\end{equation}
where $v_{min}$, $v_{c}$ and $v_{max}$ are the minimal, central and maximal velocity,
corresponding to an energy width of $\Delta \epsilon$ at a central energy of $\epsilon_c=mv_c^2/2$. 
Furthermore, the temporal extent can be estimated as 
\begin{equation}
\Delta t \approx \frac{\Delta r }{v_c} \approx \frac{\Delta \epsilon}{2\epsilon_c}\tau.
\label{temporalextent}
\end{equation}
Interestingly, from this simple analysis, we find that the photoelectron wave packet will
extend its duration linearly in $\tau$. For photoelectrons ionized by attosecond pulses, 
with a large bandwidth $\Delta\epsilon\approx\epsilon_c$, this implies that 
the temporal extent of the electron wave packet is approximately equal to its time of propagation, 
$\Delta t \approx \tau$.
In other words, we should {\it not} consider these photoelectron wave packets as attosecond quantities. 
We will return to the propagation of photoelectrons in Sec.~\ref{generalizationFirstOrder}
for a more detailed analysis,
but first we must consider the actual photoionization process in more detail. 


\subsection{Exponential turn-on of ionization}
\label{sec:expturnon}
In the previous subsection, we considered a ``flat-top'' XUV pulse shape, 
but envelopes with such sharp features represent a rather poor idealization
of the often smoothly varying physical pulses.  
Rather than turning on the ionizing field instantaneously, as was done above,  
the electric field can be smoothly increased from minus infinity 
using a ``slow'' exponential turn-on \cite{SakuraiMQM1994}. 
Even though this represents an infinitely long interaction, 
we will show that it is extremely important for a deep theoretical understanding 
of photoionization phenomena also on the attosecond timescale.
The electric field is written $\tilde E(t)=iE\exp[-i\Omega t + \beta t]$, 
including an attenuation factor, $\beta>0$, that reduces the field strength as $t\rightarrow-\infty$. 
The attenuation factor ensures that the probability for the emission of a photoelectron is zero at remote negative times. 
Inserting the attenuated field into Eq.~(\ref{firstorderpert}) yields
\begin{eqnarray}
a^{(1)}_\alpha(t) &=& \lim_{\beta\rightarrow0^+}
\frac{e}{\hbar} ~ z_{\alpha i} ~E ~ 
\left(
\frac{1}{\beta+i\Delta\omega_\alpha}\right)
e^{i\Delta\omega_\alpha t+ \beta t}
\nonumber \\
&=&
\frac{e}{\hbar} ~ z_{\alpha i} ~ E
\left(
\pi\delta(\Delta \omega_\alpha)
-i\wp \frac{1}{\Delta\omega_\alpha}
\right)e^{i\Delta\omega_\alpha t},
\label{expon}
\end{eqnarray}
with $\Delta\omega_\alpha\equiv\omega_\alpha-\omega_i-\Omega$, and where
the limit corresponds to the ``well-known theorem from complex function theory'' \cite{Mattuck}.
The notation used requires some explanation: 
We write $\delta$ for the Dirac delta function, 
and $\wp$ is written to indicate that the discontinuity of $1/\Delta\omega_\alpha$ has been removed, 
so that when integrated the result is Cauchy's principal value. 
We mention that this limiting form can be recovered 
by multiplying the denominator of Eq.~(\ref{expon}) by its complex conjugate and then identifying the limits of the delta function, 
$\beta/(\beta^2+\Delta\omega_\alpha^2)\rightarrow \pi  \delta(\Delta\omega_\alpha)$; 
and the principal value, 
$\Delta\omega_\alpha/(\beta^2+\Delta\omega_\alpha^2)\rightarrow \wp( 1/\Delta\omega_\alpha)$. 
%

Similarly, we can model a slow turn-off 
using $\tilde E(t)=iE\exp[-i\Omega t-\beta t]$ for $t>0$ as $t\rightarrow\infty$,
giving the same result as in Eq.~(\ref{expon}), but with the opposite sign of the principal value part.
Combining the slow turn-on and the slow turn-off, therefore, 
leads to symmetric cancellation of the principal value contribution. 
The resulting complex amplitude, corresponding to an interaction with 
a field of central frequency $\Omega$ for times going from 
minus infinity to plus infinity, 
shows that energy must be conserved,
\begin{equation}
S_{\alpha/i}^{(1)}=
e z_{\alpha i} ~ E ~
2\pi\delta(\epsilon_\alpha\underbrace{-\epsilon_i-\hbar\Omega}_{-\epsilon_\kappa})
,
\label{pertex3}
\end{equation}
and that only an energy-conserving state, with $\epsilon_\kappa\equiv\hbar^2\kappa^2/(2m)=\hbar\Omega+\epsilon_i$, will be populated.
This result corresponds to a Fourier transform of 
a continuously oscillating electric field, Eq.~(\ref{firstorderover}). 
We stress that it is valid as  
$t\rightarrow\infty$, which we interpret as the electric field being turned \textit{off}. 

The interesting aspect of the exponential turn-on model, Eq.~(\ref{expon}), is that it allows us to reconstruct a 
photoelectron wave packet at \textit{any} finite time, $t$, 
while the ionization is ongoing and the electric field is \textit{on}, 
\begin{eqnarray}
\bigl|\tilde\Psi^{(1)}(t)\bigl> = 
\sum_\alpha \! \! \! \! \! \! \! \! \! \int  ~ a_\alpha^{(1)}(t)~\bigl|\tilde{\alpha}(t)\bigl>=
& & \nonumber \\
\underbrace{
eE~
\Bigl[
\overbrace{
\pi \sum_c z_{ci} \left|c\right> }
^{Conserving} 
- 
i
\overbrace{
\wp
\sum_\alpha \! \! \! \! \! \! \! \! \! \int  ~
 \frac{z_{\alpha i}}{\epsilon_\alpha-\epsilon_\kappa}\left|\alpha\right> 
}^{Principal \ value}
\Bigl]
}_{\textrm{Time-independent part}}
~
\underbrace{\exp[-i \omega_\kappa t]}_{{Temporal \  phase}},
&&
\label{firstwavepacket}
\end{eqnarray}
where the total phase of the wave packet is evolving at a {\it common} angular frequency, $\omega_\kappa$,
corresponding to energy conservation. 
In Eq.~(\ref{firstwavepacket}), the time, $t$, is finite. 
This implies that all states can be populated, also those that do not conserve energy.
In fact, the time-independent part of the wave packet consists of two terms:  
the first one is associated  
to the energy-conserving states, while 
the second term is  a non-trivial principal-value superposition of non-energy-conserving states.
The latter states are out of phase by $\pm\pi/2$ compared to the energy-conserving contribution depending on whether 
the energy denominator is positive or negative.
This total phase displacement of $\pi$ can be understood in analogy with a classical pendulum,  
which is set in motion by an external oscillating force either in an over-driven or under-driven mode. 
Here, because we are dealing with quantum mechanics, we need to sum over all possible ``pendulums'' 
corresponding to the continuum states above and below the resonance located at $\epsilon_\kappa$. 

More explicitly, the energy-conserving transition goes from the initial, partial-wave state, $\ket{i}\equiv\ket{n_i,\ell_i,m_i}$, 
to a superposition of energy-conserving, partial-waves states, $\ket{c}=\ket{\kappa,\lambda_c,\mu_c}$, 
where the summation over $c$ implies that $\lambda_c=\ell_i\pm1$ and $\mu_c=m_i$ for linearly polarized light.
Similarly, the non-energy-conserving states can be written explicitly as partial-wave states,
$\ket{\alpha}=\ket{\kappa',\lambda_\alpha,\mu_\alpha}$, where $\alpha$ runs over both the bound and continuous states, 
but $\kappa' \neq \kappa$. 

\subsection{Asymptotic form of the photoelectron wave packet}
\label{afofpwp}
It is not easy to directly infer the form of the first-order wave packet from Eq.~(\ref{firstwavepacket}),
but we expect that it should be an {\it outgoing} wave.
To see this, we first note that the energy-conserving part is real and that the principal-value part is imaginary, at $t=0$. 
Next, we consider the asymptotic form of the radial wavefunction  
for a photoelectron state with $\epsilon=\epsilon_\alpha=\epsilon_{\kappa'}$, 
corresponding to quantum numbers $\alpha = [\kappa',\lambda,\mu]$, 
\begin{eqnarray}
\bigl<\vec r\bigl|
\alpha
\bigl>
&\equiv&\frac{1}{r} u_{\kappa',\lambda}(r) Y_{\lambda,\mu}(\hat r),  
\label{coulombfunctionexp0}
\end{eqnarray}
where we have separated the radial part from the angular part.
In the asymptotic limit, 
we may substitute for the approximate form of the radial part, as derived in Eq.~(\ref{coulombfunction}),
\begin{eqnarray}
\bigl<\vec r\bigl|
\alpha
\bigl>
 &\approx&
 N_{\kappa'} \frac{1}{r} 
\sin\Bigl[ \kappa' r+ \underbrace{ \frac{\ln(2 \kappa' r)}{\kappa' a_0}+\eta_{\kappa',\lambda}-\frac{\lambda\pi}{2}}_{ \Phi_{\kappa',\lambda}(r) } 
\Bigl] Y_{\lambda,\mu}(\hat r),  
\label{coulombfunctionexp}
\end{eqnarray}
where $\Phi_{\kappa',\lambda}(r)$ is the radial-dependent and $\lambda$-dependent phase-shift,  
and where the coefficient for energy re-normalization is 
$N_{\kappa'}=[2m/(\pi^2\hbar^2 \epsilon_\alpha)]^{1/4}=\sqrt{2m/(\pi\hbar^2\kappa')}$.
The asymptotic part of the bound states can be neglected, 
because they decay exponentially with the distance from the nucleus and can not affect the outgoing photoelectron.
This allows us to 
aggregate the sum over the bound state spectrum with the contribution of
the continuous spectrum in a single 
integral from minus infinity to plus infinity. 
In this way, the smoothed ``bound part'' is still exponentially damped and it will not contribute, 
but the integral is easier to handle.  

The principal-value part of the wave packet, Eq.~(\ref{firstwavepacket}), 
is evaluated by rewriting the sine-function in Eq.~(\ref{coulombfunctionexp}) 
as outgoing and incoming waves, $u^{(out/in)}_{\kappa',\lambda}(r)\sim \exp[\pm i\kappa'r]$. 
It is then possible to integrate analytically for the principal-value using Cauchy's integral theorem, 
as sketched in Fig.~\ref{complexplane}.
\begin{center}
	\includegraphics[width=0.65\textwidth]{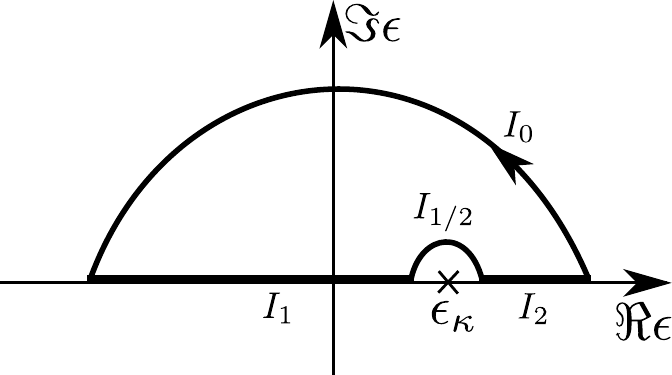}
	\captionof{figure}{
	Integration paths in the complex plane for obtaining the principal value $P=I_1+I_2$ 
	of outgoing waves: $\exp[i\sqrt{2\epsilon}r]/(\epsilon-\epsilon_\kappa)$. 
	The outer integration path vanishes, $I_0\rightarrow 0$, as the electron enters the asymptotic region, $r\rightarrow\infty$. 
	The semicircle over the pole $I_{1/2}$ becomes 
	$-i\pi$ times the corresponding residue as the arc is reduced.
	Using Cauchy's integral theorem, we write: $I_0+I_1+I_{1/2}+I_2=0$, which implies  
	that the principal value is: $P\approx -I_{1/2}=i\pi\exp[i\sqrt{2\epsilon_\kappa}r]$. 
	A similar calculation can be made for the incoming waves, but here 
	the integration paths go into the negative imaginary axis which results in an additional minus sign.	
	For simplicity, we have here omitted the phase-shifts, $\Phi_{\kappa',\lambda}(r)$ and the amplitude prefactors, 
	since they present no further complication. Atomic units were used for compactness.
	}
	\label{complexplane}
\end{center}
The result from this calculation is that the imaginary part of the wave packet 
takes a cos-like form at the energy-conserving wave number, $\kappa$, so that  
\begin{eqnarray}
\bigl<\vec r\bigl|
\wp
\sum_\alpha \! \! \! \! \! \! \! \! \! \int  ~
 \frac{z_{\alpha i}}{\epsilon_\alpha-\epsilon_\kappa}\left|\alpha\right> 
~\approx~ &&  \nonumber \\
 \frac{\pi}{r} \sum_{c} ~ z_{ci}~ N_\kappa~
\cos\left[ \kappa r+\Phi^{(c)}_{\kappa,\lambda}(r) \right] Y^{(c)}_{\lambda,\mu}(\hat r), &&
\label{evalperturbed}
\end{eqnarray}
where the sum over $c$ is needed if the wave packet populates multiple angular channels  
(with short-hand notation: $Y^{(c)}_{\lambda,\mu}$, so that $[\lambda,\mu]=[\lambda_c,\mu_c]$).
Inserting these asymptotic expressions into Eq.~(\ref{firstwavepacket}), 
we find that the photoelectron wave packet is, indeed, 
a sum of outgoing Coulomb waves for each angular channel $c$ 
\begin{eqnarray}
\tilde\Psi^{(1)}(t,\vec r) 
~\approx~ && \nonumber \\
\frac{\pi e E}{ir}   \sum_{c} z_{ci} N_\kappa 
\underbrace{ \exp\left[i\left( \kappa r+\Phi^{(c)}_{\kappa,\lambda}(r) - \omega_\kappa t \right)\right]}
_{\textrm{Outgoing Coulomb wave}}Y^{(c)}_{\lambda,\mu}(\hat r), && 
\label{firstwavepacketasymp}
\end{eqnarray} 
where the result of the long-range Coulomb interaction and of the short-range interactions are contained in 
$\Phi^{(c)}_{\kappa,\lambda}(r)$ for the wave number, $\kappa$, corresponding to energy conservation.
This intuitive form of the first-order wave packet is valid asymptotically,  
as can be identified in Fig.~\ref{Pert}, 
where we have numerically calculated a representative 
first-order wave packet for hydrogen.
\begin{center}
	\includegraphics[width= 0.85\textwidth]{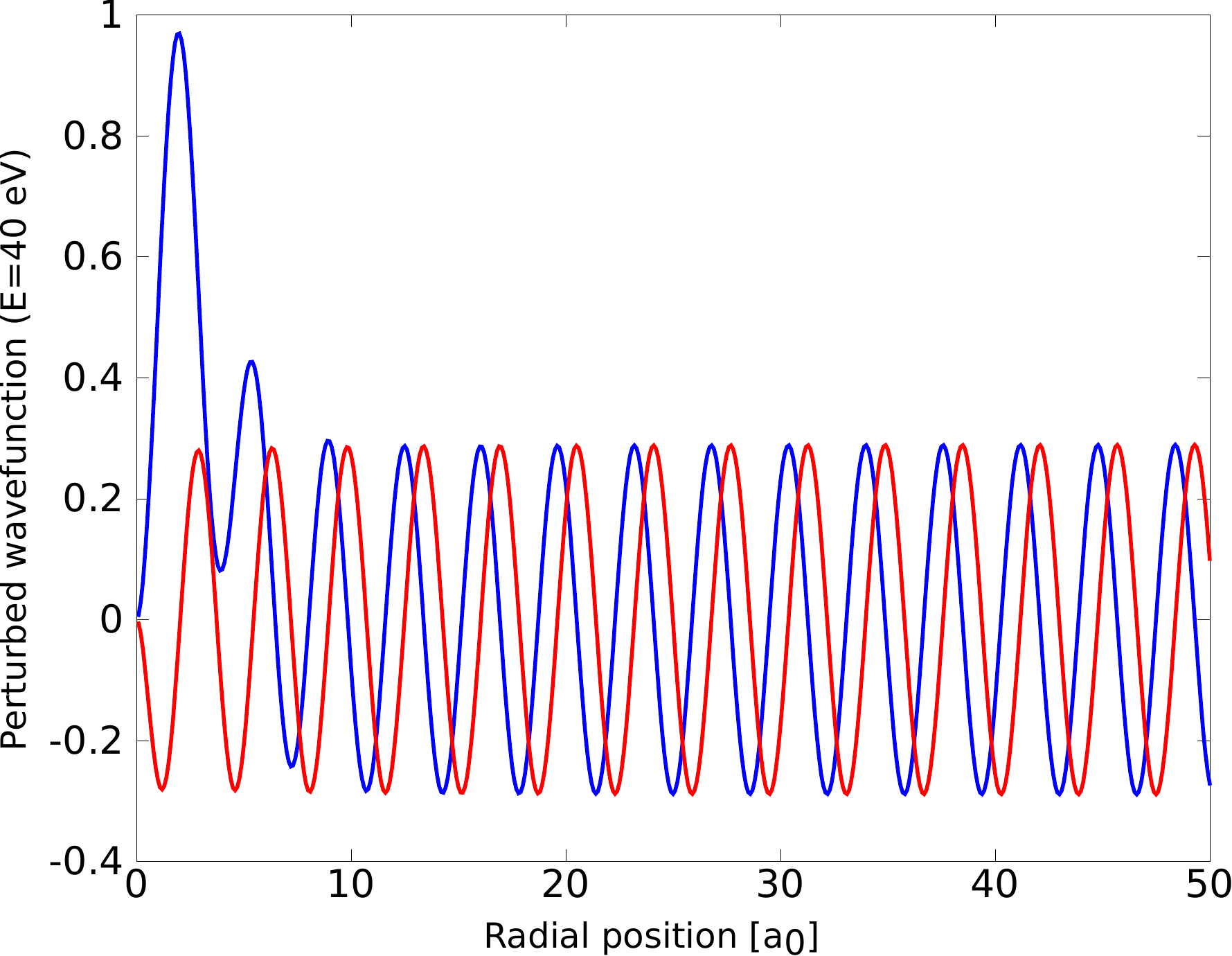}
	\captionof{figure}{
First-order radial wave packet (perturbed wavefunction) calculated numerically for Hydrogen at a kinetic energy of $40$~eV. 
The real (red) and imaginary (blue) part oscillate as an outgoing wave, 
as expected from the asymptotic approximation.
	}
	\label{Pert}
\end{center}
Our result for the asymptotic form of the wave packet, Eq.~(\ref{firstwavepacketasymp}), 
is quite general and it can be applied to many atomic systems by choosing an appropriate phase, $\Phi_{\kappa,\lambda}(r)$,
with different asymptotic phase-shifts, $\eta_{\kappa,\lambda}$, but also ionic charges, $Z$. 
Close to the core, the imaginary part of the wave packet  
will go to zero (it will \textit{not} diverge like the irregular solution \cite{Friedrich1994}).
The polarization of the bound states is clearly visible in Fig.~\ref{Pert} as a ``bump'' in the imaginary part.
Using the asymptotic approximation, we neglect this short-range feature, 
which turns out to be an excellent approximation
for laser-assisted photoionization, as we shall discuss in Sec.~\ref{sec:STPT}. 
The asymptotic form of the wave packet \cite{AymarJPB1980} can be also expressed in terms of 
the Coulomb Green's function \cite{Zon,Edwards}. 
In the special case of an hydrogenic system, exact calculations can be performed
by using compact representations of the Coulomb Green’s function \cite{MaquetPRA1977,MaquetJPB1998}. 
Furthermore, the wave packet can be related to the so-called {\it perturbed wavefunction}, 
which satisfies a corresponding inhomogeneous, time-independent differential equation \cite{DahlstromCP2012,TomaJPB2002}. 
In this sense, the wave packet that we have derived using time-dependent perturbation theory is a time-independent quantity,
which plays a key role also in time-independent perturbative methods!

\subsection{Generalization of first-order photoelectron wave packets}
\label{generalizationFirstOrder}
With the help of an integration performed in the complex energy-plane, we have just shown 
that a monochromatic ionizing XUV field, with frequency $\Omega$, leads to an 
outgoing photoelectron wave packet:  
$\bigl|\tilde \Psi^{(1)}(t;\Omega)\bigl>=\bigl| \Psi^{(1)}(\Omega)\bigl>\exp[-i\omega(\Omega)t]$, 
which oscillates at the characteristic frequency of energy conservation, 
$\omega(\Omega)=\omega_i+\Omega$, but with no other time-dependence.  
Likewise, the effective wave number of the wave packet varies with the photon energy 
so that the phase of the wave packet (in angular channel $\lambda$) is: 
\begin{eqnarray}
\arg\left[ \braket{\vec r} {\tilde \Psi^{(1)}(t;\Omega)}\right] \propto && \nonumber \\
\phi_{\Omega} + \kappa(\Omega)r + \Phi_{\kappa(\Omega),\lambda}(r) -\omega(\Omega) t, &&
\label{pephasesmore}
\end{eqnarray}
where $\phi_{\Omega}$ is the phase of the XUV field, 
and $\kappa(\Omega)$ is the wave number 
upon absorption of one such photon with frequency $\Omega$. 
Any attosecond XUV \textit{pulse} can be written as a 
linear superposition of monochromatic waves, and the corresponding
 photoelectron wave packets
can be written in terms of their time-independent (spectral) components,   
\begin{eqnarray}
\bigl|\tilde\Psi^{(1)}(t)\bigl>~&=&~
\frac{1}{2\pi}\int d\Omega ~
\bigl|\tilde \Psi^{(1)}(t;\Omega)\bigl> \nonumber \\
~&=&~
\underbrace{
\frac{1}{2\pi}\int d\Omega ~
\bigl|\Psi^{(1)}(\Omega)\bigl> ~
}_{Time-independent}
\underbrace{
\exp[-i\omega(\Omega)t]}
_{Temporal \  phase}
,
\label{eFourierFirstorder}
\end{eqnarray}
where $\tilde\Psi^{(1)}(0;\Omega)\equiv \Psi^{(1)}(\Omega)$. 
Note that the superposition presented here is different from the one in Eq.~(\ref{superposition}),
because the latter includes a superposition of both ingoing and outgoing waves, 
while Eq.~(\ref{eFourierFirstorder}) only contains outgoing waves. 
This is due to the fact that we have already imposed 
a time-boundary condition, namely 
that the laser field was not {\it on} at $t\rightarrow-\infty$. 
As a consequence, Eq.~(\ref{eFourierFirstorder}) describes the physical process at {\it all} times 
and for {\it any} pulse shape, 
without the need for {\it time-dependent} complex amplitudes to describe 
the population of the states, as would be required with Eq.~(\ref{superposition}). 
The wave packet properties of this superposition can be analysed 
using the SPA, Eq.~(\ref{spapprox}), because, in the 
limit of time going to infinity, the oscillations of the
phase factor will be fast compared to \textit{all} pre-factors \cite{CarvalhoPR2002}.  
The SPA is applied as
\begin{eqnarray}
\tilde \Psi^{(1)} (r,t)
&=&  \frac{1}{2\pi}\int d\Omega ~ a(\Omega) ~ \exp[if(\Omega;r,t)]  \nonumber \\
&\approx &  \frac{1}{2\pi}~ 
a(\Omega_s) ~ \sqrt{\frac{i\pi}{f''(\Omega_s)}}~\exp[if(\Omega_s)], 
\label{spacetime}
\end{eqnarray}
where $f(\Omega;r,t)\equiv f(\Omega)$ is the phase function; 
and $a(\Omega)$ is a pre-factor, which is a slowly varying function of $\Omega$. 
It contains both the spectral envelope function of the attosecond pulse and the dipole matrix element. 
As given in Eq.~(\ref{pephasesmore}), 
the phase factor for the wave packet (in angular channel $\lambda$) is  
\begin{equation}
f(\Omega;r,t) = \phi_{\Omega} + \kappa(\Omega) r + \Phi_{\kappa(\Omega),\lambda}(r) - \omega(\Omega)t,
\label{pephasefactor}
\end{equation} 
where $\kappa$ and $\omega$ are determined by energy conservation from an initial bound state after absorbing one photon from the $\Omega$-field.
The stationary phase equation, Eq.~(\ref{stationary}),
yields the XUV frequencies, $\Omega_s$,  
that give the dominant contribution to the electron wave packet at a given space--time position, $(r,t)$ \cite{WollenhauptPRL2002}. 
%
%
Using Eq.~(\ref{pephasefactor}), the solution to the stationary-phase equation 
gives the time when an electron with wave number $\kappa$ arrives at $r$
\begin{equation}
\frac{\partial f}{\partial \Omega} ~=~ 
\underbrace{
\frac{\partial \phi_{\Omega}}{\partial\Omega}
}_{\tau_{GD}} +
\underbrace{
\frac{\partial \kappa}{\partial\Omega}r
}_{\tau_{free}} +
\underbrace{
\frac{\partial}{\partial \Omega}\Phi_{\kappa(\Omega),\lambda}(r)
}_{\tau_\lambda+\tau_{LR}(r)}
-
\underbrace{\frac{\partial \omega}{\partial \Omega}}_{=1}t
~=~0,
\label{evalpestat}
\end{equation}
where we have identified the group delay of the attosecond pulse, $\tau_{GD}(\Omega)$; 
a free-particle trajectory $\tau_{free} = v_c r$;
the short-range Wigner delay, $\tau_\lambda$;
the long-range Coulomb delay, $\tau_{LR}(r)$; and
the actual time itself, $t\gg\tau_{GD}$.
This implies that the electron with wave vector $\kappa$ will arrive at $r$ at
\begin{equation}
t = \tau_{GD}  + \tau_{free} + \tau_{LR} + \tau_\lambda,
\label{pedelays}
\end{equation} 
which clearly depends on when the appropriate frequency ionized the atom, $\tau_{GD}$;
and the time it takes for the electron to propagate to the detector, $\tau_{free}+\tau_{LR}+\tau_\lambda$.
If the electron followed a free-particle trajectory the delay would be $\tau_{free}$, 
but due to the Coulomb interaction the electron is delayed by an additional amount $\tau_{LR}(r)+\tau_\lambda$,
corresponding to the long-range and short range corrections to the timing of the trajectory.

The ``delay'' from the long-range interaction is difficult to interpret, 
as we already discussed in Sec.~\ref{sec:coulomb}, 
but we may note that it has a negative sign asymptotically
\begin{equation}
\tau_{LR}(r) = 
\frac{\hbar}{\epsilon}\frac{1}{2\kappa}\left\{
1-\ln[2\kappa r]
\right\}\approx
-\hbar \frac{\ln[2\kappa r]}{2\kappa \epsilon}
, 
\label{longrangedelay}
\end{equation}
which implies that the electron arrives {\it faster} than if it was propagating freely!
The physical reason is that, as we discussed in relation to Eq.~(\ref{wignerweak}), 
attractive potentials lead to a greater local velocity of the electrons.
Also the short-range interaction {\it may} have a negative delay, which classically can be interpreted as the electron 
being somewhat advanced, {\it i.e.} starting at a small radial distance, $a_{eff}=-\tau_\lambda/v_c>0$, outside the origin. 
In the case of the Coulomb potential, the Wigner delays are, however, positive, 
and such simple arguments do not apply. 
In the asymptotic limit these corrections are very small compared to the free travel time and can be neglected. 
This implies that the dominant contribution to the electron wave packet at a given point in space--time $(r,t)$,  
arises from the part of the wave packet that fulfils the simple relation: $v_c t \approx r$.  
We then obtain that the probability density of the photoelectron in an integrated angular channel is
\begin{equation}
\rho(t,r)=|\Psi^{(1)}(r,t)|^2 \sim |a_\kappa|^2 \frac{m}{\hbar t},
\label{pedensity}
\end{equation}
where $a_\kappa$ is the complex amplitude of the state with wave number $\kappa$ and 
the time-dependence can be understood as a radial quantum diffusion of the wave packet, 
in good agreement with the simple classical analysis presented above in Eq.~(\ref{temporalextent}).
This mapping of complex amplitudes into space and time is the basic principle of 
the TOF detection scheme, which is used in many attosecond experiments.
In Fig.~\ref{pulseProp} we sketch the rough dynamics of photoionization. 
\begin{center}
	\includegraphics[scale= 0.85]{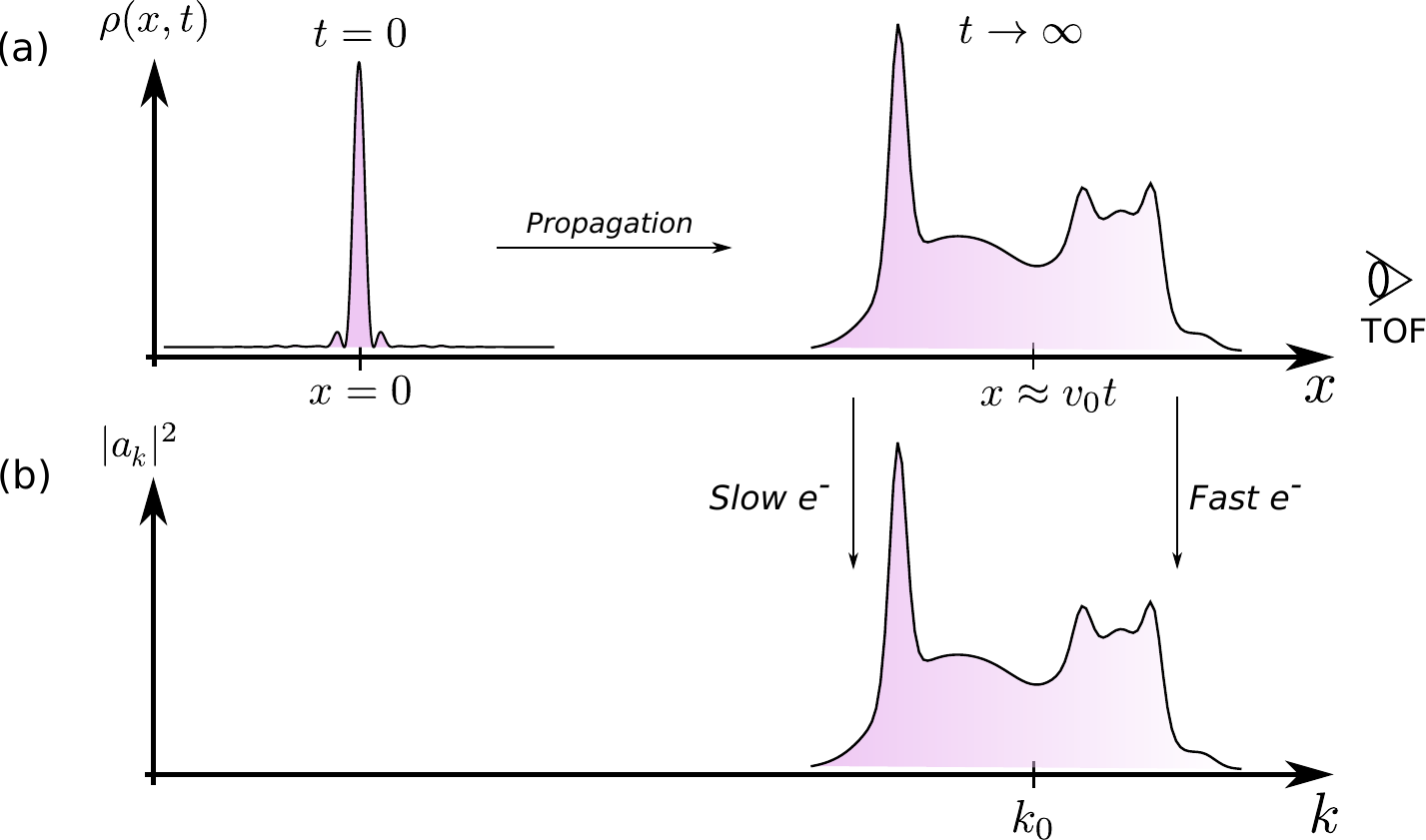}
	\captionof{figure}{ {\it Photoionization in time and space:}
(a) 
Initially, at the ionization event $t=0$, the wave packet is located at the atom. As the electron leaves the atom, the slow part of the wave packet is stretched far way from the fast part due to quantum diffusion. 
(b) The momentum distribution, $|a_k|^2$, of the wave packet is naturally accessible using a time of flight (TOF) detector, where the probability distribution of fast electrons are detected first, and the slow electrons arrive later. 
To make this example more interesting, the attosecond pulse has a ``cat'' encoded in its momentum distribution, 
this ``cat'' eventually appears in the space--time distribution. 
(A more conservative momentum distribution would correspond to the odd harmonics of HHG.)
	}
	\label{pulseProp}
\end{center}
The photoelectron momentum-to-space mapping, described above, is similar to the
frequency-to-time mapping for super-chirped light pulses, which was discussed in Sec.~\ref{spaex1}. 
For photoelectrons, the mapping occurs automatically after long enough times, 
and there is no need for a dispersive material. 
Classically, the electron mapping is \textit{obvious} since electrons 
in a classical distribution are moving at different speeds.
If we know that they all started at a distinct point in space--time, 
then we know that the fast electrons will reach the detector (at $\vec r$) first. 

In the case of an APT consisting of high-order harmonics, 
the photoelectrons from the highest harmonic will arrive first,
followed by an empty region and then the next highest harmonic, and so on. 
The fact that they were ionized at several short instances apart, 
namely at each peak of the APT, can {\it not} be observed directly in TOF spectrometer  
due to the strong quantum diffusion taking place in the flight tube. 
In some sense, the {\it signature} of the APT or SAP  
is seen already in the energy and momentum distributions because the electrons do 
arrive in bursts for each harmonic or in a continuous bunch \cite{ManstenPRL2008},
but in order to access the detailed attosecond time scale, one has to
probe the photoelectron wave-packets with a coherent IR laser field. 
Next, we shall describe different implementations of this concept.

\section{Theory of XUV photoionization \\ in the presence of an IR field}
\label{sec:STPT}
As shown in the preceding sections, 
attosecond delays in photoionization can be deduced from the energy dependence of 
the {\it phases} of the transition amplitudes associated with the process.
In the experiments reported so far, 
an auxiliary IR laser field is used as a ``clock'' to measure the delays.
Then the analysis is somewhat more complicated by the presence of this additional dressing field.
Indeed, in the presence of an IR laser field, 
the photoelectron released upon absorption of one XUV photon can absorb and emit IR photons, 
thereby changing its energy. 
Provided that the IR field is weak, 
the energy of the electron will change 
as the result of the exchange of only \textit{one} IR photon at most,
so that the overall process amounts to a two-photon transition. 
In this section, we give a theoretical background relevant to treat this class of laser-assisted photoionization, 
with special emphasis on the contribution from stimulated continuum--continuum transitions 
using second-order time-dependent perturbation theory  \cite{SakuraiMQM1994}. 
These transitions are associated to the so-called Above-Threshold Ionization (ATI) processes, 
because the electron is already free when it exchanges the second photon. 

The time-dependent dipole interaction with the fields, linearly polarized along the $z$-axis, 
is written as $\tilde V(z,t)=\tilde V_{I}(z,t)+\tilde V_{II}(z,t)$,
where we have separated the dipole interaction with the XUV field, 
$\tilde V_{I}(z,t)$ in Eq.~(\ref{dipoleoperator}), 
from the subsequent dipole interaction with the laser field,  
\begin{equation}
\tilde V_{II}(z,t)=\int d\Omega_{II}ezE_{II}(\Omega_{II})e^{-i\Omega_{II}t}/2\pi.
\label{VII}
\end{equation}
With this notation, it is implied that $\Omega\approx \Omega_I>I_p \gg \Omega_{II} \approx \omega$, 
where the former and latter frequencies correspond to the central frequency of the XUV and laser field, respectively. 
We may include a broad bandwidth, $\Delta \Omega_I$ and $\Delta \Omega_{II}$, on both fields,
provided that the bandwidths do not overlap. 
The second-order complex amplitude for the time-ordered interaction with $V_I(t)$ and $V_{II}(t)$ is:
\begin{eqnarray}
a^{(2)}_f(t) &=& 
\frac{1}{i\hbar} \int_{-\infty}^{t} dt'~ 
\sum_\alpha \! \! \! \! \! \! \! \! \!  \int ~ \bigl<\tilde{f}(t')\bigl|V_{II}(z,t')\bigl|\tilde{\alpha}(t')\bigl>a_\alpha^{(1)}(t') \nonumber \\
&\equiv&
\frac{1}{i\hbar} \int_{-\infty}^{t} dt'~ 
\bigl<\tilde{f}(t')\bigl|V_{II}(z,t')\bigl|\tilde \Psi^{(1)}(t')\bigl>,
\label{secondorderpert}
\end{eqnarray} 
where we have identified the first-order wave packet from  Eq.~(\ref{eFourierFirstorder}). 
It is of interest to compare the second-order complex amplitude, Eq.~(\ref{secondorderpert}),
with the first-order one, as given in Eq.~(\ref{firstorderpert}). 
In the latter, the electron makes a transition from the initial bound state 
into the field-free continuum, creating a first-order wave packet, $\ket{\tilde \Psi^{(1)}(t)}$. 
In the second order amplitude, this first-order wave packet can be seen as the initial state for the second interaction,
and the transition towards the final state will occur towards a different continuum state.
We can write the second-order wave packet formally as
\begin{equation}
\bigl|\tilde \Psi^{(2)}(t)\bigl> ~ \equiv ~ 
\sum_\alpha \! \! \! \! \! \! \! \! \! \int  ~
a_\alpha^{(2)}(t)~\bigl|\tilde{\alpha}(t)\bigl>,
\label{secondwp}
\end{equation}
in analogy with Eq.~(\ref{firstwp}) and our task is now to determine the second-order complex amplitudes. 
After the interactions are over,
the S-matrix for the two-photon process, 
$S^{(2)}_{f/i} = \lim_{t\rightarrow\infty}a^{(2)}_f(t)$, 
can be rewritten as: 
\begin{eqnarray}
S^{(2)}_{f/i} 
~&=&~ \frac{e}{i\hbar}~ \frac{1}{(2\pi)^2} 
	\int d\Omega_{I}~  \bigl<f\bigl|z\bigl|\Psi^{(1)}(\Omega_{I})\bigl>
      ~ \int d\Omega_{II}~ E_{II}(\Omega_{II}) 
\nonumber \\ ~&\times&~
	\underbrace{
	\int_{-\infty}^{\infty} dt ~          \exp[i(w_f-\Omega_{II}-\omega_c(\Omega_{I}))t], 
	}_{2\pi\delta(\omega_f-\omega_i-\Omega_I-\Omega_{II})}
\label{twophotonS1}
\end{eqnarray}
where we have expanded the time-dependent quantities  
as Fourier integrals over frequency, using Eq.~(\ref{eFourierFirstorder}) and (\ref{dipoleoperator}),
and where we have changed the order of integration so that all time-dependence is trapped in a simple inner integral. 
This time-integral is identified as a delta function that enforces energy-conservation of the two-photon process,
and it is now used to eliminate the spectral integral over $\Omega_{II}$ so that the final result becomes:
\begin{eqnarray}
S^{(2)}_{f/i} 
~&=&~ \frac{e}{i\hbar}~\frac{1}{2\pi} 
	\int d\Omega_{I} ~ \bigl<f\bigl|z\bigl|\Psi^{(1)}(\Omega_{I})\bigl> ~ E_{II}(\Omega_{II}'),
\label{twophotonS}
\end{eqnarray}
where $\Omega'_{II}= \omega_f-\omega_i-\Omega_I$.
This corresponds to the well-known S-matrix for two-photons, 
here derived assuming a general bandwidth on both fields.  

The two-photon S-matrix, Eq.~(\ref{twophotonS}), has an intuitive time-independent interpretation:
The integral over $\Omega_I$ can be seen as an integral over all the different 
{\it quantum paths} that can lead to the final state $\bigl|f\bigl>$.
In order to reach this specific final state,
the photon energies of the XUV and IR fields must 
satisfy a global energy conservation requirement.
If both interactions have a broad bandwidth then there is also a broad range of 
{photon-pairs} available.
This results in a``blurring'' effect that arises from the convoluted quantum paths.  
If one of the fields, say the probe, is quasi-monochromatic: $\Omega_{II} =\pm\omega$, 
which implies that $\Delta\Omega_{II}\ll|\omega|$, 
the effects of this convolution are negligible.  
Specializing to the dominant contributions associated to the transitions taking place when
 an XUV photon is absorbed first, only two distinct quantum paths are possible:
\begin{eqnarray}
S^{(a)}_{f/i} 
~&=&~ \frac{e}{i\hbar} ~ E^{(a)} ~ \bigl<f\bigl|z\bigl|\Psi^{(1)}(\Omega_{<})\bigl> \nonumber \\
S^{(e)}_{f/i} 
~&=&~ \frac{e}{i\hbar} ~ E^{(e)} ~ \bigl<f\bigl|z\bigl|\Psi^{(1)}(\Omega_{>})\bigl>
,
\label{twophotonS2}
\end{eqnarray} 
where $E^{(a)}=E_{II}(\omega)$ and $E^{(e)}=E_{II}(-\omega)=E_{II}^*(\omega)$, 
corresponding to absorption (a) and emission (e) of a laser photon from the fields 
as depicted in Fig.~\ref{arrowshigherorder}. 
Note that the intermediate XUV photon energy is different in the two paths to ensure global energy conservation.
\begin{center}
	\includegraphics[width= 0.85\textwidth]{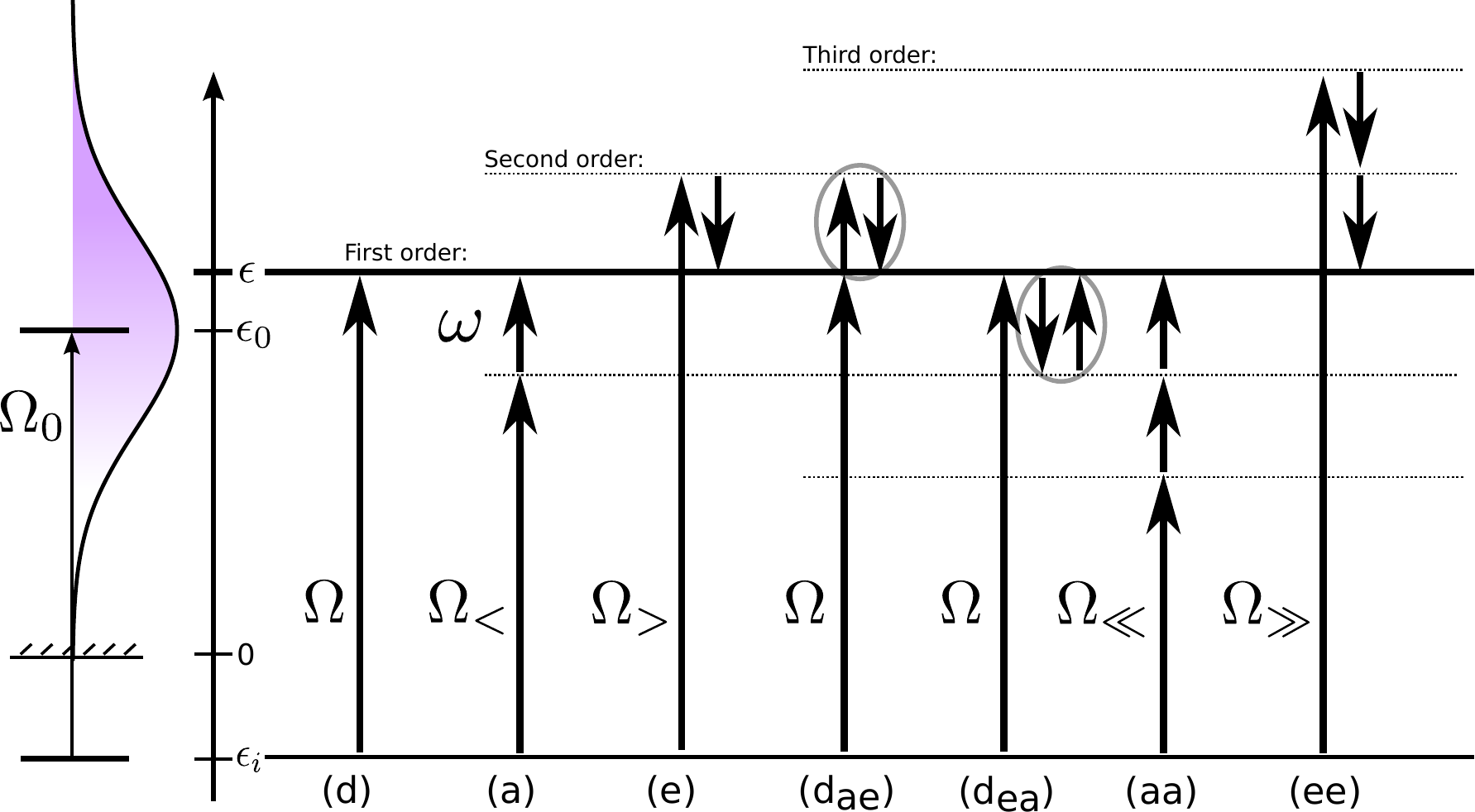}
	\captionof{figure}{ {\it Photon-diagrams for laser-assisted photoionization: }  
Single XUV-photon absorption contributes with a direct path (d) to the final state.
The dominant two-photon processes involve the absorption or emission of a laser photon
indicated by paths (a) and (e).
Higher-order processes involve the exchange of more laser photons. 
	}
	\label{arrowshigherorder}
\end{center}
Here, the photoelectron undergoes ATI transitions
that require a precise representation of the continuum states, 
which makes {\it ab-initio} computations a formidable task for multi-electron atoms or molecules.
However, as we shall show next, it is feasible to derive a convenient approximation method 
designed to evaluate the phase-shift induced by the continuum--continuum transition stage of the process.
An accurate determination of these phases are, in fact, 
essential to interpret correctly the delays as measured in recent attosecond time-delay experiments.

\subsection{Asymptotic approximation for ATI transition amplitudes}
In order to obtain an estimate for the {\it phase} of the two-photon matrix element, 
we will rely on an approximation, which utilizes the asymptotic forms 
of the final state and the first-order wave packet \cite{KlunderPRL2011,DahlstromCP2012}.
Both these wave functions are of continuum character, with positive energies,  
and their phase-shifts can be derived from their asymptotic behaviour 
in the limit of a large radial distance, $r$. 
Atomic units are used in this section for compactness: $\hbar=m=e=1/(4\pi\epsilon_0)=1$.
Specializing to the case of a hydrogenic system with nuclear charge $Z$,
the approximate form of the two-photon matrix element for a given photon-pair, ($\Omega$,$\omega$), 
becomes
\begin{eqnarray}
S^{(2)}_{\vec k/i} \approx 
i
\pi  
(8\pi)^{3/2}
E_\omega E_\Omega  \braOket{R_{\kappa,\lambda}}{r}{R_{n_i,\ell_i}} &&   \nonumber \\ 
\times
\sum_{L,M}(-i)^L e^{i\eta_L(k)}Y_{L,M}(\hat k) %
\times
\sum_{\lambda,\mu}\braOket{Y_{L,M}}{Y_{1,0}}{Y_{\lambda,\mu}}\braOket{Y_{\lambda,\mu}}{Y_{1,0}}{Y_{\ell,m_i}}  
&& \nonumber \\ \times
N_k N_\kappa\int_0^{\infty}dr 
\underbrace{
\sin[kr+\Phi_{k,L}(r)] \ r \ \exp[i(\kappa r+\Phi_{\kappa,\lambda}(r))]
}_{Asymptotic  \  radial \  functions},  &&
\label{S2assymp}
\end{eqnarray} 
where we have performed a partial-wave expansion with the final state having an asymptotic momentum, 
$\vec k$, using Eq.~(\ref{planewaveexpanded});
and where we have replaced the exact form of the radial wave functions by their asymptotic forms. 
The radial functions are written with the short-hand notation for the Coulomb phase-shift,
\begin{equation}
\Phi_{k,L}(r)  =  \frac{Z\ln(2kr)}{k}+\eta_{k,L}-\frac{\pi L}{2},
\label{longrangephase}
\end{equation} 
as defined in Eq.~(\ref{coulombphase}), where $\eta_{k,L}=\sigma_{k,L}+\delta_{k,L}$. 
Here $\sigma_{k,L}=\arg\{\Gamma[L+1-iZ/k]\}$ is the pure Coulomb phase 
with $\Gamma[z]$ being the complex gamma function; and 
$\delta_{k,L}$ is a phase-shift due to an additional short-range interaction. 
The wave numbers of the first-order wave packet, $\kappa$, and of the final state, $k$, 
satisfy energy conservation after one XUV photon:
$\epsilon_\kappa=\epsilon_i+\Omega$, 
and after the exchange of one IR photon:
$\epsilon_k=\epsilon_\kappa\pm\omega$, respectively.
Following Ref.~\cite{DahlstromCP2012}, we evaluate the radial integral and obtain:
\begin{eqnarray}
S_{\vec k/i}
 &\approx&  
-
\frac{\pi}{2}
(8\pi)^{3/2}E_{\omega} E_{\Omega} N_k N_{\kappa} 
\nonumber \\ &\times&
\frac{(2\kappa)^{iZ/\kappa}}{(2k)^{iZ/k}}
\frac{\Gamma[2+iZ(1/\kappa-1/k)]}{(\kappa-k)^{iZ(1/\kappa-1/k)}}
\nonumber \\ &\times&
\frac{1}{|k-\kappa|^2}~\exp\left[-\frac{\pi Z}{2} \left( \frac{1}{\kappa}-\frac{1}{k}\right)\right]
\nonumber \\ &\times&
\sum_{L=\ell_i,\ell_i\pm 2}  Y_{L,m_i}({\hat k})   
\sum_{\lambda=\ell_i\pm 1}\langle Y_{L,m_i}| Y_{1,0}|Y_{\lambda,m_i} \rangle  
\nonumber \\ &\times&
 \langle Y_{\lambda,m_i}| Y_{1,0}|Y_{\ell_i,m_i}\rangle
\braOket{R_{\kappa,\lambda}}{r}{R_{n_i,\ell_i}} i^{-\lambda} e^{i\eta_\lambda (\kappa)}.  
\label{Mas1}
 \end{eqnarray}
Reading this equation from below, the XUV dipole transition 
from the initial state, 
$\ket{i}=\ket{n_i,\ell_i,m_i}$, 
to the intermediate state, 
$\ket{\vec \kappa}$, is found on line five.
The fourth line contains the trivial angular part of the second dipole transition 
from $\ket{\vec \kappa}$ to $\ket{\vec{k}}$, while 
the radial part of this continuum--continuum transition is divided into two factors on line two and three. 
On the third line, the pre-exponential factor: $1/|k-\kappa|^2$, is characteristic 
for free-free transitions \cite{Mattuck}. 
If the IR frequency is relatively small as compared to the kinetic energy of the photoelectron, 
namely in the {\it soft-photon limit}, 
we have $\kappa \approx k-\omega/k$, so that $1/|k-\kappa|^2 \approx k^2/\omega^2$, 
which makes clear that the transition amplitude increases for high electron momenta, 
as well as in the soft-photon limit, $\omega\rightarrow 0$.  

Because we consider transitions between Coulomb states, 
the transition amplitude contains also a
real exponential that depends on the nuclear charge, $Z$, together with $\kappa$ and $k$. 
More interestingly, the second line is {\it complex} and it depends on the same three quantities,
thus, introducing a new {\it phase-shift} into the matrix element.   
Note that these correction factors are {\it universal} since they depend neither on the angular momentum 
nor on the short-range atomic interactions.
Finally, the first line contains some pre-factors, 
such as the electric field amplitudes and normalization constants
of the wavefunctions associated to the relevant electronic states.

To address the question of the phase of the two-photon matrix element, one notices that besides a trivial contribution from the spherical harmonic in the final state, 
$Y_{L,m_i}(\hat k)$, it contains only phase-shifts that are governed by the angular momentum $\lambda$ of the intermediate state, {\it i.e.} a state that can be reached via {\it single}-photon ionization. More precisely,  for a given transition channel (characterized by the angular momenta of the intermediate and final state: $\ell_i \rightarrow \lambda \rightarrow L$), the phase of the matrix element reduces to: 
\begin{eqnarray}
\arg [S_{\vec k/i}^{(L,\lambda,m_i)}] &=&
\pi+\arg[Y_{L,m_i}({\hat k})] + \phi_\Omega + \phi_\omega 
\nonumber \\ &-&
{\pi \lambda \over 2} + \eta_\lambda (\kappa)  + \phi_{cc}(k,\kappa) ,	
\label{phiMas1}
\end{eqnarray}
where $\phi_\Omega$ and $\phi_\omega$ are the phases of the XUV field, $\Omega$, and of the IR laser, $\omega$, 
respectively, and where 
the {\it continuum--continuum phase} can be approximated by 
\begin{equation}
\phi_{cc}^{(P)}(k,\kappa)\equiv
\arg\left\{
\frac{(2\kappa)^{iZ/\kappa}}{(2k)^{iZ/k}}
\frac{\Gamma[2+iZ(1/\kappa-1/k)]}{(\kappa-k)^{iZ(1/\kappa-1/k)}}
\right\},
\label{phiccP}
\end{equation}
corresponding to the argument of line 2 in Eq.~(\ref{Mas1}). The superscript, $(P)$, 
indicates that this result is obtained when using long-range {\it phases} of the type in Eq.~(\ref{longrangephase}). 
The continuum--continuum phases for absorption and stimulated emission of one IR laser photon are shown in Fig.~\ref{ccphases}.
\begin{center}
	\includegraphics[width= 0.85\textwidth]{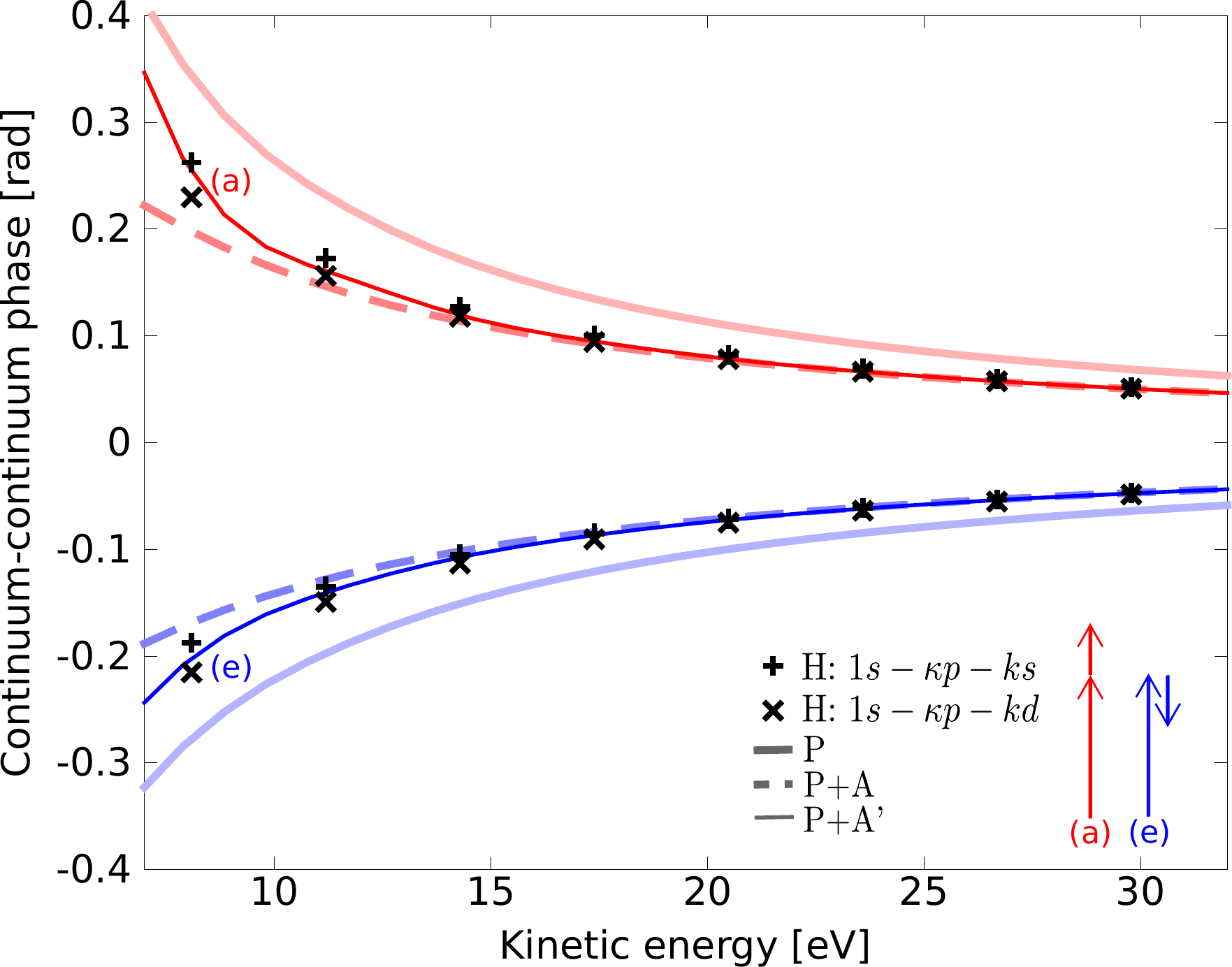}
	\captionof{figure}{ {\it Continuum--continuum phases:}
ATI phase-shifts for absorption (a) of one 800\,nm laser photon;
and for stimulated emission (e) of one laser photon, in a Coulomb potential with $Z=1$. 
The asymptotic approximation (P) provides the correct qualitative behaviour, 
while the long-range amplitude corrections (P+A and P+A') 
lead to quantitative agreement, at high enough energy, with the exact calculations in hydrogen ($+$ and $\times$). 
	}
	\label{ccphases}
\end{center}
We stress that the final state scattering phase, $\eta_L(k)$, cancels out and that it enters neither in  Eq.~(\ref{Mas1}) nor in Eq.~(\ref{phiMas1}). 

\subsubsection{Long-range amplitude corrections}

So far, we have established that the ``long-range phases'', 
{\it i.e.} the logarithmic divergence of the Coulomb phases in Eq.~(\ref{longrangephase}), 
is important for the phase of ATI transitions in an ionic potential of charge $Z$.
In order to improve our asymptotic approximation,   
we must include also long-range amplitude-variations.
We do this by returning the normalization constants $N_k$ and $N_\kappa$  
back inside the radial integral in Eq.~(\ref{S2assymp}), 
and write them as functions $N_k(r)$ and $N_\kappa(r)$ for the radial amplitude variations \cite{DahlstromCP2012}.
In analogy with the local momentum from WKB theory, as stated in Eq.~(\ref{WKBstate}), 
these effects can then be approximated by expanding the normalization factors, so that
\begin{eqnarray}
N_k(r)N_{\kappa}(r) &\equiv & 
\sqrt{\frac{2}{\pi p_k(r)}}\sqrt{\frac{2}{\pi p_\kappa(r)}}
\nonumber \\
& \approx &
\sqrt{\frac{4}{\pi^2 k \kappa}}
\Big[
1-
\underbrace{
\frac{1}{2} 
\left(\frac{1}{\kappa^2} + \frac{1}{k^2} \right) \frac{Z}{r} 
}_{{Correction \ term}}
\Big], 
\label{ampcorrasymp}
\end{eqnarray}
to first-order in the Coulomb potential. 
It is also possible to recover a similar asymptotic form of the Coulomb functions
by using the iterative formulas in Ref.~\cite{Abramowitz}, 
{\it i.e.} without the need for a semi-classical approximation.
Including this correction term, the continuum--continuum phase becomes: 
\begin{eqnarray}
\phi_{cc}^{(P+A)}(k,\kappa) \equiv  && \nonumber \\
\arg\left\{
\frac{(2\kappa)^{iZ/\kappa}}{(2k)^{iZ/k}}
\
\frac{
\Gamma[2+iZ(1/\kappa-1/k)] \ + \ \gamma(k,\kappa)
}{(\kappa-k)^{iZ(1/\kappa-1/k)}}
\right\}, &&
\label{ccphasePA}
\end{eqnarray}
where the additional term is
\begin{equation}
\gamma(k,\kappa)=
iZ \ \frac{(\kappa-k)(\kappa^2+k^2)}{2\kappa^2k^2} \ 
\Gamma[1+iZ(1/\kappa-1/k)],
\label{ccphasePAgamma}
\end{equation}
with Eq.~(\ref{ccphasePA}) corresponding to the long-range phase {\it and} amplitude data (P+A) in Fig.~\ref{ccphases}. 
We note that this amplitude correction leads to an excellent agreement with the exact calculation at high energies.  
The Taylor expansion in Eq.~(\ref{ampcorrasymp}) 
is, however, problematic because the Coulomb potential will dominate over the kinetic energy close to the core, 
$Z/r>\epsilon_k$ as $r\rightarrow 0$, which implies that the expansion is invalid for $r<Z/\epsilon_k$.
This leads to an artificial divergence of the electron wavefunction close to the core, 
which is physically unreasonable because the wavefunction should vanish at $r=0$. 

Because an accurate analytical expression for the continuum--continuum phase 
can be a valuable tool in the analysis of experimental data, 
we may try to avoid the region close to the core, 
by an {\it ad-hoc} mathematical trick, 
namely by changing the starting point of the radial integral.  
Indeed, the result presented as (P+A') in Fig.~(\ref{ccphases}) is slightly better,  
where the gamma function in Eq.~(\ref{ccphasePAgamma}) was replaced  
by an {\it incomplete} gamma function represented by an integral with an imaginary starting point, $r_0=iZ(1/\kappa^2+1/k^2)/4$.
(We note that this ``regularization procedure'' is slightly more systematic than 
the procedure presented in Ref.~\cite{DahlstromCP2012}, but the best matching point was found by trial-and-error.)
Strictly speaking, at very low kinetic energy, typically below 10~eV, 
the asymptotic approximation starts to break down. 
Physically, this is due to an increasing influence of the Coulomb potential for slow electrons. 
Mathematically, this is expected since the asymptotic expressions for Coulomb function 
have $kr$ as arguments, which implies that a small $k$ will lead to a ``good'' asymptotic wavefunction only
at a large distance $r$ from the core \cite{Abramowitz}.


\subsection{Extracting time-delay information from laser-assisted photoionization signals}
\label{sec:rabitt}
Having established the asymptotic approximation for the complex amplitudes of ATI processes,
we now turn to the {\it probability} for the emission of a photoelectron with energy $\epsilon_k=\epsilon_i+\Omega$,
as depicted in Fig.~\ref{arrowshigherorder}. 
The probability is given by the square of the sum of the amplitudes:  
\begin{eqnarray}
P_{\vec k}&\approx& |S_d+S_a+S_e|^2 \nonumber \\
&=&|S_d|^2+|S_a|^2+|S_e|^2 \nonumber \\ &+& 2\Re\left\{S_d^*(S_a+S_e)+S_a^*S_e\right\},
\label{Pdae}
\end{eqnarray}  
where $d$, $a$ and $e$ label the paths associated to the lowest-order processes: 
(d), (a) and (e) in Fig.~\ref{arrowshigherorder}. 
The total probability depends on the relative phase of all individual quantum paths
and the maximal probability for photoemission occurs when all paths are in phase, 
$\arg[S_d]=\arg[S_a]=\arg[S_e]$. 
In experiments, the phase of the two-photon amplitudes labelled (a) and (e) 
can be controlled by changing the sub-cycle delay, $\tau$, between the probe field and the attosecond pulses.  
More precisely, one controls the relative phase of the IR field, 
$\phi_{\pm\omega}\equiv\pm\omega\tau$ in Eq.~(\ref{phiMas1}), 
with respect to the group delay of the XUV pulse. 
The probe-phase dependence is $S_a\propto E_\omega\propto \exp[i\omega\tau]$ 
and $S_e\propto E_\omega^*\propto \exp[-i\omega\tau]$. 
This implies that the cross-terms in Eq.~(\ref{Pdae}) vary differently as a function of $\tau$:
(d)-(a) and (d)-(e) are modulated with periodicity $\omega\tau$ associated to the exchange of only one IR photon \cite{ChiniOE2010}; 
while the cross-term (a)-(e) is modulated with periodicity $2\omega\tau$ due to the two IR photons involved \cite{VeniardPRA1996,Muller2002}. 

The latter transitions are directly involved in the RABITT scheme, since path (d) does not contribute. 
This is because only odd XUV harmonics are used to photoionize the atom, as was indicated in Fig.~\ref{RABITT}, 
so that the signal from cross-term (a)-(e) occurs at zero-background. 
Let us emphasize that in the case of a SAP of XUV radiation, the delay-dependent modulation 
of the photoionization signal at the central energy $\Omega_0$, 
will {\it also} be given by the cross-term (a)-(e) alone, 
because the other two cross-terms cancel with each other \cite{DahlstromCP2012}.
All the cross-terms in Eq.~(\ref{Pdae}) together form the {\it on-set} of streaking,
{\it i.e.} a small constructive or destructive interference at the high or low momentum respectively, 
so that the average momentum of the photoelectron is deflected by different amounts depending on $\tau$.  
Provided that the soft-photon approximation is valid, 
the displacement of this entire structure for a laser-assisted SAP is exactly equal to that of APT \cite{DahlstromCP2012}.     
Therefore, the maximal probability for photoemission occurs when the amplitudes associated to paths (a) and (e) are in phase, 
$\arg[S_a]=\arg[S_e]$,
which is equivalent to   
\begin{eqnarray}
\phi_{\Omega_<}+\omega\tau+\eta_\lambda(\kappa_<)+\phi_{cc}(k,\kappa_<)= && \nonumber \\
\phi_{\Omega_>}-\omega\tau+\eta_\lambda(\kappa_>)+\phi_{cc}(k,\kappa_>) 
\label{findmaxP}
\end{eqnarray}
where we have used the explicit phases of the relevant two-photon matrix elements in Eq.~(\ref{phiMas1}),  
assuming one dominant intermediate angular channel, with angular momentum $\lambda$; 
and momenta $\kappa_<$ and $\kappa_>$ corresponding to absorption of photon $\Omega_<$ and $\Omega_>$, respectively.
The solution to Eq.~(\ref{findmaxP}) is: 
\begin{eqnarray}
\tau &=& 
\overbrace{\frac{\phi_{\Omega_>}-\phi_{\Omega_<}}{2\omega}}^{ \tau_{GD}} \nonumber \\
&+&
\underbrace{\frac{\eta_\lambda(\kappa_>)-\eta_\lambda(\kappa_<)}{2\omega}}_{ \tau_{\lambda}}+
\underbrace{\frac{\phi_{cc}(k,\kappa_>)-\phi_{cc}(k,\kappa_<)}{2\omega}}_{ \tau_{cc}},
\label{tauRABITT}
\end{eqnarray}
where we observe that the probe-delay, $\tau$, that maximizes the yield is a sum of three delays: 
\begin{itemize}
\item $\tau_{GD}$: the group delay of the XUV field is {\it when} the attosecond pulse arrived on target, 
{\it i.e.} when the fields $\Omega_<$ and $\Omega_>$ added constructively at the atom.
\item $\tau_\lambda$: the Wigner delay is the ``delay'' in single-photon ionization, {\it i.e.} 
an asymptotic temporal-shift of the photoelectron wave packet. 
Strictly speaking, this delay is 
interesting, {\it i.e.} giving information on the electron dynamics, only in difference with a Coulomb reference, 
{\it e.g.} hydrogen with $\eta^{(H)}_{\lambda}(\kappa)\equiv\sigma_{\lambda}(\kappa)=\arg\{\Gamma[1+\lambda-i/\kappa]\}$. 
In this way, we may say that the photoelectron is {\it delayed} by 
$\tau_\lambda-\tau^{(H)}_\lambda$, as compared to hydrogen.
\item $\tau_{cc}$: the continuum--continuum delay, 
{\it i.e.} a {\it measurement-induced delay} due the electron being probed 
by an IR laser field in a long-range potential with a Coulomb tail of charge $Z$. 
This delay can be traced back to the phase-shifts of the ATI matrix elements.  
\end{itemize}
We stress that the delays presented in Eq.~(\ref{tauRABITT}) are calculated from the {\it finite-difference} approximations 
to the actual derivatives, 
$\tau_{GD}=\partial \phi_\Omega / \partial \Omega$ and 
$\tau_{\lambda}=\partial \eta_{\kappa,\lambda} / \partial \Omega$. 
For these approximations to be valid, we must require that the spectral phases vary slowly, 
{\it e.g.} that $|\Delta\phi_\Omega| \equiv |\phi_{\Omega>}-\phi_{\Omega_<}| \ll 2\pi$, 
corresponding to a small phase variation over two laser photons, $\Delta\Omega = 2\omega$.
Using the framework presented in Sec.~\ref{sec:groupdelays}, 
one can easily show that these slowly varying phases imply that the frequency components 
of the attosecond pulse must be confined to within a fraction of the laser period. 
In this sense, ``slow reactions'', {\it e.g.} resonances, 
which induce a dramatic phase jump in the spectral domain 
are difficult to study using the conventional RABITT scheme.
In the case of streaking, the delays arise also as finite-difference approximations to the actual delays,
but here the smallest energy difference is instead one laser photon, $\Delta\Omega=\omega$ \cite{DahlstromCP2012}.   
The atomic delays, $\tau_\theta=\tau_\lambda+\tau_{cc}$, are shown in Fig.~\ref{ionizationdelays}, 
corresponding to the case of hydrogen, with $\tau_\lambda$ for a few different values of the angular momentum, $\lambda$. 
\begin{center}
	\includegraphics[width= 0.85\textwidth]{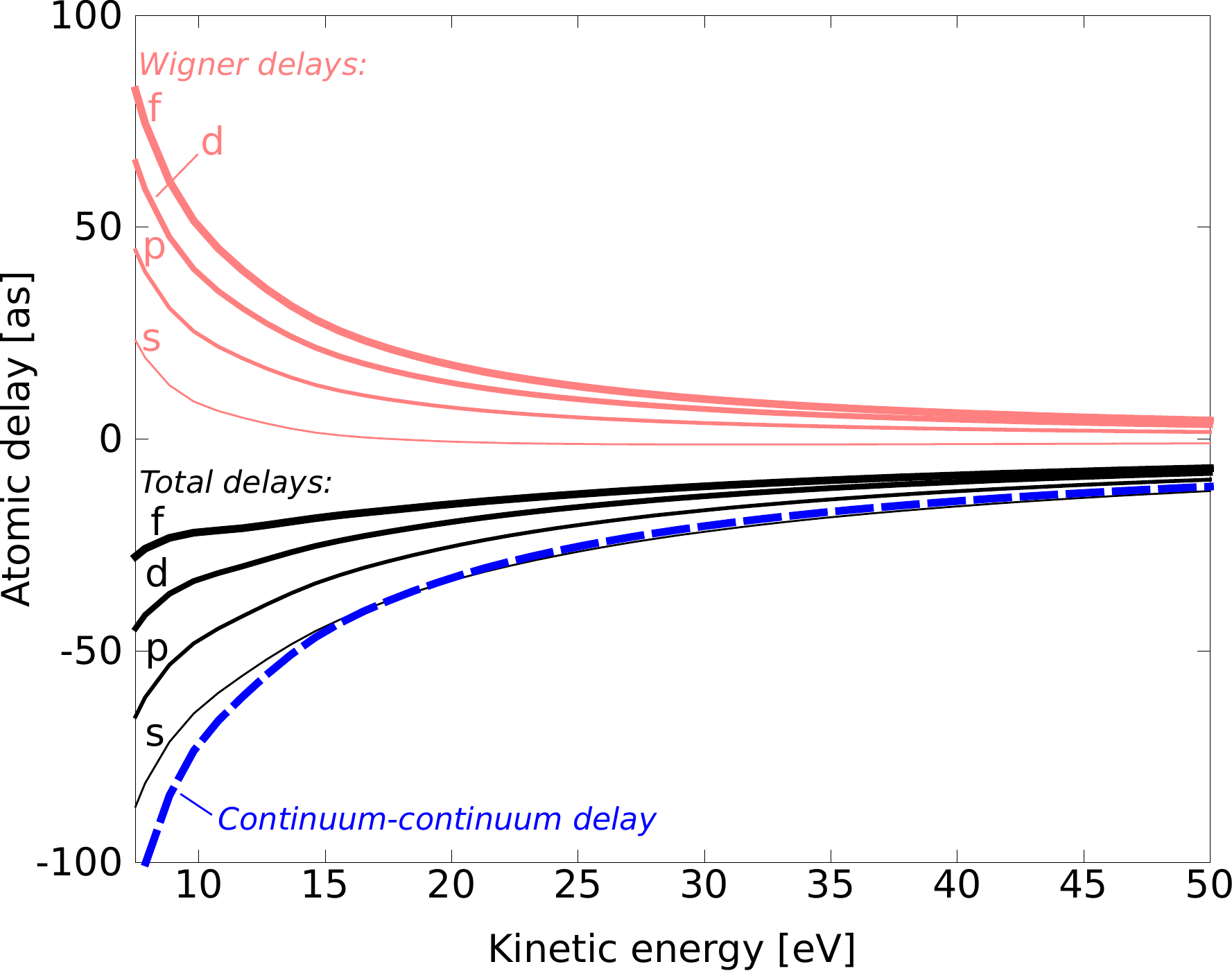}
	\captionof{figure}{   {\it Atomic delays in laser-assisted ionization:}
The Wigner delays for hydrogen with a photoelectron of $s$, $p$, $d$ or $f$-character [red, upper curves] plus  
the universal continuum--continuum delay for the laser-probing process with $Z=1$ and $\hbar\omega=1.55$~eV [blue, dashed curve],
yields the total delay in laser-assisted photoionization [black, lower curves]. 
	}
	\label{ionizationdelays}
\end{center}

In this way, it is possible to ``measure'' the group delay of the attosecond pulse provided that the atomic delay: 
$\tau_{\theta}=\tau_{\lambda}+\tau_{cc}$, can be calculated accurately and that it is subtracted from the experimental data. 
The first observation of attosecond pulse structures in 2001 was made in this way by subtracting the influence of the ``atomic phase'' as calculated by means of complex two-photon matrix elements \cite{PaulScience2001}. We stress that the exact temporal structure of the attosecond pulse is {\it not} measured directly, neither with the RABITT nor with the attosecond streak-camera technique, 
but that the influence of the atomic potential tends to decrease at high kinetic energy. 
The question of the precise determination of these ``atomic delays'' (also called ``streaking delays'')
and of their importance in RABITT and streaking measurements has motivated a
large number of theoretical studies, see Ref.~\cite{BaggesenPRL2010,YakovlevPRL2010,ZhangPRA2010,KheifetsPRL2010,ZhangPRA2011,SmirnovaPRL2011,NageleJPB2011,NagelePRA2012,MoorePRA2011}.
Much of this recent work has focused on the attosecond streak-camera technique, 
where typically non-perturbative approaches are employed to account for high-order photon processes with the probe field. 
In the next subsection, we will also comment on the influence of such an intense probe field 
with a view-point from perturbation theory using the asymptotic approximation.

Alternatively, the Wigner-like delay, $\tau_{\lambda}$, can be determined if $\tau_{GD}$ is known from independent measurements 
and if $\tau_{cc}$ can be calculated \cite{KlunderPRL2011,DahlstromCP2012}. 
However, since the precise characterization of the attosecond pulse requires the previous knowledge of $\tau_{\lambda}$, 
it would seem that we have run into an uncomfortable circular argument. 
Instead, if the ``same'' attosecond pulse is used to photoionize simultaneously different atomic states, 
information about the differences in atomic delays, $\tau_\theta=\tau_\lambda+\tau_{cc}$, can be obtained directly in the experiment, without {\it any} need to know the temporal structure of the attosecond pulse 
\cite{Schultze25062010,KlunderPRL2011}. 
Then, by simply subtracting the universal $\tau_{cc}$ as given by Eq.~(\ref{tauRABITT}), 
the $\tau_{\lambda}$-difference between the two photoelectrons is obtained. A similar separation of the time-delays as shown in Eq.~(\ref{tauRABITT}) was reported by the group of Burgd\"{o}rfer for streaking of SAP by solving numerically the TDSE \cite{NageleJPB2011,NagelePRA2012}. 
Interestingly, they also showed that $\tau_{cc}$ [there called: ``the Coulomb-laser  coupling''] 
could be calculated from a classical ensemble of electrons 
interacting simultaneously with both the laser field and the Coulomb potential. 
Prior to these successful demonstrations of the seperability of the two contributions, 
the influence of the short-range scattering phase-shifts was shown by Yakovlev and coworkers \cite{YakovlevPRL2010}
and the importance of the long-range Coulomb tail was discussed by Zhang and Thumm \cite{ZhangPRA2011}.

\subsection{Towards higher probe intensities}
So far the discussion has been conducted within a perturbative framework,
by retaining the lowest-order contributions from the IR probe field.
A natural task is then to investigate the consequences of an increased probe-field intensity. 
Numerical evaluations of the TDSE show that a strong probe field will alter the temporal information \cite{MauritssonPRA2005}. 
A modest increase of the probe intensity will, however, not alter the observed delay.
This stability of the delay at low-intensity probe fields was first pointed out by Zhang and Thumm \cite{ZhangPRA2010} 
within the Eikonal-Volkov Approximation \cite{SmirnovaJPB2006} by considering a weak probe field. 

Using perturbation theory for the interaction with the probe field makes it difficult to draw conclusions  
about the strong-probe field interaction, 
but it is possible to apply the asymptotic approximation, Eq.~(\ref{S2assymp}),
also for processes involving additional probe photons.  
In Fig.~\ref{arrowshigherorder}, the complex amplitudes involving two probe photons are labelled: 
(d$_{ae}$), (d$_{ea}$), (aa) and (ee). 
Here, (d$_{ae}$) and (d$_{ea}$) can be regarded as corrections to the direct path (d), 
because they exchange two probe photons so that the total energy is unchanged. 
Any such ``probe-photon loops'', indicated by grey rings in Fig.~\ref{arrowshigherorder}, 
involve two continuum--continuum transitions: $k \rightarrow \kappa \rightarrow k$. 
Therefore, they acquire two continuum--continuum phases: $\phi_{cc}(\kappa,k)$ and $\phi_{cc}(k,\kappa)=-\phi_{cc}(\kappa,k)$, 
which compensate each other. 
For instance, the phase of (d$_{ae}$) and (d$_{ea}$) is equal to that of the direct path (d)  
[except for trivial perturbation expansion coefficients].
The two other paths (aa) and (ee) yield a new probability modulations at $4\omega\tau$, 
due to the four photons involved in the (aa)-(ee) cross-term, and they will not disturb the lower-order modulations. 
One can show that the modulation shift of the (aa)-(ee) cross-term will be approximately equal 
to that of the (a)-(e) cross-term, so that no fundamentally new information is obtained at this higher-modulation rate, 
as reported in the experimental work by Swoboda and co-workers \cite{SwobodaLP2009}. 
In this way, including additional probe-photon loops, 
one can argue for the validity of Eq.~(\ref{tauRABITT}), 
also when higher-order contributions become important,  
at least while the probe-interactions remain in the perturbative regime
and the bound states are not too much dressed. 

In the following section, we will review the state-of-the-art experimental efforts to measure 
these delays in laser-assisted photoionization. As we will see, there are so-far few experimental results 
to compare with the theoretical predictions, and in addition, these experimental data points are 
located in highly complex regions, {\it e.g.} close to pseudo-resonances in Neon \cite{MoorePRA2011} and correlation-induced Cooper minima in Argon \cite{GuenotPRA2012}, where the SAE approximation is likely to break down.

\section{Experimental observations of attosecond delays in photoemission}
\label{sec:expobs}
  
The approximations for laser-assisted photoionization described in Sec.~\ref{sec:STPT} 
hold very well for atomic systems, where the photoelectron can be described within the SAE approximation.   
In fact, the accuracy of the asymptotic approximation was first bench-marked by comparison with exact calculations in hydrogen 
carried out by R.~Ta\"ieb \cite{KlunderPRL2011,DahlstromCP2012}. 
The experimental work on photoionization time-delays, on the other hand, 
is carried out on many-electron systems, most often noble gas atoms such as Neon and Argon.
The question then arises to which degree the SAE approximation is valid for these systems,
and especially so when the photoelectron is released, not from the outer-most orbital, but from ``inside'' the core. 
Screening effects due to electron correlation can be accounted for 
using many-body perturbation theory (MBPT), such as the random-phase approximation (RPA), 
or more elaborate methods \cite{KheifetsPRL2010,NagelePRA2012,MoorePRA2011,GuenotPRA2012},
but we have not considered such corrections here. We predict, however, that in the coming years, 
the interpretation of attosecond experiments will offer a new testing ground for many-body calculations.  
Future experiments will provide interesting opportunities to test various  
theoretical methods, such as MBPT, for light-induced electron-electron interactions 
in connection to measurements of phases and delays in complex atomic and molecular systems.   
At this point, we report that a satisfactory agreement between theory and experiment has not yet been reached,
and that more experiments are needed in ``simple'' energy regions, where perhaps more straight-forward analysis would apply.
In the following, we will proceed with a brief overview of the current experimental efforts. 
We will place special emphasis on the time-delay measurements and phase measurements using APT, 
where we have taken active part, 
but we will also discuss the experiments using SAP. 
We stress that both approaches aim at measuring the so-called atomic delays \cite{PaulScience2001}, 
and that the one-photon Wigner delay can be accessed only after subtracting the continuum-continuum delay as computed by 
either the quantum mechanical or classical approaches \cite{KlunderPRL2011,NageleJPB2011}.

\subsection{Atomic-delay measurements using APT}

A delay of 110\,as between the $3p$ and $3s$ states in Argon has been 
measured experimentally by Kl\"under and co-workers using an APT 
with a photon energy of $\sim 35$\,eV \cite{KlunderPRL2011}. 
Theoretically, this delay is identified as a {\it difference} of atomic phases between two orbitals.  
These concepts in laser-assisted photoionization 
dates back to the first measurements of attosecond pulses in 2001 using RABITT \cite{PaulScience2001,Muller2002},
but the ``atomic delays'', by themselves, had never been accessible experimentally before  
due to the simultaneously unknown GD of the attosecond pulses. 
A breakthrough in measuring these atomic delays was possible 
due to the new way that the experiment was performed,
where {\it two} RABITT scans were recorded simultaneously from two different initial orbitals.
Interestingly, in this setup, the unknown temporal structure of the attosecond pulses can be subtracted 
without ever knowing their exact shape, as can be understood by considering Eq.~(\ref{tauRABITT}).
While the experiments is distinctly different, 
the method was inspired by the previous delay-measurements using SAP \cite{CavalieriNature2007,Schultze25062010}.   
A schematic illustration of the experiment is shown in Fig.~\ref{kathrinDouble}.
\begin{center}
	\includegraphics[width= 0.85\textwidth]{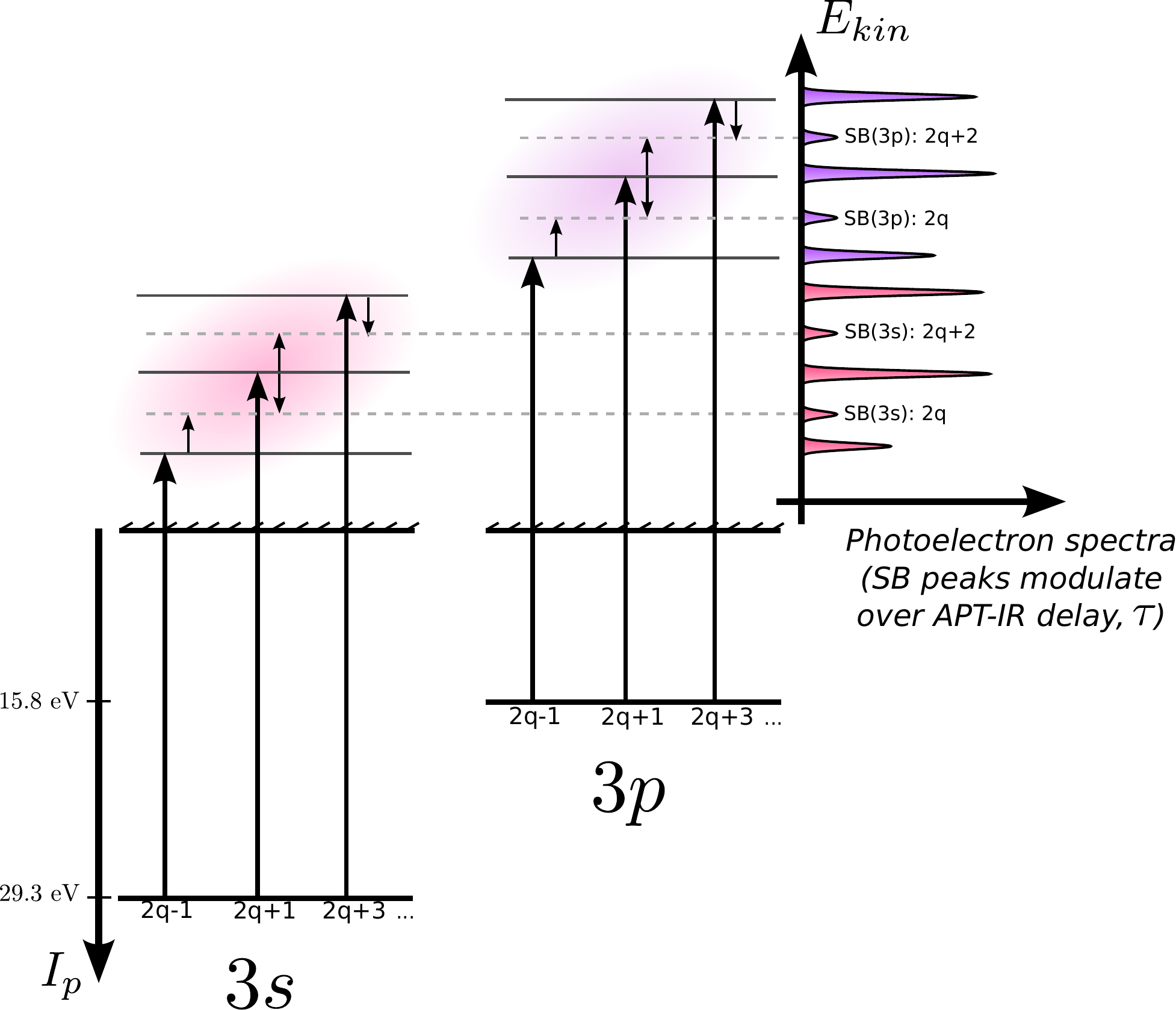}
	\captionof{figure}{ {\it Delay experiment for the $3s$ and $3p$ states in Ar using APT:} 
The APT is composed of three odd harmonics, $2q+1$, that ionize (pump) 
electrons from either initial state, $3s$ or $3p$, into the continuum. 
After absorbing a harmonic photon, the photoelectron can 
either absorb or emit a laser probe photon in order to reach an even number sideband (SB) state, labelled by $2q$. 
The SB probability oscillates with the delay APT-IR delay, $\tau$, due to interference between the two quantum paths. 
Information about the attosecond timing is found in the relative modulation offset between 
the same sideband numbers from different initial states \cite{KlunderPRL2011}. 
	}
	\label{kathrinDouble}
\end{center} 
In this work, the high-order harmonic comb was first passed through 
a thin Chrome foil acting as a band-pass to separate four odd harmonics (21-27).
In this way, only three sidebands, $\textrm{SB:}~22,~24,~26$, 
were produced when ionizing argon atoms from either orbital. 
This limiting of the harmonic comb was made to prevent different sidebands from
different orbitals to overlap in photoelectron kinetic energy.
The analysis of the experiment \cite{KlunderPRL2011} was first carried out within the SAE approximation using Hartree-Fock phase-shifts 
from the litterature \cite{KennedyPRA1972}, 
an adequate theoretical method for photoelectrons from the {\it outer} orbital, $3p$. 
On the other hand, photoelectrons from the {\it inner} orbital, $3s$, couple strongly with the outer orbital, 
and the corresponding single-photon phase-shifts are greatly altered. 
This important effect was identified by Kheifets by including RPA effects in the absorption of a single XUV-photon. 
Interestingly, these photoelectrons exhibit a correlation-induced minimum in the one-photon ionization step,
{\it i.e.} a complex kind of Cooper minimum \cite{CooperPR1962}, 
with a corresponding peak of several hundreds of attoseconds in the atomic delay \cite{GuenotPRA2012}. 
This is, indeed, a challenging spectral region, where several quantum paths can interfere, 
and the results become difficult to interpret theoretically.

The signal can also be strongly altered by atomic resonances.
In the spectral domain, an APT corresponds to harmonics that can be ``aimed'' towards specific energy regions of interest. 
The frequency of the harmonics from the HHG process can be tuned naturally 
by changing the fundamental (driving) laser pulse frequency, $\omega+\delta\omega$, 
so that the high-order harmonic frequencies increase or decrease by $\delta\Omega=(2q+1)\delta\omega$. 
Swoboda {\it et al.} found that the phase of the modulation of the lowest-lying sideband in Helium  depended critically on the frequency of the harmonics \cite{SwobodaPRL2010}. 
In Fig.~\ref{swobodaResonance}, an illustration of the principle of the experiment is shown.
\begin{center}
	\includegraphics[width= 0.85\textwidth]{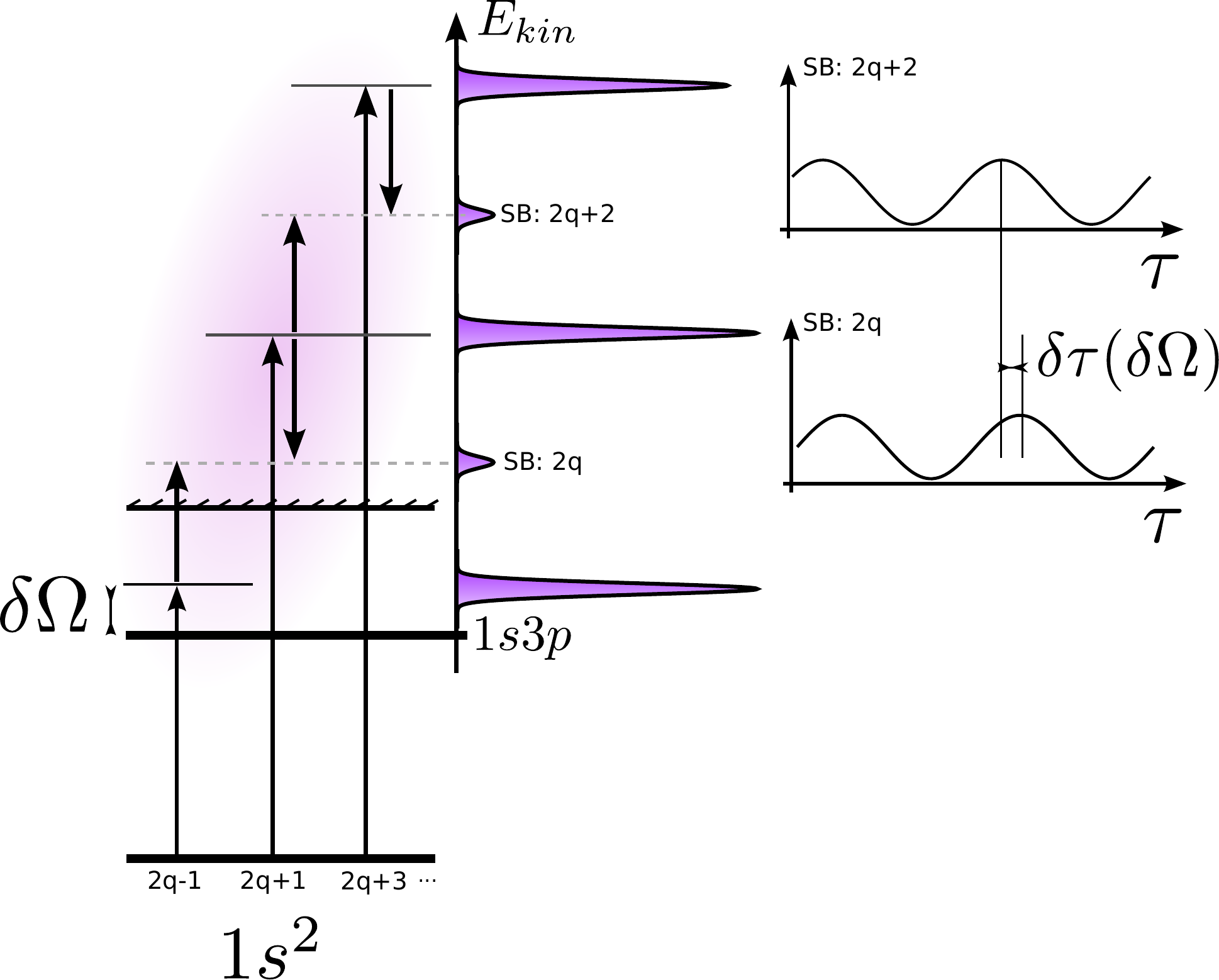}
	\captionof{figure}{ {\it Phase induced by the $1s3p$ state in He using APT:}
The phase in photoionization, $\delta\tau$, 
depends on the detuning, $\delta\Omega$, 
between the harmonic, $(2q-1)\omega=15\omega$, and the bound state, $1s3p$. 
The effect arises from the $\pi$-jump that occurs in the two-photon amplitude,  
when passing the resonance. 
The higher-lying sideband, $\textrm{SB:}~2q+2$, acts as an important ``reference clock'' in this experiment, 
which is mostly insensitive to small frequency changes of the harmonic fields \cite{SwobodaPRL2010}.
	}
	\label{swobodaResonance}
\end{center}
The observed modulation-shifts were attributed to a phase effect occurring over the intermediate {\it bound } state, $1s3p~ ^1P_1$, 
which was ``on resonance'' with a below-threshold harmonic, $15\hbar\omega<I_p^{(He)}$. 
The modulation-shift was then compared to the corresponding case in Argon,
where the harmonic instead directly creates a photoelectron, $15 \hbar  \omega>I_{p}^{(Ar)}$, and where the bound-states do not contribute considerably. 
By tuning the harmonic over the sharp resonance, an abrupt $\pi$-shift is expected. The experimental $\pi$-shift is, however, ``smoothed'' due to the finite duration of the laser probe field and of the APT ($\sim 30$~fs), which is much shorter than that of the long-lived bound state. In order to fully resolve the abrupt rise of the delay due to pure atomic effects, one needs harmonics that are more narrow in frequency than the atomic resonance, 
{\it i.e.} an APT which is longer that the lifetime of the state. 
Similar phase-shifts have also been found and analysed at short-lived complex continuum resonances in Nitrogen molecules 
\cite{HaesslerPRA2009,CaillatPRL2010}. In this case, the life-times are more comparable to the duration of the APT, implying that observed modulation-shifts contain some information of about the lifetime of the resonances.

\subsection{Atomic-delay measurements using SAP}

It was for a long time believed that the so-called ``attosecond streak-camera'' \cite{ItataniPRL2002},
provided a direct link between the time-domain and the energy-domain, 
as often illustrated by measuring the oscillations of few-cycle optical light pulses \cite{GoulielmakisScience2004}. 
This stand-point needed to be revised after an experiment by Schultze and co-workers, where it was stated that 
``the zero of time in atomic chronoscopy is currently tainted with an error of up to several tens of attoseconds'' 
\cite{Schultze25062010}, 
due to a measured  time-delay of $\sim 20$\,as between the streaked photoelectrons 
from the $2p$ and $2s$ shells in Neon at an XUV photon energy of 100\,eV.  
The key to obtaining this intriguing experimental result was to collect simultaneously {\it two} streaked photoelectron spectrograms from two different initial atomic states, thus accessing the {\it difference} in streaking  delay between the two processes. Interestingly, by measuring this difference in timing, the unknown GD of the attosecond pump pulse cancels out, 
because it is the same in both processes, as can be identified using Eq.~(\ref{tauRABITT}).
The experimental results motivated at great deal of further theoretical work, 
where it was concluded that the long-range Coulomb tail is responsible for a 
considerable part of the observed delay in streaking 
\cite{ZhangPRA2010,ZhangPRA2011,NageleJPB2011}. 
In addition to this long-range effect, the laser-induced polarization of the atom can lead to 
a shift of the streaking delay if the probe field is too strong 
\cite{SmirnovaJPB2006,BaggesenPRL2010}.
Extensive theoretical work including many multi-electron effects have so far only  
accounted for a fraction of the experimental time-zero shift \cite{KheifetsPRL2010,MoorePRA2011}.
The theory of the attosecond streak-camera has been refined to account for the long-range Coulomb tail by matching the asymptotic wavefunction to the appropriate scattering phase \cite{SmirnovaPRL2011}, 
but these modifications are controversial as they go against Eq.~(\ref{tauRABITT}), as is also stressed in Ref.~\cite{NagelePRA2012}. 
More experimental results from other systems than Neon, 
are required to compare with these new theoretical models. 

In conclusion, the attosecond pump--probe schemes using a SAP or APT, Fig.~\ref{pulsesketch}~(b) and (c) respectively,
can be used to obtain information about the probe field-dressed system, 
and not directly the probe-free system. 
To explore properties of the unprobed system
using experimental data where the probe is present, 
 a theoretically reliable way of accounting for the probe must be found \cite{KlunderPRL2011, NageleJPB2011}.
By decreasing the intensity of the probe field,
the streaking modulation is reduced and one arrives in a regime where standard perturbation theory is applicable \cite{DahlstromCP2012}. This presents a large advantage for quantitative theoretical work, since it can be based on well-established MBPT.

\subsection{Other attosecond time-delay experiments}
\label{sec:attoselfprobeexp} 
Finally, in this subsection we briefly discuss other attosecond delay experiments that are {\it not} directly related to single-photon ionization by an XUV field.
This important class of attosecond experiments are conducted in 
the HHG cell directly, and they are sometimes referred to as {\it self-probing}
systems, see Ref.~\cite{HaesslerJPB2011} for a recent tutorial on self-probing of molecules with HHG. 
In these experiments, the electron wave packet in the continuum is thought of as a probe for the entire system, 
holding promise of simultaneous temporal and spatial resolution of molecular dynamics \cite{Worner2010}. 
%
In this subsection, we will restrict ourselves to one of the simplest self-probing systems, namely, atoms undergoing HHG in two-colour laser fields consisting of a strong fundamental, $\omega$, and a much weaker, parallel second harmonic, $2\omega$. An interesting aspect of this two-colour setup, is that the presence of the second harmonic breaks the inversion symmetry of the HHG process, which leads to the production of odd {\it and} even harmonics  \cite{MauritssonPRL2006,FrolovPRA2010,Shafir2010}. 
A phase-difference (or time-delay) in the intensity modulation of different even harmonics was observed as a function of the delay between the $\omega$ and $2\omega$ fields. Early experimental and theoretical work suggested that the phase of this modulation could be used to characterize the emission times of attosecond pulses \cite{DoumyPRL2009,DahlstromPRA2009,DudovichNP2006}. Experimentally, it was eventually found that the phase behaved in an  unexpected way close to the harmonic cut-off \cite{HePRA2010}, 
which, after some more detailed calculations \cite{DahlstromJPB2011}, lead to a new interpretation of the observed delays 
with, instead, a connection to the Keldysh tunnelling parameter \cite{KeldyshJETP1965}. 
%

The delay in {\it tunnelling} ionization has been studied by the group of Keller using angular streaking, often referred to as the ``attoclock'' 
\cite{EckleScience2008,EckleNP2008,PfeifferNP2011}. In this  setup, a close-to-circularly polarized femtosecond laser field 
is used to tunnel-ionize electrons. Subcycle resolution is obtained by studying angular displacements of the final momentum of the photoelectrons. Using this angular streaking method a vanishing tunnelling delay has been confirmed. We need to stress the differences between the two kinds of streaking: The conventional form of streaking requires a SAP and a femtosecond probing laser field with linear polarization, 
while the angular streaking requires only a single femtosecond pulse with a tailored polarization. In regular streaking, temporal information about laser-assisted, single-XUV-photon ionization is gained; 
while in the case of angular streaking, the temporal aspects of strong-field (multi-photon) tunnelling is studied. The theoretical work presented in Ref.~\cite{PfeifferNP2011} highlights the difficulties and prospects of self-probing systems involving electron continuum probes initiated through strong-field tunnelling.

\section{Conclusions and outlook}
\label{sec:conclusion}

Attosecond science is attracting a great deal of attention because it promises  
accurate control and probing of electron processes in atomic and molecular systems in real time. 
In this tutorial, we have presented a detailed theoretical analysis of photoionization of neutral atoms 
by attosecond XUV pulses and coherent XUV high-order harmonics. 
We have found that, indeed, attosecond pulses can be used to initiate photoionization at well-defined times, 
but to probe such an ultra-fast event requires careful considerations about the quantum mechanical interactions  
between the photoelectron and the remaining ion.  
Ultimately, it is commonly believed, that the probing process will be carried out using a second attosecond pulse, 
but such experiments are demanding due to the need for two-photon processes in the XUV range. 
We refer the reader to Ref.~\cite{TzallasNature2011} for a state-of-the-art experimental scheme  
using XUV for both pumping and for probing the processes on a time-scale bordering the attosecond domain. 

At the present time, 
the temporal aspects of photoionization are more often probed by a phase-locked, IR-laser field. 
This has the advantage of a much stronger interaction with the photoelectron, 
in fact, here the interaction can be so strong that many such laser photons are absorbed. 
This situation is called ``streaking'' because it can shift the photoelectron momentum distribution
to higher or lower values depending on the subcycle delay between the attosecond pulses and the probe field.  
This highly non-linear process is difficult to handle with high-fidelity in theory, 
and this is the main reason why we have considered instead the interaction with a moderate probe field,
which may induce exchange of a single IR-laser probe photon at most. 

Such moderate probe fields are the standard choice in the temporal probing of APT using the RABITT scheme.  
We have demonstrated \cite{DahlstromCP2012} that this interferometric interpretation is 
valid also for SAP in the {\it on-set} of streaking. 
Using this ``photon-picture'', we have identified the phases of the lowest-order quantum paths and
we have shown that the same temporal information about the photoionization process is gained 
with either pulse structure, SAP {\it or} APT. 

Placing an interpretation on the ``delay'' in laser-assisted photoionization, 
is not as straight-forward as indicated by the early streak-camera formalism \cite{MairessePRA2005}.  
This is because the strong-field approximation is {\it not} a reliable approximation 
for laser-assisted photoionization.
Instead, it turns out that the long-range interaction 
between the photoelectron and the remaining ion leads to a shift of the time-delays \cite{ZhangPRA2010}, 
which often dominates over the one-photon, Wigner-like delays of the photoelectron \cite{YakovlevPRL2010}. 
Using time-dependent perturbation theory, 
we have shown that these ``measurement artifacts'' can be separated from 
the Wigner-like delays, as written explicitly in Eq.~(\ref{tauRABITT}), 
in the case of both APT \cite{KlunderPRL2011} and SAP \cite{NageleJPB2011}.   
Furthermore, we have shown that these temporal artifacts  
can be traced back to phase-shifts in ATI transition-matrix elements between continuum states, 
hence the name ``continuum--continuum phases'' \cite{DahlstromCP2012}. 
We have here presented both correction curves and analytical expressions 
for the purpose of correcting for such artifacts in the quest for the {\it true} delay in single-photon ionization.      

In this tutorial, we have limited ourselves to the SAE approximation, 
with a single photoelectron in a static atomic potential.  
While this approach is certainly valid for hydrogenic systems \cite{DahlstromCP2012}, 
it is likely to break down in many-electron atoms, when electron--hole interactions are strong
and when the electron relaxation plays an important role for the dynamics. 
Here, accurate calculations that include multi-electron screening effects of the ion are required. 
Although a few full-scale simulations have been made 
including many-electron dynamics \cite{KheifetsPRL2010,MoorePRA2011}, 
a conclusive agreement between experiment and theory is yet to be presented. 
This calls for {\it both} more experimental data and detailed theoretical work, 
and then a careful comparison of the two.  
In this way, the time-delay experiments provide an important testing ground 
for more exact analysis and interpretations in attosecond physics and strong-field physics. 
Only when understanding these detailed ultra-fast phenomena, 
will ``attophysics'' truely provide a route to 
accurate control and probing of electron processes in atomic and molecular systems in real time.

\subsubsection*{Acknowledgements:}
This research was supported by the Marie Curie program ATTOFEL (ITN), the European Research Council (ALMA), the Joint Research Programme ALADIN of Laserlab-Europe II, the Swedish Foundation for Strategic Research, the Swedish Research Council, the Knut and Alice Wallenberg Foundation, the French ANR-09-BLAN-0031-01 ATTO-WAVE program, COST Action CM0702 (CUSPFEL). 
Parts of the computations have been performed at IDRIS, and part of the work was carried out at the NORDITA workshop: ``Studying Quantum Mechanics in the Time Domain'', in Stockholm. 
I (J.M.D.) would like to thank 
University Lecturer Johan Mauritsson 
for guidence and discussions during my PhD-studies in Lund and for his comments on the manuscript; 
CNRS Research Director Richard Ta\"ieb 
for performing the exact computations on hydrogen in Fig.~\ref{ccphases} and for this comments on the manuscript;
Diego Gu\'enot 
for stimulating discussions and for help with preparing Fig.~\ref{harmonics};
and also Dr.~K.~Kl\"under, Dr.~M.~Gisselbrecht and the attophysics-group in Lund. 
I thank Dr.~T.~Carette for proof-reading the manuscript and 
I thank Professor Eva Lindroth for discussions about electron-correlation effects.

\bibliographystyle{unsrt}

\hyphenation{Post-Script Sprin-ger}




\end{document}